\documentclass[a4paper,fleqn,usenatbib]{mnras}

\usepackage{newtxtext,newtxmath}

\usepackage[T1]{fontenc}
\usepackage{ae,aecompl}

\usepackage{graphicx}	
\usepackage{amsmath}	

\newcommand{\oversim}[2]{\protect{\mbox{\lower0.5ex\vbox{%
  \baselineskip=0pt\lineskip=0.2ex
  \ialign{$\mathsurround=0pt #1\hfil##\hfil$\crcr#2\crcr\sim\crcr}}}}}
\newcommand{\simgreat}{\mbox{$\,\mathrel{\mathpalette\oversim>}\,$}} 
\newcommand{\simless} {\mbox{$\,\mathrel{\mathpalette\oversim<}\,$}} 
\newcommand{\simprop} {\mbox{$\,\mathrel{\mathpalette\oversim\propto}\,$}} 

\newcommand{\removed}[1]{}

\newcommand{\PhiN}{\phi_\text{N}}
\newcommand{\Phiph}{\Phi_\text{ph}}

\newcommand{\mcl}{M_\mathrm{oc}}
\newcommand{\rpl}{r_\mathrm{Pl}}
\usepackage{lineno}

\title[Milgrom's pr{\'a}h]{Asymmetrical tidal tails of open star
  clusters: stars crossing their cluster's pr{\'a}h\thanks{See the
    Acknowledgements for the meaning of 'pr{\'a}h'.} challenge Newtonian
  gravitation}

\author[Kroupa et al.]
{Pavel Kroupa$^{1,2}$\thanks{pavel.kroupa@mff.cuni.cz;
    pkroupa@uni-bonn.de},
 Tereza Jerabkova$^{3}$\thanks{Tereza.Jerabkova@eso.org},
 Ingo Thies$^2$,
  Jan Pflamm-Altenburg$^2$,
  \newauthor
  Benoit Famaey$^4$,
Henri M.J. Boffin$^3$,
  J\"org Dabringhausen$^1$,
Giacomo Beccari$^3$,
\newauthor
Timo Prusti$^5$,
  Christian Boily$^5$,
  Xufen Wu$^{7,8}$, 
Jaroslav Haas$^1$,
Hosein Haghi$^6$,
\newauthor
  Akram Hasani Zonoozi$^{2,6}$,
Guillaume Thomas$^{9,10}$,
Ladislav {\v S}ubr$^1$,
Sverre J. Aarseth$^{11}$
\\
$^1$Charles University in Prague, Faculty of Mathematics and Physics,
Astronomical Institute, V Hole\v{s}ovi\v{c}k\'ach 2, CZ-180
00 Praha 8, Czech Republic\\
$^2$Helmholtz-Institut f\"ur Strahlen- und Kernphysik, University
of Bonn, Nussallee 14-16, D-53115 Bonn, Germany\\
$^3$European Southern Observatory, Karl-Schwarzschild-Strasse 2, 85748
Garching bei M\"unchen, Germany\\
$^4$Observatoire astronomique de Strasbourg, 11, rue de l'Université
F-67000 Strasbourg, France\\
$^5$European Space Research and Technology Centre (ESA ESTEC),
Keplerlaan 1, 2201 AZ Nordwijk, Netherlands\\
$^6$Department of Physics, Institute for Advanced Studies in Basic
Sciences (IASBS), PO Box 11365-9161, Zanjan, Iran\\
$^7$CAS Key Laboratory for Research in Galaxies and Cosmology,
Department of Astronomy,
University of Science and Technology of China, \\ Hefei, 230026, P.R. China\\
$^8$School of Astronomy and Space Science, University of Science and Technology of China, Hefei 230026, China\\
$^9$Instituto de Astrofísica de Canarias, E-38205 La Laguna, Tenerife, Spain\\
$^{10}$Universidad de La Laguna, Dpto. Astrofísica, E-38206 La Laguna, Tenerife, Spain\\
$^{11}$Institute of Astronomy, University of Cambridge, Madingley Road, Cambridge CB3 0HA, UK\\
}

\pubyear{20xx}

\begin{document}
\label{firstpage}
\pagerange{\pageref{firstpage}--\pageref{lastpage}}
\maketitle

\begin{abstract}
  After their birth a significant fraction of all stars pass through
  the tidal threshold (pr{\' a}h) of their cluster of origin into the
  classical tidal tails. The asymmetry between the number of stars in
  the leading and trailing tails tests gravitational theory.  All five
  open clusters with tail data (Hyades, Praesepe, Coma Berenices,
  COIN-Gaia~13, NGC~752) have visibly more stars
  within~$d_{\rm cl}\approx 50\,$pc of their centre in their leading
  than their trailing tail. Using the
  Jerabkova-compact-convergent-point (CCP) method, the extended tails
  have been mapped out for four nearby~$600-2000\,$Myr old open
  clusters to $d_{\rm cl}>50\,$pc.  These are on near-circular
  Galactocentric orbits, a formula for estimating the orbital
  eccentricity of an open cluster being derived.  Applying the Phantom
  of Ramses code to this problem, in Newtonian gravitation the tails
  are near-symmetrical.  In Milgromian dynamics (MOND) the asymmetry
  reaches the observed values for $50 < d_{\rm cl}/{\rm pc} < 200$,
  being maximal near peri-galacticon, and can slightly invert near
  apo-galacticon, and the K\"upper epicyclic overdensities are
  asymmetrically spaced.  Clusters on circular orbits develop orbital
  eccentricity due to the asymmetrical spill-out, therewith spinning
  up opposite to their orbital angular momentum. This positive
  dynamical feedback suggests Milgromian open clusters to demise
  rapidly as their orbital eccentricity keeps increasing.  Future work
  is necessary to better delineate the tidal tails around open
  clusters of different ages and to develop a Milgromian direct
  $n$-body code.
\end{abstract}


\section{Introduction}
\label{sec:introd}
A galaxy is not populated by stars randomly, because observations show
stars to form predominantly as binary systems in embedded clusters
which emanate from sub-pc density maxima of molecular clouds
\citep{Kroupa95a, Kroupa95b, Porras+03, LL03, MarksKroupa11,
  Megeath+16, Dinnbier+22a}. The properties of these change with
galactocentric distance as a result of the varying local gas density
(\citealt{PflammKr08,Pflamm+13, Miville+17, Djordjevic+19, Wirth+22}).
Depending on the number of embedded clusters and their masses, a
molecular cloud can spawn an OB association upon the expulsion of
residual gas (e.g., \citealt{DW20, Dab+21}). The kinematical state of
the molecular cloud (e.g., it may be contracting if it formed from
converging gas flows) defines if the OB association is expanding,
still or contracting, or it may even be a mixture of these states
(e.g., \citealt{Wright+19, Kuhn+20, Armstrong+22} for evidence for
complicated kinematics). Thus, the expansion due to gas-expulsion of a
very young cluster can be masked by the still-ongoing radial infall of
very young stars from the surrounding molecular cloud. Overall, the
embedded clusters expand due to expulsion of their residual gas to the
sizes of the observed open clusters \citep{BK17}.

Each young open cluster is therefore expected to be surrounded by a
{\it natal cocoon} of stars. This natal cocoon consists~(i) of coeval
stars lost through the gas expulsion process forming a tidal tail~I
\citep{KAH,MoeckelBate10,DK20a,DK20b}, and~(ii) of stars that are
nearly co-eval and which formed in other embedded clusters within the
same molecular cloud \citep{DK20a,DK20b}.  The shape and extend of~(i)
and~(ii) can be used to age-date the cluster \citep{Dinnbier+22b}.
Evidence for natal cocoons around young ($30-300\,$Myr aged) open
clusters have been found and named ``coronae'' \citep{Meingast+20} or
``halos'' \citep{Bouma+21}.  Extreme examples of such natal cocoons
are the recently discovered few-dozen-Myr old relic filaments that are
interlaced with very young open clusters \citep{Jerabkova+19,
  Beccari+20}.  As the open cluster ages, the natal cocoon expands and
thins out while the classical kinematically cold tidal tail (referred
to as ``tidal tail~II'') develops over time through
energy-equipartition driven evaporation \citep{Henon69,
  Baumgardt+02,HeggieHut03}. For velocities $\approx 1\,$km/s of the
outgoing stars and at an age $>200\,$Myr, the natal cocoon has largely
dispersed from the $\approx 200\,$pc region around the cluster, and the
evaporating open cluster has grown a detectable classical cold
kinematical tidal tail~II \citep{DK20a,DK20b}.

The above processes contribute, together with resonances and
perturbations, to a complex distribution of Galactic disk stars in
phase space. The purpose of this contribution is to consider the
particular process of how stars spill from their open cluster into the
Galactic field in view of the newly-discovered asymmetry of the tidal
tail\footnote{Unless otherwise stated, ``tidal tail'' will refer to
  the classical tidal tail~II.} around the Hyades open cluster by
\cite{Jerabkova+21a}.  The asymmetry is a potentially decisive
diagnostic as to the nature of gravitation which drives cluster
dissolution through the process of energy equipartition within the
cluster.  As demonstrated in an accompanying publication
(Pflamm-Altenburg et al., in prep.), in Newtonian gravitation and for
a smooth Galactic potential, the leading and tidal tail must be
symmetric within Poisson noise due to the finite number of stars. The
ESO/ESA team (Jerabkova et al., in prep.; \citealt{Boffin+22}) has now
used the Gaia eDR3 to map out the extended tidal tails of three
further open clusters (Praesepe, Coma Berenices, NGC~752) with ages in
the range $600-2000\,$Myr.  As calculated by Pflamm-Altenburg et al.
(in prep.), the null hypothesis that the tails of Hyades and NGC~752
show the Newtonian symmetry, is in tension with the data with more
than 6.5~and 1.3~sigma confidence, respectively (the leading tail
having more stars in both cases).  An asymmetry can arise through a
very specific perturbation \citep{Jerabkova+21a}, but the same
perturbation cannot have affected the Hyades and NGC~752 in a similar
manner. Here the alternative hypothesis is tested if Milgromian
gravitation, which is non-linear and thus leads to a lopsided
equipotential surface around an open cluster \citep{Wu+10, Wu+17}, can
account for the amplitude and sign of the observed asymmetries.  This
contribution constitutes a first explorative step towards relaxational
stellar dynamics in Milgromian gravitation such that this topic can
only be superficially skimmed, pointing out where a future significant
research effort is needed to deepen our understanding of the observed
phenomena on star-cluster scales.

In the following, the formation and properties of tidal tails and the
observed extended tidal tails of four open clusters are described
(Sec.~\ref{sec:tidaltails}). Milgromian gravitation is introduced in
Sec.~\ref{sec:Mil}, with the here-applied simulation method of tidal
tails. The models are documented in Sec.~\ref{sec:models}.
Sec.~\ref{sec:results} contains the results, with Sec.~\ref{sec:efmix}
outlining the formation and evolution of embedded clusters through the
Newtonian into the Milgromian regime.  The conclusions with a
discussion and an outline for future work are provided in
Sec.~\ref{sec:concs}.

\section{Open clusters and their tidal tails}
\label{sec:tidaltails}

The tidal tail of a star cluster contains fundamental information on
the nature of gravitation, and a brief discussion of its formation and
evolution around an initially virialised star cluster in Newtonian
gravitation is provided (Sec.~\ref{sec:fillup}). This is followed by a
discussion of the extraction and properties of real tidal tails
(Sec.~\ref{sec:theTails}).

\subsection{The decomposition of open star clusters into their galaxy}
\label{sec:fillup}

\cite{BM03} performed direct Newtonian $n$-body models of initially
virialised clusters of different initial masses $M_{\rm oc, 0}$. These
orbit in a spherical logarithmic Galactic potential.  During the first
$\approx 50\,$Myr a cluster looses about 30~per cent of its initial
mass due to stellar evolution, assuming it to be populated with a
canonical stellar initial mass function (IMF, \citealt{Kroupa01}).  After these
$\approx 50\,$Myr, an open cluster evolves through
two-body-relaxation-driven evaporation that continuously pushes stars
across its tidal pr{\'a}h, thereby populating the classical tidal
tail~II.  The cluster thus slowly dissolves with a near-constant
mass-loss rate,
$\dot{M}_{\rm oc} \approx 0.7\,M_{\rm oc, 0} / T_{\rm diss, 0}$.  The
lifetime of an open cluster starting with a canonical IMF (average
stellar mass $\bar{m} \approx 0.55\,M_\odot$, table~4-1 in
\citealt{Kroupa+13}) can be approximated to be (eq.~7 in
\citealt{BM03})
\begin{equation}
  {T_{\rm diss} \over {\rm Myr}}  \approx 0.86 \, G^{-{1\over 2}} \, \left[   {M_{\rm oc}
      \over \bar{m} \,  {\rm ln}\left(0.02\,M_{\rm oc}/\bar{m} \right)} \right]^{0.79},
\label{eq:lifet}
\end{equation}
where
$G \approx 0.0045 \, M_\odot^{-1} \, {\rm pc}^3 \, {\rm Myr}^{-2}$ is
the gravitational constant.  Thus, a very young open cluster weighing
$M_{\rm oc, 0}=1300\,M_\odot(=M_{\rm oc})$ dissolves within
$T_{\rm diss,0}\approx 2\,$Gyr.  The remaining life-time of an
  already evolved open cluster (with
  $\tau_{\rm oc}/T_{\rm diss, 0} \simgreat 0.3$, $\tau_{\rm oc}$ being
  its current astrophysical age) can be approximated by
  Eq.~\ref{eq:lifet} with $\bar{m} \approx 0.75\,M_\odot$ because it
  has lost most of its lowest-mass stars. The present-day
  Hyades ($M_{\rm oc} \approx 275\,M_\odot$) will thus dissolve in
  about 800~Myr.  These estimates provide a useful orientation of the
  life-times of open clusters but are uncertain since they rely on
  simulations made for significantly more massive cluster models not
  orbing near the mid-plane of the Galaxy, and since individual open
  clusters containing $n<10000$ stars follow evolutionary tracks that
  increasingly diverge from each other for smaller $n$ due to the
  chaotic nature of small-$n$ dynamics.

\begin{figure}
        \centering
        \includegraphics[width=\hsize]{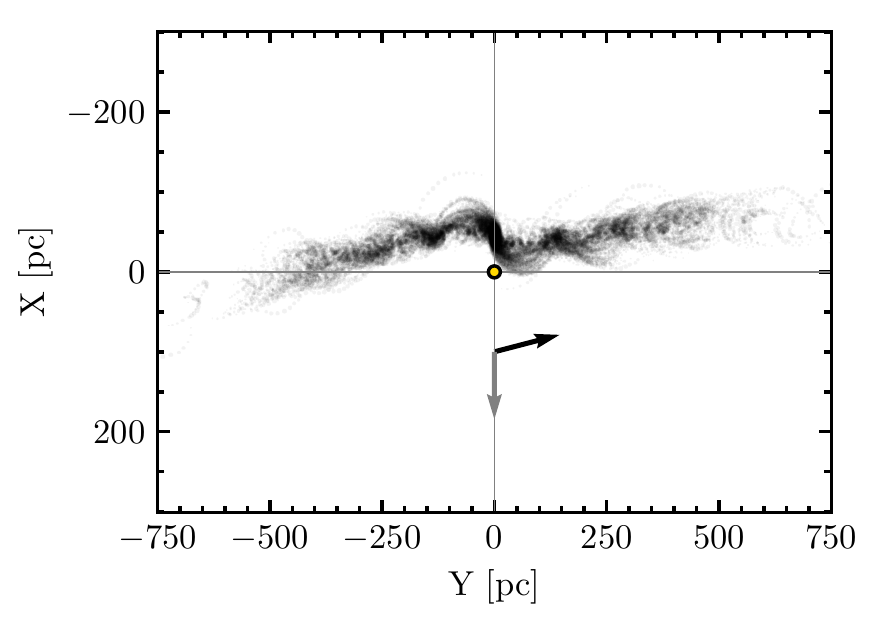}
        \caption{ Direct Newtonian $n$-body computation of a
          Hyades-like star cluster with initial mass
          $M_{\rm oc, 0}=1235\,M_\odot$ and initial half-mass radius
          $r_{{\rm h}0}=2.6\,$pc in a realistic Galactic
          potential. Stellar particles are shown for a series of
          snapshots at 620, 625, 630, 635, 640, 645, 650, 655, 660,
          665, 670, 675, 680, 685, 690, and 695 Myr~in Galactic
          Cartesian coordinates. The Sun is marked as a yellow
          point. The grey arrow points to the Galactic centre, and the
          black arrow is the cluster velocity vector in the
          corresponding coordinates, with the Z-axis pointing towards
          the north galactic pole. The time stacking of snapshots
          shows the movement of individual stars to and from the
          K\"upper epicyclic overdensities. The realistic star cluster
          trajectory with excursions out of the Galactic plane does
          not significantly affect the physical appearance of the
          tails, while projection effects as seen from the Sun do
          (model~M1 and figure with permission from
          \citealt{Jerabkova+21a}).}
        \label{fig:YX_M}
\end{figure}

  The escape process of stars from their cluster is complex. Stars
  that are {\it ejected} typically leave the cluster faster than the
  velocity dispersion after an energetic close encounter with a binary
  in the cluster. Energy-equipartition-driven {\it evaporation}, on
  the other hand, produces stars that have very small velocities
  relative to the cluster's centre of mass. This two-body-relaxation
  driven process dominates by far the flux of stars from the cluster,
  at least after the initial binary population has been mostly
  dynamically processed \citep{Kroupa95c}.  Evaporation can be
  visualised by noting that at any time the velocity distribution
  function of stars in the cluster is approximately Maxwellian. The
  high-velocity tail is lost across the pr{\'a}h and is constantly
  refilled stochastically through two-body relaxation.  Even after
  being formally unbound (having a positive energy) a star can orbit
  many times around and even through the cluster such that each
  cluster is surrounded by a ``halo of lingering stars'' waiting to
  exit \citep{FH00}. Whether the star exits near the inner L1~Lagrange
  point, or the outer~L2 point, is random as it results from the
  accumulation of many uncorrelated stellar-orbital perturbations
  (Pflamm-Altenburg et al., in prep.).  At galactocentric distances
  significantly larger than the tidal radius of the cluster,
  $r_{\rm tid}$ (Eq.~\ref{eq:rtid} below), the cluster-centric
  potential is symmetric in Newtonian gravitation and~L1 and~L2 are
  equidistantly placed at a distance $r_{\rm tid}$ from the cluster
  centre-of-mass (CCoM) along the line joining the
  galactic-centre--CCoM. Due to this symmetry, the leading and
  trailing tails contain the same number of stars to within Poisson
  fluctuations.  This follows from detailed calculations of stellar
  orbits (\citealt{Just+09}; Pflamm-Altenburg et al., in prep.) as
  well as in the standard linearised treatment of the tidal field
  \citep{ChumakRastorguev06a,Ernst+11}.  The expected symmetry of the
  tails is demonstrated in Fig.~\ref{fig:YX_M} for a Hyades-like open
  star cluster.

Since the dispersion of velocities of the evaporated stars is
comparable to the velocity dispersion of the cluster, the leading and
trailing tidal tails are kinematically cold.  Each star that leaves
the cluster is on its own rosette orbit about the galaxy. While the
rosette orbits belonging to one of the two tails are all
next-to-equal, they are phase shifted relative to each other and form
a complex superposition pattern of stellar density along the tidal
tail. The motion of the stars can be mathematically approximated as
epicyclic motions relative to the local guiding centre which is the
CCoM, a detailed analysis being provided by \cite{Just+09}.  The stars
accumulate where their velocities relative to the local circular
velocity are slowest and form regularly-spaced {\it K\"upper epicyclic
  overdensities} along the tails \citep{Kuepper+08, Just+09,
  Kuepper+12}.  For clusters not on circular orbits, the same holds,
except that the form of the tails becomes more complicated and
time-dependent \citep{Kuepper+10}.  The spacing of the K\"upper
overdensities is a sensitive function of the gravitational potential
of the galaxy and of the mass of the cluster which defines the
velocity dispersion of the evaporating stars: a larger escape speed
implies a larger velocity dispersion and less-well defined and more
distantly-spaced overdensities \citep{Kuepper+15}.  In Newtonian
gravitation, the K\"upper overdensities are symmetrically and
periodically spaced along the tidal tails.  The K\"upper overdensities
have been detected but not recognised as such around the globular
cluster Pal~5 (\citealt{Odenkirchen+01, Odenkirchen+03}, sec.~7.3
therein), but have been discussed as such by \cite{Erkal+17}[see also
Sec.~\ref{sec:mass_veldisp}]. A fully-dissolved massive-born cluster
left the stellar stream GD-1 that has been observed to harbour strong
evidence for regularly-spaced K\"upper overdensities \citep{Ibata+20},
and \cite{Jerabkova+21a} discovered the overdensities for the first
time around an open cluster (the Hyades). Apart from being
gravitational probes, in association with the current mass of the
cluster, the tidal tails can test whether the IMF is a probabilistic
or an optimal distribution function \citep{WJ21}. Once the entire
population of stars in the tidal tails is known, then the rate of
dissolution of an open cluster can be tested for -- can the Newtonian
expectation (Eq.~\ref{eq:lifet}) be confirmed?

The open cluster leaves, as its remnant, a high-order multiple stellar
system, often made of similar-mass stars \citep{Fuente98, Fuente98b,
  Angelo+19}, or a dark cluster dominated by stellar remnants
\citep{BK11}. The tidal stream becomes indistinguishable from the
Galactic field population as it spreads in length and thickens through
perturbations \citep{ChumakRastorguev06a}, a process not yet well
understood.  Given that embedded clusters loose between~50 to~80~per
cent of their stars through gas expulsion \citep{KB02, Brinkmann+17},
it is to be expected that about 20~per cent to a half of all stars
would have been released into a galactic field through classical tidal
tails~II. The spilling-out of stars from their clusters is thus
important for defining the stellar phase-space distribution function
of a galaxy.


\subsection{Extraction from the field and the discovery of the tidal tail
  asymmetry}
\label{sec:theTails}

Detecting and interpreting the properties of tidal tails is easiest
for clusters on circular orbits. By the nature of their origin from
the inter-stellar medium of their host galaxy, young (age
$\tau_{\rm oc}\simless 100\,$Myr) to intermediate-aged
($100 \simless \, \tau_{\rm oc}/{\rm Myr}\, \simless 1500$) open
clusters are on near-circular orbits within the disk of their galaxy
\citep{ChumakRastorguev06, Carrera+21}.  With the Gaia mission, open
star clusters have become prime targets to analyse tidal tails.

The tidal tails around open clusters are difficult to extract from the
field population of the Galaxy, since the tail stars are already part
of the field population and because they comprise a small fraction of
stars in a given volume near an open cluster. Thus, returning to the
example of Fig.~\ref{fig:YX_M}, the open cluster with
$M_{\rm oc, 0}=1235\,M_\odot$ will have~$\approx 1180$~stars in the
tails at an age of $\approx 650$~Myr, given the cluster looses~30~per
cent of its mass due to stellar evolution and the average stellar mass
is $\approx 0.5\,M_\odot$. From Fig.~\ref{fig:YX_M}, each tail can be
approximated as a cylinder with a radius of~$30\,$pc and a length
of~$600\,$pc corresponding to a volume of $1.7 \times 10^6\,$pc$^3$.
The stellar number density in the Solar neighbourhood is about one
star per~pc$^3$ such that the one-sided tail volume contains about
$1.7 \times 10^6\,$ field stars while consisting of~590 ex-cluster
stars.

The members of a dissolving cluster that are still close to the
cluster have nearly parallel velocity vectors such that their proper
motion directions, when plotted as great circles on the celestial
sphere, intersect at two opposite convergent points. Using this
convergent point (CP, \citealt{Smart1939,Jos1999,vanLeeuwen09}) method
to identify co-moving stars in the Gaia DR2, the parts of the tails
closest to their clusters have been found around the Hyades
\citep{Roeser+19,MeingastAlves19}, the Praesepe \citep{RS19} and Coma
Berenices \citep{Tang+19, Fuernkranz+19}. These three clusters have
similar intermediate-ages (Table~\ref{tab:clusterdata}) such that the
stars born in them do not stand-out from the field population in a
colour-magnitude diagram.  They are close-by to the Sun, within
190~pc, such that the Gaia data are now allowing the tidal tails to be
extracted from the background field population.  Using the CP method,
the tails of the Hyades (fig.~3 in \citealt{Roeser+19}, fig.~1 in
\citealt{MeingastAlves19}), Coma Berenices (fig.~7 in
\citealt{Tang+19}, fig.~2 in \citealt{Fuernkranz+19}) and the Praesepe
(fig.~2 in \citealt{RS19}) have been mapped out to distances of
about~$d_{\rm cl}\approx 30$ to~$170\,$pc from their clusters. All
leading tails are more populated than the trailing tails. In addition,
\cite{Bai+22} report the tidal tails of the open cluster COIN-Gaia~13
which is about 513~pc distant, has an age of about~250~Myr and a mass
of about $439\,M_\odot$. Its leading tail also appears to contain more
stars than its trailing tail (their fig.~4).  Given the importance of
mapping out tidal tails for the fundamental questions noted above, it
would be desirable to trace the tails to their end tips. But the CP
method cannot pick-up stars in the extended tails because their space
motions differ systematically from the CCoM.

The new {\it compact convergent point method} (CCP) introduced by
\cite{Jerabkova+21a} uses the information that a tidal tail is a
causally correlated but extended kinematical structure.  The
Jerabkova-CCP method uses direct $n$-body computations of each open
cluster to quantify a transformation which maps the extended tidal
tail phase-space structure into a compact configuration for stars that
belong to the structure.  This allows extraction of strong candidates
from the background field population even to the tips of the tidal
tails. The K\"upper overdensities were found for the first time for an
open cluster using this method \citep{Jerabkova+21a}.  The CCP method
is not very sensitive to the exact dynamical age of the so-designed
model since the tails develop slowly and on a much longer time-scale
than the present-day half-mass crossing time in the cluster (being
about $10\,$Myr for the Hyades).

How confidently are the extended tidal tails mapped out using the
  new CCP method?  The CCP method uses the older CP method as a
  benchmark. The CP method is essentially a zeroth-order approximation
  as it assumes the tail stars to share the same space motion as the
  bulk of the cluster.  The CCP method relies on an $n$-body model of
  the star cluster to have been evolved to its present-day position
  and velocity such that the calculated tidal tails are used to filter
  out, from the Gaia data, those stars that match, within a tolerance,
  the phase-space occupied by the model tails.  In this procedure the
  zeroth order approximation in the model needs to match the
  CP-derived tidal tails exactly, therewith enforcing a strict anchor
  for the CCP method.  Due to uncertainties in the orbit determination
  the final model needs to be adjusted in position and velocity (the
  same shift in 6D space for all stars in the model) to agree with the
  real cluster position.  This adjusted model is then compared to the
  tidal tails recovered with the CP method. The details are documented
  in \cite{Jerabkova+21a}.  It will be important to continue testing
  the CCP method with the newer Gaia DR3 and DR4 including
  complementary data (e.g., radial velocities, stellar spin rates,
  chemical tagging) and further $n$-body models to more
  comprehensively quantify the degree of uncertainty, bias and
  large-scale completeness in the derived tidal tails.

\begin{figure*}
        \centering
\includegraphics[scale=1.4]{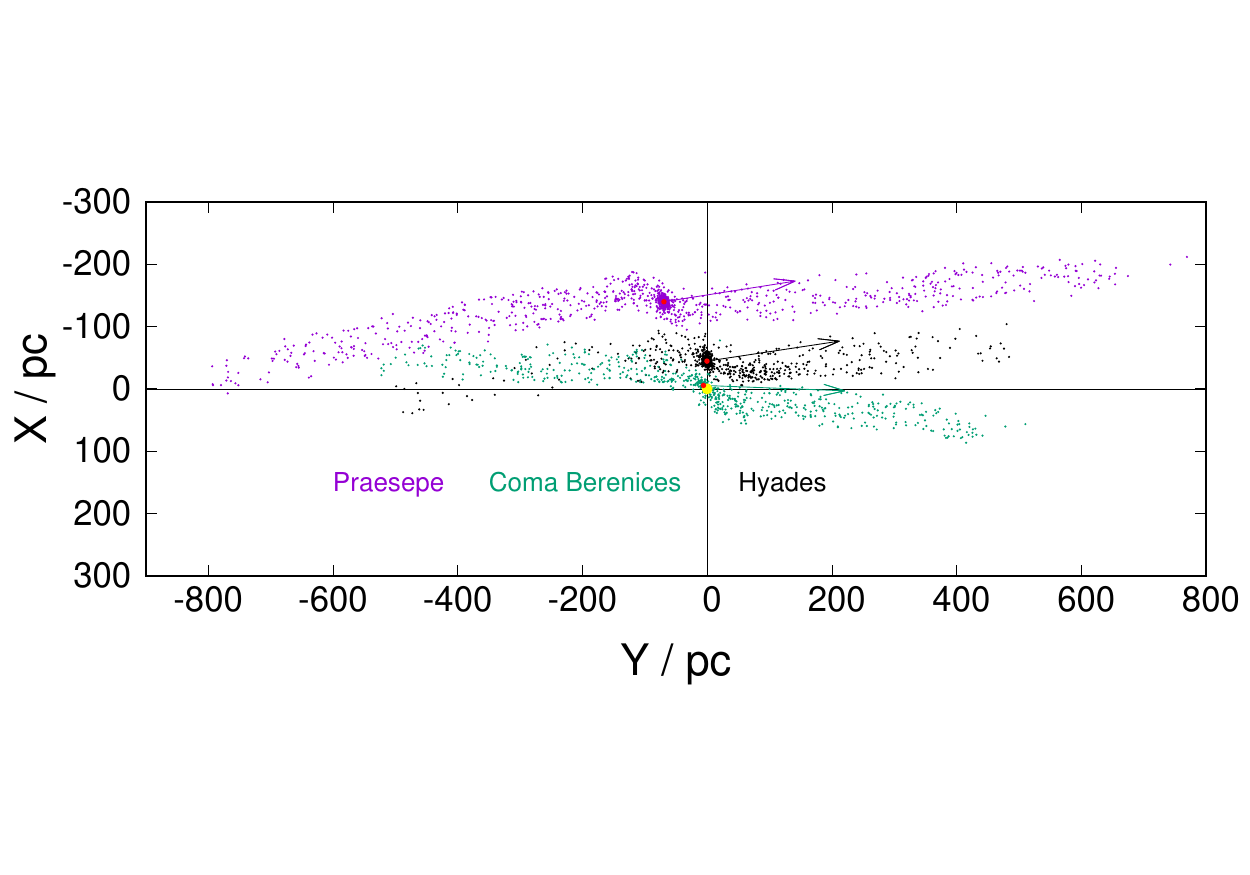}\\[-6.1cm]
\includegraphics[scale=1.4]{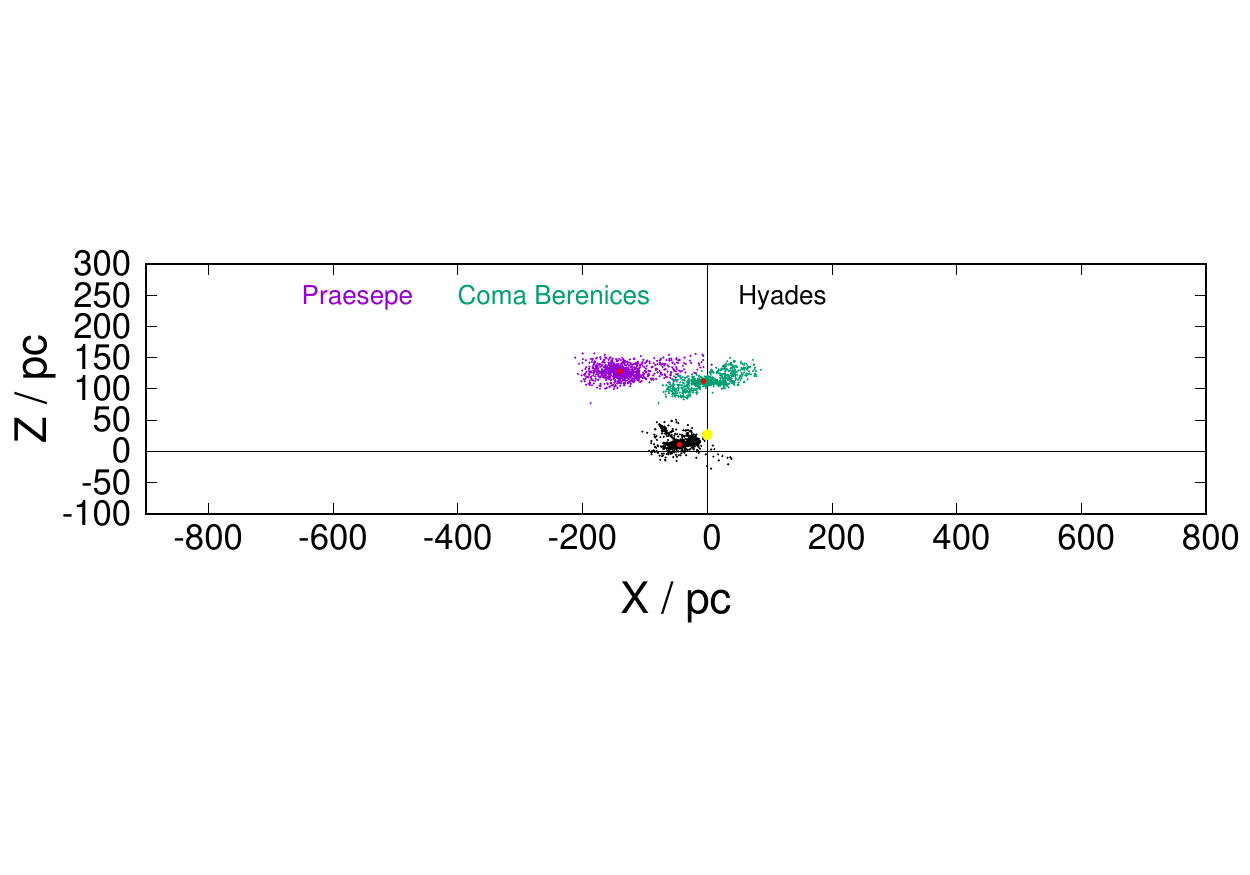}\\[-7cm]
\includegraphics[scale=1.4]{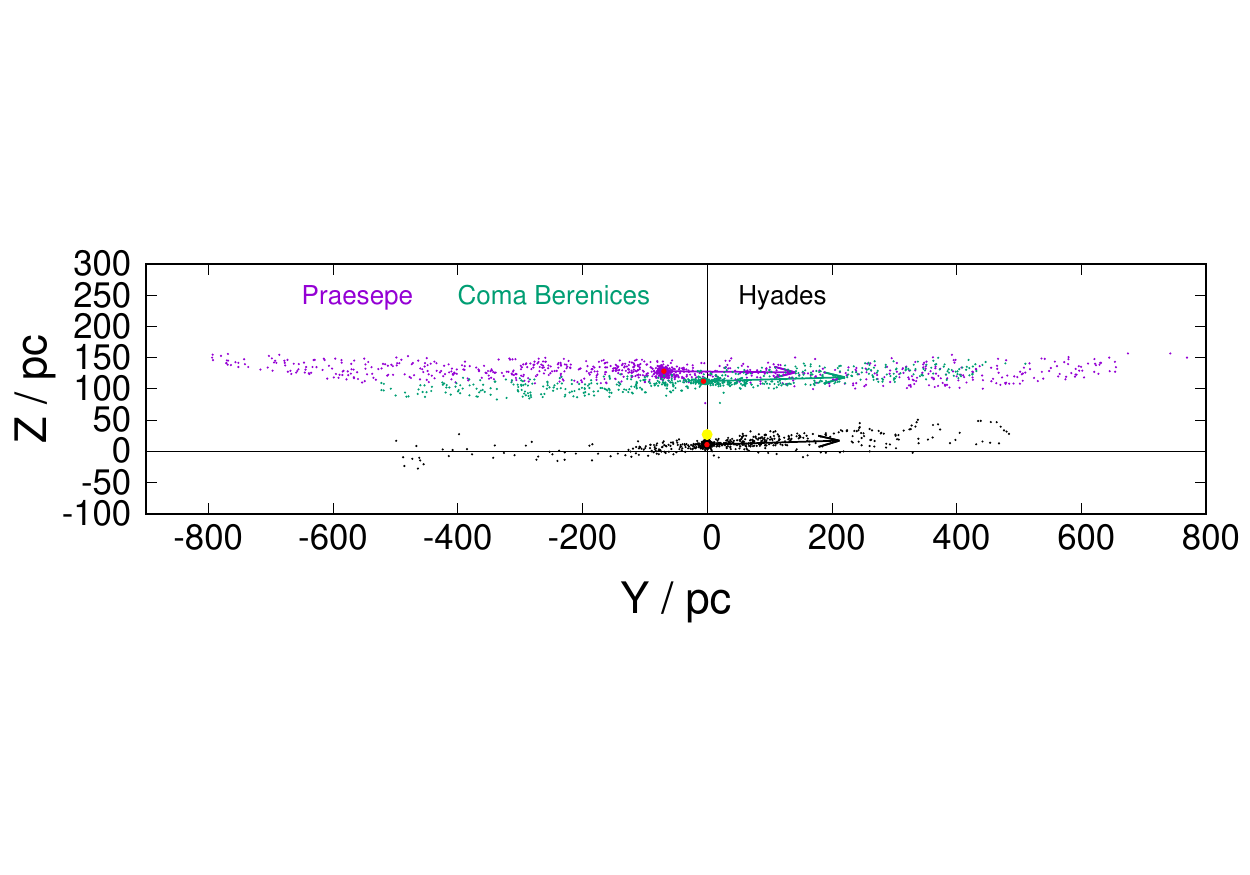}\\[-3.7cm]
\caption{The classical tidal tails~II extracted from the Gaia eDR3
  using the CCP method by \citet{Jerabkova+21a} for the Hyades, and
  for Coma Berenices and Praesepe by (Jerabkova et al., in prep.), in
    the three projections in Galactic Cartesian coordinates. The
    Galactic centre is towards positive $X$ and Galactic rotation
    points towards positive $Y$.  The coordinate system is anchored at
    $(X, Y, Z) = (0, 0, 0)\,$pc and the Sun is the filled yellow
    circle at $(0, 0, +27)\,$pc. The centre of each cluster is
    indicated by the filled red dot. Note that Coma Berenices lies
    almost directly above the Sun towards the Galactic north pole. The
    arrows show the full orbital and peculiar motions of each cluster
    with a length corresponding to $V_{\rm X, Y, Z}$ in
    Table~\ref{tab:clusterdata}. The middle panel does not show the
    velocity arrows as these are directed mostly into the plane.}
        \label{fig:realtails}
    \end{figure*}

  The CCP method has been applied to the the similarly-aged
  ($600-800\,$Myr) open clusters Hyades \citep{Jerabkova+21a},
  Praesepe and Coma Berenices (Jerabkova et al., in prep.). The
  extracted tidal tails are shown in
  Fig.~\ref{fig:realtails}. \cite{Boffin+22} applied a slightly
  amended Jerabkova-CCP method to the $\approx 1.75\,$Gyr old open
  cluster NGC~752, finding the tails to be extended for at least
  $260\,$pc from tip to tip and the leading tail to also contain more
  stars than the trailing one.

  In summary, it is noteworthy that all five open clusters (Hyades,
  Praesepe, Coma Berenices, COIN-Gaia~13 and NGC~752) which have tail
  data have the leading tail more populated than the trailing tail
  within a cluster-centric distance of $d_{\rm cl}\approx 50\,$pc.  To
  avoid the halo of unbound stars lingering around each cluster and
  numerical resolution contraints, we here only study the tail
  asymmetry in the distance range $50< d_{\rm cl}/{\rm pc} < 200$ for
  the three similarly-aged clusters Hyades, Praesepe and Coma
  Berenices with corresponding data (Sec.~\ref{sec:properties}), but
  the caveat be added that the results on Praesepe and Coma Ber are
  preliminary. The NGC~752 cluster is much older, further from us and
  the tail data only reach to $d_{\rm cl}\approx 130\,$pc, such that
  this cluster is not currently accessible to the MOND models with the
  available approximative means, as discussed below.

\subsection{Properties of the Hyades, Praesepe, Coma Berenices,
  NGC~752}
\label{sec:properties}

In order to take the first step towards constraining the possible
origin for the observed tail asymmetry evident in the previous work
using the CP method (Sec.~\ref{sec:theTails}) and the new results
applying the CCP method (Fig.~\ref{fig:realtails}), the known data on
the four clusters with CCP-extracted tail data are collated in
Table~\ref{tab:clusterdata}.

The table lists the positions and velocities of the four clusters, and
also different estimates for their orbital properties in the Galactic
potential. The true orbits are somewhat uncertain because the Galactic
potential is uncertain. Two methods are applied to estimate the
orbital eccentricity for each cluster: The first method
(Pflamm-Altenburg et al., in prep.) assumes the Galactic potential as
given in \cite{Allen1991}.  Each cluster is integrated backwards with
a time-symmetric Hermite method \citep{Kokubo+98} for the nominal age
as given in Table~\ref{tab:clusterdata}.  For this purpose, the
observed postions and velocities of the star clusters in the
equatorial system are converted into a Galactic inertial rest frame
using a Solar position of $(-8300\,\mathrm{pc}, 0, 27\,\mathrm{pc})$
and a velocity of
$11.1\,\mathrm{km/s}, 232.24\,\mathrm{km/s}, 7.25\,\mathrm{km/s}$, as
in \cite{Jerabkova+21a}.  This provides the most recent peri- and
apo-galacticon distances, $R_{\rm peri}, R_{\rm apo}$, respectively,
and the maximum excursion from the Galactic mid-plane, $Z_{\rm
  max}$. The orbital eccentricity, $e$, follows from
\begin{equation}
  e =  \left( R_{\rm apo} - R_{\rm peri} \right)
  / \left( R_{\rm apo} + R_{\rm  peri} \right)\,.
   \label{eq:ecc}
\end{equation}
An alternative estimate of the orbital eccentricity, $e_{\rm snap}$, follows from the
current position and velocity data of the cluster, i.e., from the
present-day snapshot. The method assumes the Galaxy has a flat
rotation curve, the details being  provided in Appendix~A.

%
\begin{table*}
\begin{tabular}{c|c|c|c|c}
  Name        & Hyades       &     Praesepe            &  Coma Berenices  & NGC~752       \\
  \hline  
  alt.  names &  Mel 25      &     Mel 88, M44, NGC 2632  &  Mel 111         & Mel~12        \\
  RA(J2000)          & 04h 31min 56.4s (1)&     08h 40min 12.9s (3) &  12h 25min 06s (4)  & 01h 56min 39.21s (11)  \\
  DEC(J2000)          & 17.012$^\circ$ (1)& 19.621 (3) &
                                                         26.100$^\circ$ (4) & 37.795$^\circ$ (11) \\
  $\varpi / \rm mas$  & 21.052 (2)& 5.361 (3) & 11.640 (2)& 2.281 (11) \\
  $d / \rm pc$        &   47.5     &    186.5      &   85.9      & 438        \\
  $\mu_{\alpha*} / \rm mas\,yr^{-1}$  & 101.005 (2)& -36.090 (3) & -12.111 (2) & 9.77 (11)\\
  $\mu_{\delta} / \rm mas\,yr^{-1}$  & -28.490 (2)& -12.919 (3)& -8.996 (2) & -11.78 (11)\\
  $v_\mathrm{los} / \rm km\,s^{-1}$  & 39.96 (2)& 35.84 (2)& -0.52 (2) & 8.2 (11)\\

  \hline

    $(V_{\rm X}, V_{\rm Y}, V_{\rm Z}) / {\rm pc\,Myr}^{-1}$ 
      &  (-32.01,  212.37,  6.13)
      &(-32.56,  216.53,  -2.74)
      & (8.92, 231.78, 6.29)
      &(-8.37, 221.02, -13.10)\\
  $V_{\rm tot} / \rm pc\,Myr^{-1}$  & 220.3 &219.0 & 232.0 & 221.6\\

  $(X, Y, Z) / {\rm pc}$ 
              & (-8344.44,  0.06, 10.22)
              &(-8441.57,  -68.90,  127.03)
             & (-8306.71,  -5.91,  112.44)
          & (-8294.05, 275.07, -158.408)\\

  $R/$pc  &8344.4     &8441.9   &8306.7  &8298.6 \\
  $R_{\rm apo}/$pc & 9013.4     &9060.4   &8937.4  &8604.7 \\
  $R_{\rm peri}/$pc &7311.7    &7375.0    &8217.7  &7746.8 \\
$Z_{\rm max}/$pc & -63.8  & 130.8   &   -136.0   & 213.4   \\
  
  $e$   &0.104  & 0.103  &0.042  &0.052 \\
  $e_{\rm snap}({\rm app})$   &0.115  & 0.114  & 0.038  & 0.0567 \\
  $e_{\rm snap}({\rm num})$   &0.105  & 0.103  & 0.041  & 0.0538 \\

  ${\rm d}R_{\rm GC}/{\rm d}t$ (sign)   &$+1$  & $+1$  & $-1$  & $+1$ \\ 
  
\hline

  $\tau_\mathrm{oc} / \rm Myr $ & 580--720 (10)& 708--832 (6) &700--800 (9)&
                                                          $\approx 1.75$ (11)\\
  $M_\mathrm{oc} / M_\odot$ & 275 (5)& 311 (6)& 112 (8)& 379 (11)\\
  $r_\mathrm{h} / \mathrm{pc} $  & 4.1 (5)& 4.8 (6) & 3.5 (7)& 5.3 (11) \\
  $r_\mathrm{tid} / \mathrm{pc}$  &9.0 (5) & 10.77 (6)& 6.9 (9)& 9.4 (11)\\
$r_{\rmn M}$/pc &0.57 &0.61 &0.36 &0.67 \\
  $N_\mathrm{tot}$  & 862 (1) & 1170 (12) & 730 (12) &640 (11) \\
  $N_\mathrm{tidal}$  & 541 (1) & 833 (12)& 640 (12)& 298 (11)\\
  $N_\mathrm{lead}$  & 351 (1) & 384 (12)& 348 (12) &163 (11)\\
  $N_\mathrm{trail}$  &190 (1) &449 (12)& 292 (12)& 135 (11)\\
  $N_\mathrm{lead}(50-\gamma\,\rm pc)$  &162($\gamma=200$)
                             &87($\gamma=200$) &133($\gamma=200$) &56($\gamma=130$)\\
  $N_\mathrm{trail}(50-\gamma\,\rm pc)$  &64($\gamma=200$) &
                                                             140($\gamma=200$) &111($\gamma=200$) &43($\gamma=130$)\\
  $q_{\rm 50-\gamma\,pc}$         &$2.53 \pm 0.37$($\gamma=200$) &$0.62\pm 0.08$($\gamma=200$)
                                           &$1.20\pm0.15$($\gamma=200$)&$1.30\pm 0.24$($\gamma=130$)\\            
  
  \hline

  $a_\mathrm{int} / \mathrm{pc}\,\mathrm{Myr}^{-2}$  & 0.037 & 0.030 & 0.021 & 0.030\\
  $a_\mathrm{ext,kin} / \mathrm{pc}\,\mathrm{Myr}^{-2}$  & 7.5 & 7.4& 7.5& 7.5\\
  $a_\mathrm{ext,bary} / \mathrm{pc}\,\mathrm{Myr}^{-2}$  & 4.1 & 4.1& 4.2& 4.3\\
  $\sigma_\mathrm{M,iso,los} / \rm km\,s^{-1}$  &  0.69 & 0.71& 0.55& 0.75\\
  $\sigma_\mathrm{M,ef,los} / \rm km\,s^{-1}$  &  0.29 & 0.28& 0.20& 0.28\\
  $\sigma_\mathrm{N,los} / \rm km\,s^{-1}$  &  0.22 & 0.22 & 0.15& 0.23\\
\end{tabular}
\caption{
  The present-day cluster parameters for the Hyades, Praesepe,
  Coma Berenices and NGC~752.  These are from top to bottom:
  alternative name, right ascension (RA), declination (DEC), both
  epoch J2000), and
  parallax, $\varpi$, of the cluster's centre, it's distance from the
  Sun, $d$, and proper motion in RA, $\mu_{\alpha*}$, in DEC,
  $\mu_{\delta}$, and line of sight velocity, $v_\mathrm{los}$.
  Assuming the Galactic potential of \citet{Allen1991} and the Solar
  position ($R_{{\rm GC}\odot}=8\,300\,$pc) and velocity vector as
  given in \citet{Jerabkova+21a}: the velocity components
  $V_{\rm X}, V_{\rm Y}, V_{\rm Z}$ and the total velocity or speed,
  $v_{\rm tot}$, current Galactocentric distance, $R$, its last apo-,
  $R_{\rm apo}$, and last peri-galactic distance, $R_{\rm peri}$, the
  last maximum orbital excursion away from the Galactic midplane,
  $Z_{\rm max}$, the orbital eccentricities, $e$ (Eq.~\ref{eq:ecc},
  obtained by full orbit integration in the Galactic potential from
  \citealt{Allen1991}), $e_{\rm snap}({\rm app})$ (obtained from the
  current postion and velocity vector and the approximative solution
  for a flat rotation curve with $v_\mathrm{circ}=220\,$km/s,
  see Appendix~A), and $e_{\rm snap}({\rm num})$ (obtained from the
  current postion and velocity vector and the Newton-Raphson solution,
  see Appendix~A), (Sec.~\ref{sec:properties}), and the sign of the
  Galactocentric radial velocity component ($+1=\,$receding,
  $-1=\,$approaching the Galactic centre).  The constrained age range
  is given by $\tau_{\rm oc}$ and the stellar mass is $M_{\rm oc}$
  within the tidal radius, $r_{\rm tid}$ (Eq.~\ref{eq:rtid}). The
  half-mass radius is $r_{\rm h}$. The MOND radius
  (Eq.~\ref{eq:rmond}) is $r_{\rm M}$.  The number of stars found
  using the Jerabkova-CCP method in the cluster and in the tidal
  tail~II is, respectively, $N_{\rm tot}$, $N_{\rm tidal}$.
  $N_{\rm lead}$ and $N_{\rm trail}$ stars are in the leading and
  trailing tails, respectively. Within the distance $50-200\,$pc,
  $N_\mathrm{lead}(50-200\,\rm pc)$ of these stars are in the leading
  tail and $N_\mathrm{trail}(50-200\,\rm pc)$ are in the trailing
  tail, the ratio of these being given by $q_{\rm 50-200\,pc}$. The
  following numbers assume the cluster has a present-day stellar mass
  $M_{\rm oc}$: The internal acceleration is $a_{\rm int}$
  (Eq.~\ref{eq:aint}), the external acceleration (Eq.~\ref{eq:ext}) is
  $a_{\rm ext,kin} = a_{\rm bulge} + a _{\rm disk} + a_{\rm halo}$
  (adopting here $v_{\rm circ}=250\,$km/s in Eq.~\ref{eq:ext}; this
  includes the total gravitational acceleration in Milgromian or
  equivalently in Newtonian gravitation with the dark matter halo such
  that the rotation curve is as shown in Fig.~\ref{fig:rc}) and
  $a_{\rm ext,bary} = a_{\rm bulge} + a _{\rm disk}$ (i.e., only
  Newtonian-baryonic, without the phantom dark matter halo,
  Fig.~\ref{fig:rc}). The line-of-sight (1D) velocity dispersion in
  Milgromian gravitation is $\sigma_\mathrm{M,iso,los}$ assuming the
  cluster is isolated (Eq.~\ref{eq:vdispMONDiso}) and
  $\sigma_\mathrm{M,ef,los}$ assuming it is situated within an EF
  given by $a_{\rm ext, kin}$ (Eq.~\ref{eq:vdispMONDef}).  The
  Newtonian line-of-light (1D) velocity dispersion is
  Eq.~\ref{eq:vdispMONDef} with $G=G_{\rm eff}$.  {\it References}:
  (1) \citet{Jerabkova+21a},
  (2) \citet{GDR2_Hyades},
  (3) \citet{cantat-gaudin2018a},
  (4) \citet{dias2014a},
  (5) \citet{Roeser+11},
  (6) \citet{RS19},
  (7) \citet{krause2016a},
  (8) \citet{KH07},
  (9) \citet{Tang+19},
  (10) \citet{Roeser+19},
  (11) \citet{Boffin+22},
  (12) Jerabkova et al. (in prep.).
}
\label{tab:clusterdata} 
\end{table*}

In order to focus on the parts of the tidal tails closest to the
clusters but sufficiently far to avoid the halo of lingering stars
(Sec.~\ref{sec:fillup}) and to be outside the tidal radius
(Eq.~\ref{eq:rtid}) of the models developed below, the number of stars
is counted in the leading and trailing tails in the distance range
50--200~pc from the clusters. That the K\"upper overdensities are
located within this distance in the real open clusters (e.g., the
first K\"upper overdensity is expected to be, in Newtonian-models, at
about $d_{\rm cl}=130\,$pc for the Hyades, Fig.~\ref{fig:YX_M}), does
not affect the ratio since the stars drift along the tail and through
the overdensity which only very slowly shifts closer to the cluster as
the cluster evaporates (Sec.~\ref{sec:fillup}).  The number ratio of
leading to trailing tail stars in the distance range $50-200\,$pc,
$q_{\rm 50-200\,pc}$, is listed in Table~\ref{tab:clusterdata}.  The
Hyades and NGC~752 show a similar asymmetry, with the leading tail
containing, respectively, $2.53 \pm0.37$ and $1.30\pm0.24$~times as
many stars than the trailing tail between, respectively,
$d_{\rm cl}=50$~and~200~pc and 50~to 130~pc from the CCoM. This is in
6.5~sigma tension for the former with Newtonian models, assuming the
Hyades orbits in a smooth axisymmetric \cite{Allen1991} Galactic
potential (Pflamm-Altenburg et al., in prep.).

The observed degree of asymmetry for the Hyades can be obtained
through an on-going encounter with a dark lump of mass
$\approx 10^7\,M_\odot$ \citep{Jerabkova+21a}.  This is a reasonable
hypothesis to explain the asymmetry for one cluster. But such a dark
lump is not observed in the form of a molecular cloud (fig.~11 in
\citealt{Miville+17}; the Sun and Hyades cluster lying within the
local cavity of low-density, high-temperature plasma of radius
$\approx 150\,$pc, \citealt{Zucker+22}), and if it were a dark-matter
sub-halo, a similar type of encounter, with a similar geometry and
timing, would have had to have happened also for NGC~752
simultaneously at the present time in the immediate vicinity of the
Sun. This is unlikely (with dark matter not having been detected and
probably not existing \citealt{Kroupa15, Roshan+21b, Asencio+22}).
With no corresponding perturbation in the local phase-space
distribution of field stars having been reported, a perturbation being
the origin for the observed asymmetries is not further considered in
the following.

\section{Milgromian gravitation and open star clusters}
\label{sec:Mil}

Given that the observed asymmetry between the leading and trailing
tails of open clusters appears to be difficult to be explained in
Newtonian gravitation, the asymmetry problem is now studied in a
modern non-relativistic theory of gravitation.  This section contains
a brief introduction to \underline{M}ilgr\underline{o}mia\underline{n}
\underline{D}ynamics (MOND), the revised Poisson equation
(Sec.~\ref{sec:genPois}), its implication for the equipotential
surface around a gravitating body and for open star clusters
(Sec.~\ref{sec:oc}).

\subsection{The generalised Poisson equation and the PoR code}
\label{sec:genPois}

The standard (Newtonian) Poisson equation,
\begin{equation}
  \vec{\nabla} \cdot \left[ \vec{\nabla} \PhiN \right] = 4\pi G \rho_\text{b},
\label{eq:Poisson}
\end{equation}
allows the position vector, $\vec{R}$, dependent phase-space baryonic
mass density, $\rho_{\rm b}(\vec{R})$, of the cluster plus hosting
galaxy to generate the Newtonian potential, $\phi_{\rm N}(\vec{R})$,
the negative gradient of which provides the acceleration at $\vec{R}$.

\cite{Mil83} extended the non-relativistic formulation of
Newton/Einstein beyond the Solar System by invoking dynamics data
which had become available for disk galaxies a few years prior to
1983, but decades after \cite{Einstein16}.  \cite{Mil83} conjectured
that gravitational dynamics changes to an effectively stronger form
when the gradient of the potential falls below a critical
value\footnote{Throughout this text and depending on the context, the
  unit for velocity is pc/Myr or km/s, noting that
  $1\,$pc/Myr\,$=0.9778\,$km/s$ \;\approx 1\,$km/s to a sufficient
  approximation.}, $a_0 \approx 3.8\,$pc/Myr$^2$, which appears to be
a constant in the Local Universe but may be related to an energy scale
of the vacuum and to the expansion rate of the Universe (for reviews
see \citealt{SandersMcGaugh02, Scarpa2006, Sanders07, FamMcgaugh,
  Milgrom14, Trippe14, Sanders15, Merritt20, BanikZhao21}, the latter
containing discussions of relativistic formulations that are
consistent both with the cosmic microwave background (CMB) and the
speed of gravitational waves.).\footnote{The interpretation of MOND as
  a generalised-inertia theory is an alternative to interpreting MOND
  as a theory of gravitation, and is related to Mach's principle, but
  has not been developed as a computable theory \citep{Milgrom14,
    Milgrom22}. \cite{Loeb22b} points out an implication for space
  travel.}

Energy- and momentum conserving time-integrable equations of motion of
non-relativistic gravitating bodies became available through the
discovery of ``a quadratic Lagrangian'' (AQUAL) as a generalisation of
the non-relativistic Newtonian Lagrangian by \cite{BM84}. This lead to
the non-linear AQUAL-MOND Poisson equation that is related to the
well-known p-Laplace operator,
\begin{equation}
\vec{\nabla} \cdot 
\left[
\mu\left( {|\vec{\nabla}\Phi| \over a_0} \right) \, \vec{\nabla}\Phi
\right] = 4\,\pi\,G\,\rho_{\rm b},
\label{eq:genPois}
\end{equation}
where $\mu(x) \rightarrow 1$ for
$x= |\vec{\nabla}\Phi| / a_0 \rightarrow \infty$ and
$\mu(x) \rightarrow x$ for $x \rightarrow 0$, the transition function,
$\mu(x)$, being derivable from the quantum vacuum \citep{Milgrom99}.
Observational data show the value of Milgrom's constant $a_0$ to be
$0.9 \times 10^{-10}< a_0/({\rm m/s}^2) < 1.5 \times 10^{-10}$
(\citealt{things}, i.e., $a_0 \approx 3.8\,$pc/Myr$^2$).

Eq.~\ref{eq:genPois} allows $\rho_{\rm b}(\vec{R})$ to generate the
full Milgromian potential, $\Phi(\vec{R})$, the negative gradient of
which provides the acceleration as in standard gravitation. Solving
this generalised Poisson equation in regions of high acceleration
recovers the Newtonian potential ($\mu = 1$ retrieving
Eq.~\ref{eq:Poisson}), while in low-acceleration regions where the
gradient of the potential is smaller than $a_0$, the acceleration
comes out to be stronger. As a consequence of the non-linear nature of
the generalised Poisson equation the boundary conditions differ to
those in Newtonian gravitation and $\Phi$ depends on the external
acceleration from the mass distribution in the neighbourhood.  The
rotation curve of a non-isolated disk galaxy of baryonic mass
$M_{\rm b}$ within an external field (EF) will decrease with
increasing galactocentric distance, while it is constant for an
isolated (no EF) disk galaxy of the same baryonic mass
\citep{Haghi+16, Chae+20}. In other words, the isolated galaxy has a
larger effective-Newtonian gravitational mass\footnote{Note that in
  MOND there is no ``gravitating mass'' beyond the mass contributed by
  the particles in the standard model of particle physics, and
  ``effective-Newtonian gravitational mass'' is not a true gravitating
  mass but a mathematical formulation assuming Newtonian
  gravitation.}, $M_{\rm grav, iso}$, than the non-isolated galaxy
($M_{\rm grav, niso}$), generating around itself at large $|\vec{R}|$
a spherical logarithmic potential $\Phi(\vec{R})$. In both cases
though, $M_{\rm grav}>M_{\rm b}$ with
$M_{\rm grav,iso}>M_{\rm grav, niso}$. This effect on $\Phi$ is called
the ``external field effect'' (EFE) and causes self-gravitating
systems to become effectively Newtonian when the external acceleration
is stronger than the internal acceleration. The EFE has been
observationally confirmed with more than 5$\,\sigma$ confidence
(\citealt{Chae+20, Chae+21}, see also \citealt{Haghi+16, Hees+16}).
The external field is also present in Newtonian gravitation but
because the potentials add linearly, it can be subtracted and has no
influence on the internal dynamics of a self-gravitating system
falling within the external field. In Milgromian gravitation on the
other hand, due to the generalised Poisson equation
(Eq.~\ref{eq:genPois}), the internal dynamics depends on this external
field by non-linearly changing the effective-Newtonian gravitational
masses of the constituents in dependence of their location within the
system. The strong equivalence principle is thus not obeyed in
Milgromian gravitation and the internal physics of falling bodies
depend on their gravitational environment.  \cite{Oria+21} discuss
the EFE in the context of the mass distribution in the Local
Cosmological Volume.

Eq.~\ref{eq:genPois} can be readily solved using well-known methods if
the distribution of matter, $\rho_{\rm b}$, can be approximated as a
continuum, which is equivalent to the system in question being
``collision-less'', or synonymously, the system having a two-body
relaxation time longer than a Hubble time \citep{Kroupa98, FK11,
  MH11}. Galaxies are collision-less systems such that
particle-mesh-grid-based methods can be used to calculate the flow of
phase-space mass density. In Newtonian gravitation, the distribution
of discrete particles can be combined in discrete spatial grid cells
in which the mass density, $\rho_\text{b}(\vec{R})$, is thusly
defined.  The standard Newtonian potential, $\PhiN(\vec{R})$, can
thereupon be computed by solving the Poisson equation
(Eq.~\ref{eq:Poisson}) on the discrete grid using different efficient
techniques. Once the potential is known in each cell, the Newtonian
acceleration, $\vec{a}(\vec{R},t) = -\vec{\nabla} \PhiN(\vec{R},t)$,
can be computed for each grid point and thus for each particle at its
location by interpolation, to finally advance each particle through
position-velocity space by one step in time.  This process is repeated
after each time step to move a Newtonian system forwards in time.

A collision-less $n$-body code based on solving Eq.~\ref{eq:genPois}
on a spherical-coordinate grid for isolated systems was developed by
\cite{LondrilloNipoti09}. This code has been applied to a number of
star-cluster-relevant problems including the EFE \citep{Haghi+09,
  Sollima+12a, WuKroupa13, WuKroupa18, WuKroupa19} but is not usable
here as the spherically-symmetrical grid does not allow the tidal
tails to be followed in sufficient detail.

The computational burden for calculating $\Phi(\vec{R})$ has been
reduced significantly by the development of a quasi-linear formulation
of MOND (QUMOND, \citealt{QUMOND}), which requires to solve only
linear differential equations, with one additional algebraic
calculation step, allowing a quick and efficient implementation in
existing particle-mesh $n$-body codes.  The generalised Poisson
equation in QUMOND is
\begin{equation}
\nabla^2 \Phi (\vec{R})=4 \pi G \rho_\text{b}(\vec{R}) + 
\vec{\nabla} \cdot \left[ \widetilde{\nu}\left(|{\vec{\nabla}} 
      \PhiN|/a_0\right) \vec{\nabla} \PhiN(\vec{R}) \right]\, ,
\label{eq:pois}
\end{equation}
where $\widetilde{\nu}(y) \rightarrow 0$ for $y \gg 1$ (Newtonian
regime) and $\widetilde{\nu}(y) \rightarrow y^{-1/2}$ for $y \ll 1$
(Milgromian regime; note that $\mu$ and $\widetilde{\nu}$ are
algebraically related, see \citealt{FamMcgaugh,Milgrom14}) with
$y=|{\vec{\nabla}} \PhiN|/a_0$.  This means that the total Milgromian 
gravitational potential, $\Phi = \PhiN + \Phiph$, can be divided into
a Newtonian part, $\PhiN$, and an additional phantom part,
$\Phiph$.  The matter density distribution, $\rho_\text{ph}(\vec{R})$,
that would, in Newtonian gravitation, yield the additional potential,
$\Phiph(\vec{R})$, and therefore obeys
$\nabla^2 \Phiph(\vec{R}) = 4\pi G \rho_\text{ph}(\vec{R})$, is known
in the Milgromian context as the phantom dark matter (PDM) density,
\begin{equation}
	\rho_\text{ph}(\vec{R}) = \frac{1} {4\pi G}  \vec{\nabla} \cdot 
\left[   \widetilde{\nu}\left(|\vec{\nabla} \PhiN(\vec{R})|/a_0\right)\vec{\nabla} \PhiN(\vec{R}) 
\right]  \,.
\label{eq:rho_phantom}
\end{equation}
PDM is not real matter but a mathematical description helping to
compute the additional gravity in Milgrom's formulation, giving it an
analogy in Newtonian dynamics.  In the context of the Local
Cosmological Volume, \cite{Oria+21} calculate regions with
$\rho_{\rm ph} < 0$ that may be identifiable with weak lensing
surveys. This phantom mass density does not take part in the
time-integration and does not introduce Chandrasekhar dynamical
friction into the system.  In the above terminology,
$M_{\rm oc, grav}=M_{\rm oc} + M_{\rm oc,ph} > M_{\rm oc}$.  The PDM
density that would source the Milgromian force field in Newtonian
gravity can thus be calculated directly from the known baryonic
density distribution, $\rho_\text{b}(\vec{R})$.  A grid-based scheme
can be used to calculate $\rho_\text{ph}(\vec{R})$ from the discrete
Newtonian potential, $\PhiN^\text{i,j,k}$ (see eq.~35 in
\citealt{FamMcgaugh}, and also \citealt{things}; \citealt{QUMOND};
\citealt{PRG}; \citealt{Lueghausen2014a}, \citealt{Lueghausen2014b}
and \citealt{Oria+21}). Once this source of additional acceleration is
known, the total (Milgromian) potential can be computed readily using
the Poisson solver already implemented in the grid-based code.

The above QUMOND technique was implemented independently by
\cite{Lueghausen2014b} and \cite{Candlish+15} into the existing {\sc
  RAMSES} code developed for Newtonian gravitation by
\cite{Teyssier2002}. {\sc RAMSES} employs an adaptively refined grid
structure in Cartesian coordinates such that regions of higher density
are automatically resolved with a higher resolution.  It computes the
Newtonian potential $\phi_\text{N}(\vec{R})$ from the given baryonic
mass-density distribution, $\rho_\text{b}(\vec{R})$, i.e., it solves
the discrete Poisson equation (Eq.~\ref{eq:Poisson}). In the Phantom
of Ramses (PoR) code, \cite{Lueghausen2014b} added a subroutine
which, on the adaptive grid, computes the PDM density from the
Newtonian potential (Eq.~\ref{eq:rho_phantom}) and adds the
(mathematical) DM-equivalent density to the baryonic one. The Poisson
equation is then solved again to obtain the Milgromian potential,
$\Phi = \phi_\text{N}+\Phi_\text{ph}$, i.e., the ``true'' (Milgromian)
potential, which is used to integrate the stellar particles in time
through space (each stellar particle experiencing the acceleration
$\vec{a}_{\rm star}(\vec{R}) = -\vec{\nabla}\Phi(\vec{R})$) and to
solve the Euler equations for the dynamics of the gas.

Although any interpolating function can in principle be chosen, the
PoR code adopts as the default interpolating function
\begin{equation}
\widetilde{\nu}(y) = -{1 \over 2} + \left( {1\over 4} + {1 \over y}
\right)^{1\over 2} \, ,
\label{eq:interpol}
\end{equation}
and has already been used on a number of problems (for the manual and
description see \citealt{Nagesh+21}).

By employing a Cartesian grid, the PoR code is suited to study the
leakage of stellar particles across a cluster's pr\'ah and to follow
the tidal tails, with an application to this problem by
\cite{Thomas+18} accounting for the observed length asymmetry between
the leading and trailing tail of the globular cluster Pal~5.

\subsection{Open clusters}
\label{sec:oc}

\subsubsection{General comments}
\label{sec:oc_gen}

In contrast to galaxies, open star clusters are ``collisional''
systems in which the two-body relaxation time is significantly shorter
than one Hubble time, $\tau_{\rm H}$, i.e., in which the thrive
towards energy-equipartition plays a decisive evolutionary role. For
open clusters the dissolution time (Eq.~\ref{eq:lifet}) is shorter
than the Hubble time, $T_{\rm diss} < \tau_{\rm H}$.  The evolution of
an open cluster therefore needs to be calculated through the direct
star-by-star accelerations. Due to the simplifying linear-additivity
of forces in Newtonian gravitation, advanced computer codes have been
developed for this purpose \citep{Aarseth99, Aarseth+08, Aarseth10,
  amuse2013,Wang+20}. With these codes, it has been possible to solve
the problem why the binary-star fraction in the Galactic field is
50~per cent, while it is about 100~per cent in star-forming regions
\citep{Kroupa95a, Kroupa95b, MarksKroupa11}, to infer that the
Pleiades cluster formed from a binary-rich Orion-Nebula-Cluster-like
precursor \citep{KAH} with multiple stellar populations
\citep{Wang+19}, and how intermediate-massive clusters are affected by
a realistic high initial binary population \citep{Wang+21}.  But at
the present there are no codes that allow this to be done in
Milgromian gravitation.

\subsubsection{Open clusters in Milgromian gravitation:
  the phantom extends the pr{\'a}h}
\label{sec:oc_MOND}
Despite it not being possible at this time to do exact calculations of
the dynamical evolution of open clusters in Milgromian gravitation,
some estimates are possible, given the existing tools.

An open cluster (or satellite galaxy) at a position vector $\vec{R}$ in
the Galactocentric reference frame will be subject to an EF,
\begin{equation}
  \vec{a}_{\rm ext,kin} = -{v_{\rm circ}^2 \over R} \, {\vec{R} \over R}
  \approx -G\, {M_{\rm MW, grav}(R) \over R^3} \, \vec{R},
\label{eq:ext}
\end{equation}
where $M_{\rm MW,grav}(R)$ is the effective-Newtonian gravitational
mass of the here-assumed-spherical Galaxy within $\vec{R}$,
$M_{\rm MW,grav}(R)$ being obtained from dynamical tracers, such as
the rotation curve within $R$.  For example, for the Hyades
$|\vec{a}_{\rm ext,kin}| \equiv a_{\rm ext,kin} \approx
7.5\,$pc/Myr$^2$ $\approx 2.0\,a_0$
($M_{\rm MW, grav}\approx 1.15 \times 10^{11}\,M_\odot, R \approx
8300\,$pc for circular velocity $v_{\rm circ}=250\,$pc/Myr).  As a
consequence, the dynamics of open star clusters is much richer than in
Newtonian gravitation, since, for example, a star cluster will change
its self-gravitational energy and thus its two-body relaxational
behaviour as it orbits within a galaxy due to the changing external
field from the galaxy \citep{WuKroupa13}.  In
Table~\ref{tab:clusterdata} $a_{\rm ext, kin}$ (Eq.~\ref{eq:ext}) and
$a_{\rm ext, bary}$ are documented for the four open clusters,
referring to the EF generated by the total gravitating mass
($M_{\rm MW, grav}$, baryonic plus dark matter for the rotation curve
as shown in Fig.~\ref{fig:rc}) and only the Newtonian acceleration
through the baryons (replacing $M_{\rm MW, grav}$ in Eq.~\ref{eq:ext}
by $M_{\rm bulge+disk}$ ), respectively.

The open clusters in Table~\ref{tab:clusterdata} have an internal
Newtonian acceleration \citep{Haghi+19}
\begin{equation}
a_{\rm int} =  G\, {M_{\rm oc} \over 2\,r_{\rm h}^2} \ll a_0\,.
\label{eq:aint}
\end{equation}
As already noted by \cite{Mil83} the open clusters in the Solar
neighbourhood are in the EF dominated regime with
$a_{\rm int} \ll a_{\rm ext, kin} \approx 2\,a_0$. Their internal
dynamics can therefore roughly be approximated by Newtonian
gravitation with a larger effective Newtonian constant
(Eq.~\ref{eq:Geff} below).

Because the effective-Newtonian gravitational mass of an open cluster
is larger than its mass in stars,
$M_{\rm oc,grav}>M_{\rm oc}$,\footnote{The ratio
  $M_{\rm oc,grav}/M_{\rm oc}$ depends on the EF
  (Sec.~\ref{sec:genPois}).} the escape speed from the open cluster is
increased in comparison to the Newtonian/Einsteinian case where
$M_{\rm oc,grav}=M_{\rm oc}$.  Detailed calculations have shown the
zero-equipotential surface around the cluster to be lopsided about the
position of the density maximum of the cluster due to the
non-linearity of the generalised Poisson equation
(Fig.~\ref{fig:rforce} below, \citealt{Wu+10, Wu+17, Thomas+18}), such
that the escape of stars may be directionally dependent. But also the
internal stellar-dynamical exchanges differ from Newtonian
gravitation.  One can visualise this problem as follows: An isolated
star~A of mass $m_{\rm A}$ generates around itself a logarithmic
Milgromian potential $\Phi_{\rm A}$ with a gravitating phantom mass,
$m_{\rm ph, A} \gg m_{\rm A}$, at large distances. Placing another
star~B near it reduces $m_{\rm ph A}$ by virtue of the EF from
star~B. A third star~C will thus experience an acceleration from
star~A and~B which is not the vectorial sum of their Newtonian
accelerations and with a magnitude and direction which depends on the
separation of~A and~B. Gravitational dynamics thus has parallels to
quark dynamics, as has been pointed out by \cite{BM84}.  The larger
effective-Newtonian gravitational masses of the stars lead to stronger
two-body relaxation while the larger effective-Newtonian gravitational
masses of the whole cluster increases the barrier for escape. It is
therefore unclear for the time being how real open clusters evolve in
Milgromian gravitation.

The isolated open cluster would have a 1D (line-of-sight)
velocity dispersion (assuming spherical symmetry and an isotropic
velocity distribution function)
\begin{equation}
\sigma_{\rm M,iso,los} \approx \left( {4 \over 81} \,
  G\,a_0\,M_{\rm oc} \right)^{1 \over 4}\, ,
 \label{eq:vdispMONDiso} 
 \end{equation}
with the 1D velocity dispersion of an EF-dominated open cluster being
\begin{equation}
\sigma_{\rm M,ef,los} \approx 3^{-1/2} \, \left(   {G' \,M'  \over 2\,r_{\rm
    h} } \right)^{1 \over 2} \, .
 \label{eq:vdispMONDef} 
\end{equation}
In this equation we can write either
\begin{equation}
  \begin{gathered}
  G' = G_{\rm eff}  = {\rm G \over \mu(x)}   \quad {\rm with} \quad M'
  = M_{\rm oc} \, ,\\
  \quad {\rm or} \quad \\
  G' = G \quad {\rm with} \quad
  M' =  M_{\rm oc, grav} =  {\rm M_{\rm  oc} \over \mu(x)} \, ,
\label{eq:Geff}
\end{gathered}
\end{equation}
being, respectively, the boosted gravitational constant or the
effective-Newtonian gravitational mass of the open cluster, with
$\mu(x)$ being given by Eq.~\ref{eq:mu} below (e.g.,
\citealt{McGaughMilgrom13}).\footnote{Analytical estimates of the
  velocity dispersion of star clusters that have an internal
  acceleration comparable to the external one
  ($a_{\rm int} \approx a_{\rm ext} \approx a_0$) are not possible.
  From Eq.~\ref{eq:aint} such clusters would need to have
  $M_{\rm oc}\approx 10\,000\,M_\odot$. \cite{Haghi+19} published
  interpolation formulae obtained from fits to Milgromian simulations
  of dwarf galaxies, but these cannot be applied to the star-cluster
  regime.} For the open clusters in Table~\ref{tab:clusterdata}, either
$G_{\rm eff} \approx 1.5 \, G$ or
$M_{\rm oc, grav} \approx 1.5 \, M_{\rm oc}$.

Table~\ref{tab:clusterdata} provides estimates of $a_{\rm int}$ and of
$\sigma_{\rm M,iso,los}$, assuming the clusters are isolated, and of
$\sigma_{\rm M,ef,los}$ using the above analytic estimate. The
variation of $a_{\rm ext,bary}$ and $a_{\rm ext;kin}$ as an open
cluster oscillates through the Galactic mid-plane adds to the
complexity, but these variations are of second order only.

The data in Table~\ref{tab:clusterdata} show that open clusters in
isolation would be significantly supervirial by a factor of 2.3–3.4
relative to the Newtonian expectation.  This is indeed the case in
observations and is equivalent to the Milgromian expectation that
$M_{\rm grav, oc} > M_{\rm oc}$.  For the Hyades,
$\sigma_{\rm 3D, obsI}\approx 0.8 \pm 0.15\,$km/s, \citep{Roeser+11}
and $\sigma_{\rm 3D, obsII}\approx 0.7 \pm 0.07\,$km/s
\citep{OE20}. For the Praesepe, the observational results are meagre,
but \cite{RS19} report $\sigma_{\rm 3D, obs} \approx 0.8\,$km/s (their
sec.~2.3).  For both clusters, $\sigma_{\rm 3D, obs}$ is larger by a
factor of approximately two than the expected Newtonian value
(Table~\ref{tab:clusterdata}).  The standard explanation is that this
discrepancy is due to additional mass in stellar remnants and a larger
measured velocity dispersion due to unresolved multiple stars (c.f.,
\citealt{Gieles+10} for $\approx10\,$Myr old clusters).  Open clusters
are known to have a deficit of white dwarfs compared to the expected
number \citep{Fellhauer+03} so the contribution by remnants is
unclear, and the contribution to the velocity dispersion in proper
motion and line-of-sight velocity measurements needs to be assessed
with realistic initial binary populations \citep{Dab+21}.  In
Milgromian gravitation, the EF-dominated estimate
($\sigma_{\rm M,ef,los}$), on the other hand, is larger than the
Newtonian value by only $\approx25\,$per cent, as is confirmed with
the PoR simulations in Sec.~\ref{sec:mass_veldisp} below. It is
unclear, however, how a fully collisional treatment of an open cluster
embedded in an external field would enhance the velocity dispersion
(see Sec.~\ref{sec:qosc}), so that these estimates have to be taken
with great caution.  Geometric factors (the density profile) need to
be taken into account and the observed values bear some contamination
by field stars.
 
Incidentally, \cite{Sollima+12b} analyse the dynamical masses of six
globular clusters finding them to be about~40~per cent larger than the
stellar-population masses assuming a canonical IMF. The authors
include modelling the binary-star population, do not exclude a
possible systematic bias, and attribute the larger dynamical masses to
retained stellar remnants or the presence of a modest amount of dark
matter. For a canonical IMF the remnants, if retained, contribute
  about 22~per cent of the mass (e.g. fig.~5 in \citealt{Mahani+21}),
  although larger fractions are possible as a result of dynamical
  evolution (see fig.~11 and~12 in \citealt{BM03}; \citealt{BK11}).
The authors do not discuss Milgromian implications, and it is noted
here that the larger effective-Newtonian gravitational masses of the
six clusters are qualitatively consistent with the general expectation
from Milgromian gravitation. A detailed analysis of the velocity
dispersion of the outer halo globular cluster Pal~14 by
\cite{Sollima+12a} leads to the cluster being reproducible with
Newtonian gravitation.  However, Milgromian solutions may be possible
if the stellar population lacks low-mass stars. This is possible if
Pal~14 formed with an IMF lacking low-mass stars as is expected at
low-metallicity \citep{Marks+12, Kroupa+13, Jerabkova+18, Yan+21}.  In
addition, the initially mass segregated cluster is likely to have lost
low-mass stars through early expulsion of residual gas during its
formation phase \citep{Marks+08, Haghi+15}.

Based on the suggested test for the validity of MOND by
\cite{Baumgardt2005}, the outer halo globular cluster NGC~2419 has
been much debated concerning how it fits-in with Milgromian
gravitation \citep{Ietal11, Ibata+11b, Ietal13}. According to
\cite{Ietal11, Ibata+11b} and \cite{Derakshani14}, Newtonian models
significantly better represent the observed luminosity and
line-of-sight-velocity dispersion profiles than Milgromian ones, while
\cite{Sanders12} developed Milgromian models that match the observed
cluster {assuming a strong degree of radial anisotropy.  This cluster
  has a half-mass radius which is about seven times larger than usual
  globular clusters and is therefore unusual.  The phase-space
  distribution function of stars within the cluster is likely to be
  affected by the Milgromian phase transition as the cluster orbits
  from the inner Newtonian Galactic regime into the outer Milgromian
  regime \citep{WuKroupa13}. This leads to an anisotropic velocity
  distribution function of stars in the cluster. Also, violent
  expulsion of residual gas in Milgromian gravitation when the cluster
  formed generates a strong anisotropic velocity distribution function
  \citep{WuKroupa18, WuKroupa19}, and may have changed the mass
  function of stars in the cluster if it was mass-segregated at birth
  (cf. \citealt{Haghi+15}). At birth, the low-metallicity cluster is
  likely to have been mass-segregated and to have had a top-heavy IMF
  \citep{Marks+12, Kroupa+13, Jerabkova+18, Yan+21} through which the
  innermost regions of the cluster could contain a large fraction of
  stellar remnants \citep{Mahani+21} affecting the velocity dispersion
  profile in the way observed (larger in the innermost region,
  dropping outwards).  More inclusive modelling of this cluster is
  clearly needed to advance our knowledge of this enigmatic object.
 
Returning to the tidal tails, as a consequence of the non-linear
generalised Poisson equation, the Milgromian potential, $\Phi$, around
an open cluster situated within the thin disk of the Galaxy is
lopsided. It generates a stronger restoring force towards the cluster
on its far side.  The external field, $\vec{a}_{\rm ext, kin}$, is
directed towards the Galactic centre across the cluster
(Eq.~\ref{eq:ext}). It adds to the radial cluster-centric acceleration
on the far side of the cluster relative to the Galactic centre, and it
opposes the cluster-centric acceleration field on the near side. Due
to the non-linearity of the generalised Poisson equation
(Eq.~\ref{eq:genPois}, \ref{eq:pois}), the Milgromian restoring force
towards the cluster's centre comes out to be stronger on the far side
and weaker on the near side.  The cluster's potential is therefore
lopsided \citep{Wu+08, Wu+10, Wu+17, Thomas+18}.  A visualisation of
this radial acceleration field is provided by Fig.~\ref{fig:rforce},
where the radius, within which the radial acceleration around a
point-mass cluster is well approximated by Newtonian gravitation, is
the MOND radius,
\begin{equation}
r_{\rm M} = \left( {G\,M_{\rm oc} \over a_0} \right)^{1 \over 2} \, .
\label{eq:rmond}
\end{equation}
 
While being only a point-mass approximation of the cluster,
Fig.~\ref{fig:rforce} thus provides the likely reason why the trailing
tail contains fewer stars in the five observed open clusters
(Sec.~\ref{sec:theTails}).  If the cluster generates a constant and
isotropic flux of escaping stars through the MOND radius, the higher
escape threshold at the far side will deflect a fraction of these back
to the cluster such that they are likely to exit on the near side to
populate the leading tail.

\begin{figure}
        \centering
        \includegraphics[width=\hsize]{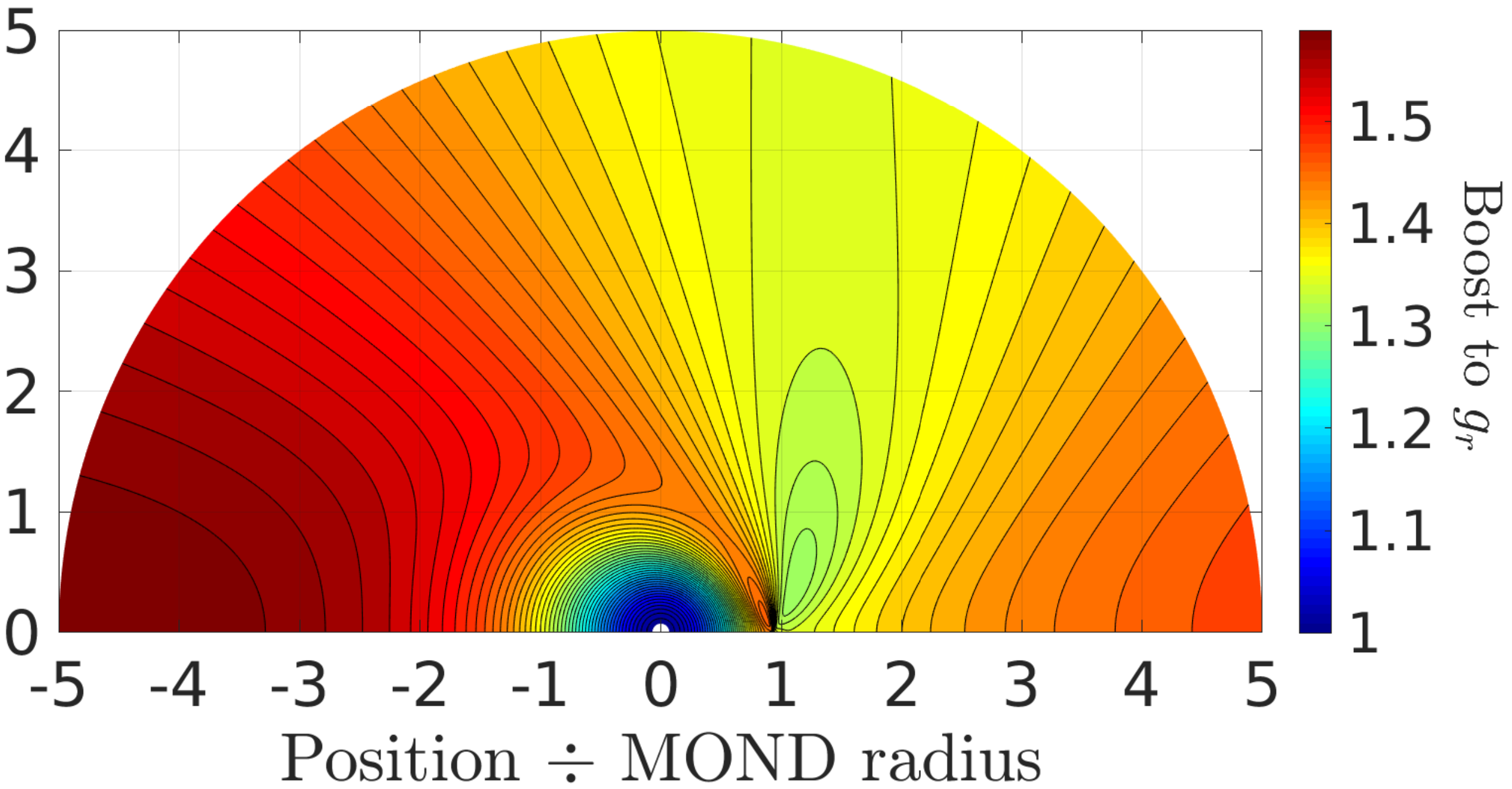}
        \caption{The radial acceleration-field around an open
          cluster, approximated here as a point-mass, is enhanced in
          Milgromian dynamics, as shown by the ``boost'' of the radial
          component of the gravitational acceleration above the
          Newtonian expectation.  Positions on the $x, y$-axes are in
          units of the MOND radius $r_{\rm M}$ (Eq.~\ref{eq:rmond}).
          The Galactic centre is in the direction of positive $x$ and
          is 8.3~kpc distant and a constant $a_{\rm ext}= a_0$
          (Eq.~\ref{eq:ext}) pointing towards the Galactic centre is
          assumed across the cluster. The distance orthogonal to this
          direction, which is the direction of a circular orbit, is
          shown on the $y$-axis. Thus, a star on the far side of the
          cluster at $x=-4, y=0$ experiences a radial acceleration
          towards the cluster centre at $x=0, y=0$ which is about
          15~per cent larger than a star at $x=4, y=0$ on the
          near-side of the cluster.  For numerical reasons, accurate
          results within $\approx 0.1 r_{\rm M}$ are not possible, a
          region where Newtonian and Milgromian gravity should be
          almost identical (for details on the calculation, see
          \citealt{BanikKroupa19a}; reproduced with kind permission
          from \citealt{BanikKroupa19a}).}
        \label{fig:rforce}
\end{figure}

In order to make a first step towards quantifying the population of
escaping stars in the leading versus the trailing tidal tail of an
open cluster, the self-consistent code PoR is applied.

\section{The models}
\label{sec:models}

Given the non-availability of a Milgromian relaxational (direct
$n$-body) code, a very rough approximation is made to obtain a first
insight as to whether Milgromian gravitation might lead to the
observed asymmetry in the tidal tails of open star clusters. For this
purpose the collision-less (no two-body relaxation) PoR code
(Sec.~\ref{sec:genPois}) is applied.  As a consistency check,
Newtonian models of the equivalent Milgromian ones are compared using
this same code to test if the former lead to symmetrical tidal tails.
Do the Milgromian models develop an unambiguous asymmetry?  Is this
asymmetry equivalent to the observed asymmetry? And does the asymmetry
vary with time?

The underlying energy equipartition process is entirely missing such
that the flux of stellar particles across the pr{\'a}h will not be
correct: in the simulations, stellar particles leave the cluster by
acquiring energy through the time-dependent tidal field as well as
through artificial heating due to the limited numerical resolution and
not due to two-body encounters. We concentrate on the differential
effect (Milgrom vs Newton) and therefore only on the asymmetry which
is governed by the asymmetry of the radial acceleration towards the
cluster between the far and near sides (Fig.~\ref{fig:rforce}), since
the PoR code self-consistently quantifies the shape of the cluster
potential.

The computations are performed with the PoR code by applying the
``staticparts patch'' \citep{Nagesh+21}, which is available at
\url{https://bitbucket.org/SrikanthTN/bonnpor/src/master/}.  This code
allows a subset of dynamical particles to be integrated in time within
the background potential of another subset of particles treated as
static, i.e., which are not integrated over time. Both Newtonian and
Milgromian models have been computed with the exact same numerical
procedure. To keep the models as consistent as possible, the selection
of the Newtonian model has not been done by switching off the
Milgromian option of PoR but by setting the Milgromian constant $a_0$
to 30 decimal orders of magnitude below the canonical value of
$a_0=1.2\times10^{-10}\,\mathrm{m\,s^{-2}} \approx 3.8\,$pc/Myr$^2$.

In the following, the Galactic potential used (Sec.~\ref{sec:gal}) and
the cluster models (Sec.~\ref{sec:clmods}) are described.

\subsection{The galactic potential}
\label{sec:gal}

For a comparison of equivalent Newtonian vs Milgromian models, it is
necessary to insert the same cluster model on the same initial orbit
into a galaxy which has the same rotation curve at the position of the
cluster.  Thus, the tidal effects in the two galaxy models are
comparable, one being purely baryonic in Milgromian dynamics, the
other being Newtonian with the same baryonic component and with an
additional spherical dark matter (DM) component.  To achieve a
rotation curve which matches that observed for the Galaxy nearby to
the Sun, the baryonic component is modelled as two radial exponential
and vertical sech$^2$ disks: Component~I has an exponential
scale-length of 7~kpc and an exponential scale height of 0.322~kpc
making up 17.64~per cent of the total baryonic mass.  Component~II has
a scale-length of 2.15~kpc and a scale height of 0.322~kpc making up
82.64~per cent of the total baryonic mass. We emphasise that this is
not meant to be an exact representation of the Galaxy but merely a
description sufficiently realistic to provide a radial and
perpendicular approximation of the Galactic accelerations near the
Solar circle. The density distribution of the DM component is defined
to match the Milgromian rotation curve shown in
Fig.~\ref{fig:rc}. Thus the Galaxy is automatically on the
radial-acceleration relation (RAR, \citealt{RAR, Lelli+17}) and on the
baryonic Tully-Fisher relation (BTFR, \citealt{BTFR, BTFRb, Lelli+16})
in both cases.  The DM halo is truncated outside 10 kpc to save
computing ressources. This is physically justified since the cluster
models orbit the Galactic center at $R_{\rm gal}\approx 8.3\,$kpc,
i.e., well within the truncation radius. In Newtonian gravity, the
field of a spherically symmetrical shell cancels out inside the shell,
so any DM component outside the orbit of any cluster particle can be
ignored.  This is, however, not true for the non-spherical disc
potential, therefore a disc template with a maximum radius of about
50~kpc is employed. Since both galaxy models only serve as a constant
background potential, no velocity component needs to be added, but
rather the galaxy+DM particles are treated as static gravity sources.

\begin{figure}
\begin{center}
\includegraphics[width=0.48\textwidth]{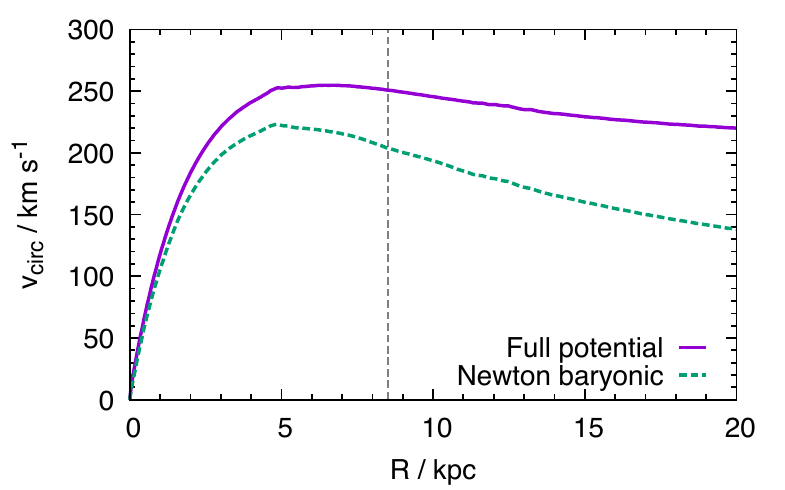}
\caption{The rotation curves for the total gravitational potential
  (Milgrom or Newton plus DM halo, solid purple line) and the
  Newtonian baryonic component (dashed green line). The vertical
  dashed line marks the orbital distance of the model clusters from
  the Galactic centre. }
\label{fig:rc}
\end{center}
\end{figure}

The models invoke 30~million static particles to provide a smooth
background potential. In the Newtonian DM case another 10 million
static particles are used for the DM halo.\footnote{The rotation curve
  (Fig.~\ref{fig:rc}) is higher by about 10~per cent than in the real
  Galaxy that has $v_{\rm circ}\approx 220\,$km/s because the
  initialisation through a particle dark matter halo (for the
  Newtonian models) has this amount of uncertainty. We leave the
  models at the slightly higher $v_{\rm circ}$ and adopt this higher
  value also for the Milgromian models as the model clusters need to
  be computed in a comparable tidal field.}  The minimum and
maximum grid refinement levels are set to 7 and 21, respectively. In the
computations, however, the maximum refinement level actually reached
is only~17. Given a total box length of 256~kpc this corresponds to a
spatial resolution of about 1.95~pc. The dynamical subset consists of
the stellar particles of the model clusters introduced in the next
section.

\subsection{Milgromian and Newtonian models of open clusters}
\label{sec:clmods}

In the Solar neighbourhood, open clusters are well described by a
Plummer phase-space distribution function \citep{Roeser+11,RS19},
which corresponds to a King model with a concentration parameter
$W_{\rm o}\approx 6$ \citep{Kroupa08}.\footnote{That the well-observed
  open and globular clusters, which are the simplest (coeval and
  equal-metallicity) stellar populations, are next-to-perfectly
  described by the Plummer phase-space distribution function
  \citep{plummer, HeggieHut03, AHW74}, which is the simplest analytical solution of
  the collision-less Boltzmann equation, is interesting and deserves
  emphasis. Note that the Milgromian models retain a Plummer density
  distribution as they evolve (Fig.~\ref{fig:dens} below).}  The
spherical and isotropic PoR cluster models are therefore set up as
Plummer phase-space distribution functions.

The Newtonian Plummer models are initialised following \cite{AHW74}. The
Milgromian Plummer models are constructed as follows. First, a Newtonian
Plummer model is set up using the Newtonian method. A Milgromian
velocity scaling factor $f_{\rm M}$ is introduced as
\begin{equation}
  f_{\rm M}=\sqrt{\nu(y)},
\end{equation}
with $\nu(y) = \sqrt{1/4 + 1/y} + 1/2, y = g_{\rm N}/a_0$, being the
Milgromian-Newtonian transition function, and
$g_\mathrm{N}=|\vec{a}_\mathrm{N}|$ with
$\vec{a}_\mathrm{N}=\vec{a}_\mathrm{i,N}+\vec{a}_\mathrm{ef,N}$ is the
Newtonian acceleration obtained by adding the internal and external
Newtonian accelerations.  The particle velocities
$\vec{v}_\mathrm{i,M}$ are then derived from the Newtonian model
velocities, $\vec{v}_\mathrm{i,N}$, via
\begin{equation}
  \vec{v}_\mathrm{i,M}=f_{\rm M} \, \vec{v}_\mathrm{i,N} \, .
\end{equation}
The Newtonian internal acceleration $\vec{a}_\mathrm{i,N}$ is taken
directly from the Newtonian Plummer model, while the Newtonian
external field component, $\vec{a}_\mathrm{ef,N}$, is related to the
actual Galactic external field, $\vec{a}_\mathrm{ext,kin}$, via
\begin{equation}
  \vec{a}_\mathrm{ef,N}=\mu(x)\,\vec{a}_{\rm ext,kin}
\end{equation}
with
\begin{equation}
  \mu=x/(x+1), \quad x=a_{\rm ext,kin}/a_0
\label{eq:mu}  
\end{equation}
(sec.~4.2.2. in \citealt{Lueghausen2014b}), being the inverse
transition function and
$a_{\rm ext,kin}=|\vec{a}_{\rm
  ext,kin}|=v_\mathrm{circ}^2/R_\mathrm{gal}$ in the direction to the
Galactic centre. Since $a_{\rm ext, kin}\approx 7.5\,$km/s$^2$
(Table~\ref{tab:clusterdata}), $\mu \approx 0.66$ in the present
context.  

The real open clusters (Table~\ref{tab:clusterdata}) have masses of a
few hundred~$M_\odot$ and half-mass radii
($r_{\rm h} \approx 1.3\times r_{\rm pl}$) near~$4\,$pc which is a
regime that the particle-mesh models cannot resolve adequately because
the density contrast to the surrounding field is too small and
available computational resources constrain the reachable refinement
levels.  The approach taken here is to start with a large
$M_{\rm oc,0}=2\times10^4\,M_\odot$ and to initialise a set of cluster
models with decreasing mass but with the same theoretical Plummer
radius, $r_{\rm pl}=10\,$pc ($r_{\rm h}\approx 13\,$pc).  While the
initial Plummer model is set up with this Plummer radius, a numerical
alteration to the particle distribution is needed to avoid strong mass
loss at the beginning: A numerical model inserted within the disk of
the galaxy does not correspond to the analytical Plummer phase space
distribution function which is bounded in mass but unbounded in radial
extend.  Particles that are at too large cluster-centric distances are
unbound in the numerical model.  Particles outside $r=r_{\rm pl}$ are
therefore re-positioned to be inside $r_{\rm pl}$ and the kinetic
energy is reset to a value corresponding to a particle orbiting inside
$r_{\rm pl}$. This stabilises the cluster against initial mass loss,
however, it also effectively compactifies the cluster. After a brief
period of settling, the cluster models stabilise at an effective
Plummer radius of about $r_{\rm npl}\approx 4.5\,$pc initially. The
models remain excellent approximations to the Plummer model
(Fig.~\ref{fig:dens} below) and the Plummer radius increases with time
as the cluster loses mass and thus becomes less bound.

The models are created to have a velocity dispersion that reaches
  that of the Hyades cluster, in order to achieve a comparable
  dynamical state. 
  The full computed set comprises Milgromian and Newtonian models with
  $M_{\rm oc, 0}=2500, 3500, 5000, 6000, 7000, 8500, 10000, 12000,
  14000, 20000\,M_\odot$.  Each model is composed of $0.1\,M_\odot$
  equal mass particles such that the Plummer phase-space distribution
  function is well sampled and different initial random number seeds
  do not affect the results. The models are computed over $1\,$Gyr.
  Only five in each set (with $M_{\rm oc, 0}=2\,500$, $3\,500$,
  $5\,000$, $7\,000$, $10\,000\,M_\odot$) are analysed in more detail,
  the heavy models being not representative of the observed open
  clusters by having too large velocity dispersions
  (Fig.~\ref{fig:sigma} below).  The initial analytical Plummer 3D
  velocity dispersion in Newtonian gravitation \citep{HeggieHut03,
    Kroupa08},
\begin{equation}
\sigma_\mathrm{ch}= \left( \frac{3\pi}{32}\frac{G\,M_{\rm oc,
      0}}{\rpl} \right)^{1\over 2} \, ,
\label{eq:plummer}
\end{equation}
is, for the full set of models, in the range
$\approx 0.86 - 2.43\,$km/s for
$r_{\rm pl}=r_{\rm npl}\approx 4.5\,$pc, being near the range of the
real open clusters (Table~\ref{tab:clusterdata}, see
Fig.~\ref{fig:sigma} below for the velocity dispersions
at~500Myr). Since the effective-Newtonian gravitational mass of an
open cluster in the Solar neighbourhood is
$M_{\rm oc, grav} \approx 1.5\, M_{\rm oc}$, it follows that the
Milgromian 3D velocity dispersion of the models should be
\begin{equation}
\sigma_{\rm ch, Mil} \approx 1.2 \, \sigma_{\rm ch} \, . 
\label{eq:sigma_Milgrom}
 \end{equation}

The tidal radius of a cluster can be approximated as
\begin{equation}
r_{\rm tid} \approx \left( {M_{\rm oc} \over K\,M_{\rm MW,
        grav} } \right)^{1/3}  \, R,
\label{eq:rtid}
\end{equation}
where $K=2$ for the logarithmic ($K=3$ for a point-mass) galactic
potential, and
$M_{\rm MW, grav} = \left( R / G \right) \, v_{\rm circ}^2$
($M_{\rm MW, grav}\approx 1.15\times 10^{11}\,M_\odot$ within
$R=8\,300\,$pc for the MW with $v_{\rm circ}=250\,$pc/Myr). Since the
initial tidal radii of the models are in the range
$18 \simless r_{\rm tid, 0}/{\rm pc} \simless 37$, the shape of the
zero-potential surface which defines the cluster pr\'ah is well
resolved such that the anisotropy of the flux of escaping particles
should adequately approximate the true anisotropy allowing a
physically correct comparison of the Newtonian (symmetrical surface,
anisotropy expected to be negligible) with the Milgromian
(non-symmetrical surface, Fig.~\ref{fig:rforce}) models.

The real clusters (Table~\ref{tab:clusterdata}) are currently on
orbits that are nearly circular with $0.03 < e < 0.12$. The
inclination angles, $\,\iota\,$, relative to the Galactic mid-plane
are also small
(${\rm tan}\left(\,\iota\,\right) = z_{\rm max} / R, \; \iota <
1\,$deg).  Given the explorative nature of this work, the models are
initialised on circular orbits within the mid-plane of the galactic
model.  Future work will address more realistic, slightly inclined
orbits. The current orbits end up being slightly eccentric in the
actual discretised galaxy potential.  In order to track the orbit of
the cluster, the density centre of each model cluster is calculated
using the density measurement method described by
\cite{CasertanoHut85} and the density centre position formula by
\cite{vHoerner1963}. Given an ensemble of $n$ stellar particles, the
local density around any stellar particle~i of mass $m_{\rm i}$ is
defined in the volume of the distance to its~$j$th neighbour,
\begin{equation}
  ^{j\!\!}\rho_{\rm i}=\frac{{j}-1}{V(r_{{\rm i},j})}\,m_{\rm i} \, ,
\end{equation}
where $r_{{\rm i},j}$ is the distance from stellar particle~i to
its~$j$th neighbour and
$V(r_{{\rm i},j})=4 \, \pi \, r_{{\rm i},j}^3/3$ is the volume of the
enveloping sphere.  The galactocentric position of the density centre
is then defined as the density-weighted average of the positions of
the stellar particles,
\begin{equation}
 ^{j\!\!}\vec{R}=\frac{\sum_{\rm i =1}^{n_{\rm tid}} \vec{R}_{\rm i}\;
   ^{j\!\!}\rho_{\rm
     i}}{\sum_{\rm i=1}^{n_{\rm tid}} \;  ^{j\!\!}\rho_{\rm i}},
\label{eq:RGC}
\end{equation}
where $n_{\rm tid}$ is the number of stellar particles within the
initial tidal radius, $r_{\rm tid,0}$ (as a simplification this radius
cutoff is kept constant), and $\vec{R}_{\rm i}$ is the galactocentric
position vector of particle~i.  Following the suggestion by
\cite{CasertanoHut85}, $j=6$ is used in this study.  It emanates
  that the orbits are not perfectly smooth (Fig.~\ref{fig:indq} ) as a
  consequence of the live dynamical computation of the cluster
  centre. For the calculation of the cluster specific angular
  momentum, its velocity dispersion and the asymmetry of the tidal
  tails, the position of the density maximum is used as the reference.

\section{Results}
\label{sec:results}

\subsection{The cluster profile, evolution of model mass and velocity
  dispersion}
\label{sec:mass_veldisp}

As noted in Sec.~\ref{sec:clmods}, the observed density profiles of
open clusters are well fit by the Plummer profile. The present models
are initialised as Plummer phase-space density distribution functions,
but do they retain Plummer profiles as they evolve?  The numerical
density profile of the Milgromian $5\,000\,M_\odot$ cluster at an age
of~500~Myr is shown in Fig.~\ref{fig:dens}.  The numerical model is
well represented by an analyitcal Plummer density distribution, as is
the case for all the other models.

\begin{figure}
\begin{center}
  \includegraphics[width=0.48\textwidth]{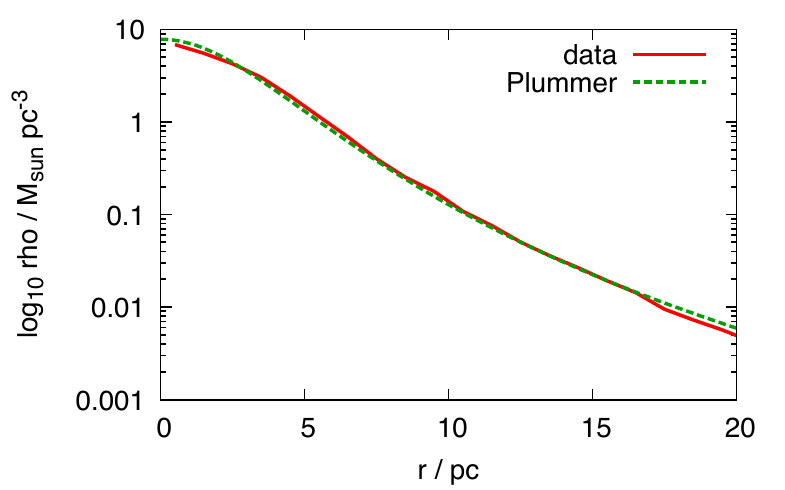}
  \vspace{-5mm}
\caption{The volume density Plummer fit (green dashed line) vs. the
  density for the $M_{\rm oc, 0}=5000\,M_\odot$ Milgromian cluster
  (solid red line, model cM3) at~500~Myr. Note that the simulated model is
  next-to-perfectly represented by the Plummer model with Plummer
  radius $r_{\rm npl}=4.9\,$pc.
}
\label{fig:dens} 
\end{center}
\end{figure}
The models dissolve by loosing stellar particles as is shown in
Fig.~\ref{fig:number}.  As for real star clusters, the low-mass models
dissolve more rapidly than the more massive ones. Compared to
realistic Newtonian calculations using the direct $n$-body method
which have $n(t)$ decrease following a concave curve (i.e., with a
slight slow-down of the evaporative stellar loss with decreasing
$n(t)$, e.g., fig.~1 in \citealt{BM03}), the present models show an
increasing rate of loss of stellar particles with time. This occurs
because the adaptive-mesh method reduces the refinement with
decreasing density that compromises the accuracy of tracing the
forces, leading to an artificial speed-up of mass loss. This is not a
problem for the purpose of the present study which is concerned with
the asymmetry of the tidal tails which probes the asymmetry of the
potential generated by the cluster--Galaxy pair, but implies that the
lifetimes of the present models cannot be applied to real open
clusters, although they allow an assessment of the relative lifetimes
between the Milgromian and Newtonian cases.  Beside the effect of the
adaptive mesh, the different shape of the mass evolution obtained here
(convex) compared to collisional $n$-body simulations (concave) is due
to two-body relaxation in the collisional $n$-body simulations leading
to cluster core contraction as energy-equipartition-driven mass loss
occurs, thus increasing the binding energy of the cluster and its
resistance to tidal effects (e.g., \citealt{BM03}).

The more rapid dissolution of the Milgromian models compared to the
Newtonian ones (Fig.~\ref{fig:number}) can be understood as follows:
Given a small loss of mass, $\delta M_{\rm oc}$, the change in binding
energy of a Newtonian star cluster is
$\delta E_{\rm bind, N} \approx \left(2/r_{\rm grav} \right)\,G\,
M_{\rm oc}\,\delta M_{\rm oc}$, where $r_{\rm grav}$ is the
gravitational radius of the cluster.  The here-relevant open clusters
(Table~\ref{tab:clusterdata}) are near-Newtonian but with a larger
effective gravitational constant, $G_{\rm eff} \approx 1.5 \, G$
(Eq.~\ref{eq:Geff}).  Thus, for the same mass loss of
$\delta M_{\rm oc}$, a Milgromian open cluster of the same mass and
radius suffers a reduction of its binding energy relative to that of
the Newtonian case,
$\delta E_{\rm bin,M} / \delta E_{\rm bind,N} \approx G_{\rm eff} / G
\approx 1.5$.  The Milgromian models can be understood to dissolve
more rapidly as a consequence of more rapidly loosing their binding
energy (see also Sec.~\ref{sec:efmix}).

\begin{figure}
\begin{center}
\includegraphics[width=0.48\textwidth]{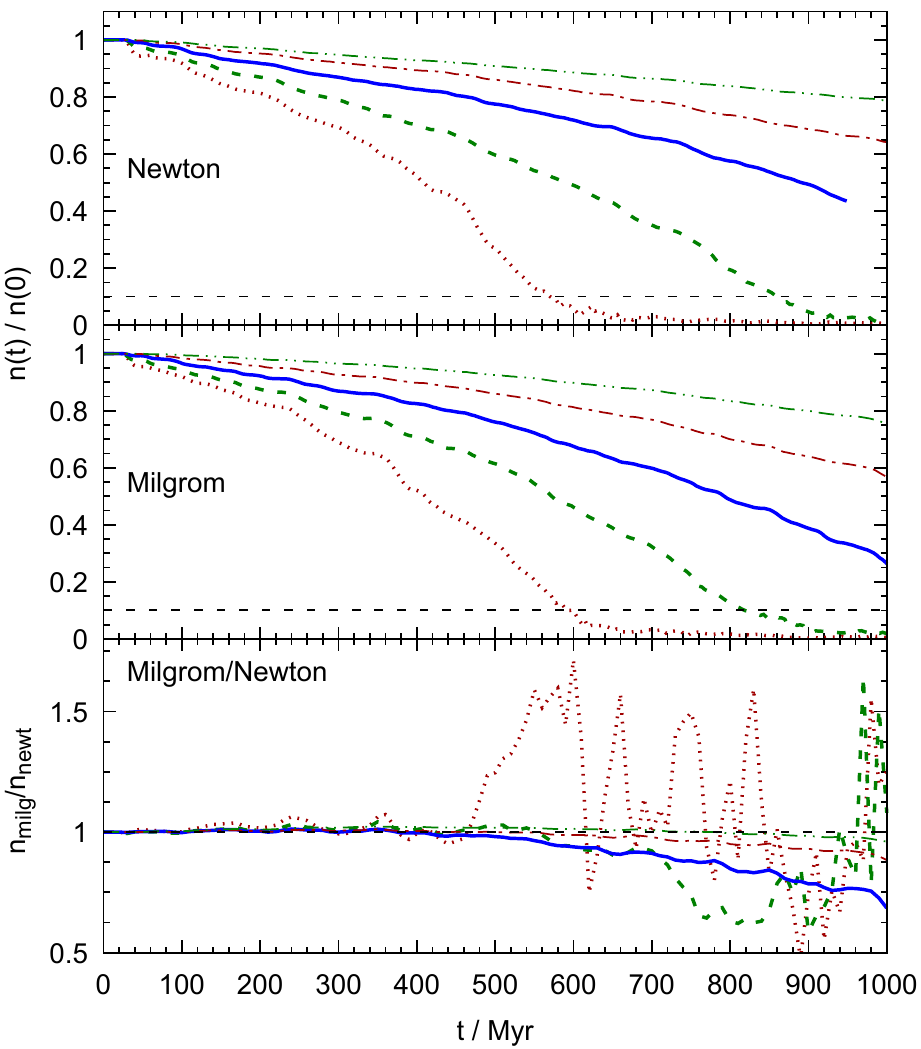}
\caption{The relative number of stellar particles within the initial
  tidal radius (Eq.~\ref{eq:rtid}) as a function of time in the five
  models with an initial mass of $M_{\rm oc, 0}=2\,500$, $3\,500$,
  $5\,000$ (blue), $7\,000$ and $10\,000\,M_\odot$ (bottom to
  top). Top panel: Newtonian models, middle panel: Milgromian models,
  bottom panel: ratio of the two.  Note that the Milgromian models
  dissolve faster than the Newtonian models. The fluctuations in
    the ratio (bottom panel) is large for the least-massive models
    (red dotted lines) due to the small number of particles and the PoR
    code not well-resolving their weak potentials. The general trend
    is for the less-massive Milgromian models to dissolve faster than
    the more massive ones, and, at a given mass, for them to dissolve
    faster than their Newtonian counterparts. The horizontal dashed
  lines in the upper two panels correspond to the 10~per cent
  dissolution threshold.}
\label{fig:number}
\end{center}
\end{figure}

The Milgromian open clusters are expected to be super-virial when
compared to their Newtonian counterparts
(Table~\ref{tab:clusterdata}).  Thus, it is of interest to consider
how super-virial the here-computed Milgromian cluster will appear to a
Newtonian observer.  For the purpose of comparing the velocity
dispersion of the Newtonian and Milgromian models at the same stage of
dissolution, it is useful to estimate their dissolution time.  Because
the more massive models are not computed to complete dissolution, the
lifetimes of the Newtonian model clusters are estimated as follows:
First, the decay times of the two least massive models (2500 and
$3\,500\,M_\odot$) down to 1/10th of the initial mass are extracted
from the data.  These are defined as their lifetimes,
$t_{\rm life, N}$, being 570 and $854\,$Myr,
respectively.\footnote{Note that an open cluster with
  $M_{\rm oc}=2\,500\,M_\odot$ would, in Newtonian gravitation,
  dissolve in $T_{\rm diss} \approx 3\,$Gyr, Eq.~\ref{eq:lifet}.}
Given the mass ratio of 0.714, this corresponds to an exponent of
about~1.2,
$t_{\rm life, N}/{\rm Myr} \approx 570 \, \left( M_{\rm oc, 0}/
  \left(2\,500\,M_\odot\right) \right)^{1.2}$.  The resulting
approximate lifetimes are then $t_{\rm life, N}\approx 570\,$Myr for
$2\,500\,M_\odot$, $860\,$Myr for $3500\,M_\odot$, $1\,300\,$Myr for
$5\,000\,M_\odot$, $2\,000\,$Myr for $7\,000\,M_\odot$, and
$3\,000\,$Myr for $10\,000\,M_\odot$. For the Milgromian models we
obtain likewise
$t_{\rm life, M}/{\rm Myr} \approx 595 \, \left( M_{\rm oc, 0}/
  \left(2\,500\,M_\odot\right) \right)^{0.95}$. Note that the
Milgromian model with the initial mass $M_{\rm oc}=2500\,M_\odot$
dissolves slightly later (at $595\,$Myr) than the Newtonian model of
the same initial mass ($570\,$Myr). Studying Fig.~\ref{fig:number} one
can see that the red dotted curve has significant fluctuations such
that this difference is not significant. The here estimated PoR
lifetimes merely serve to define the time-scale over which the
velocity dispersion is calculated, and also as an indication of the
ratio, $t_{\rm life, M}/t_{\rm life, N}$, as a consequence of loss of
particles across the pr\'ah through non-relaxational processes
(probably mostly orbital precession, as discussed in
Sec.~\ref{sec:qosc}). The elapsed relative time is then divided by the
lifetime for each cluster model. For the least massive cluster, the
existing simulation data correspond to as much as 1.8~lifetimes since
the simulation ran for $1000\,$Myr and thus continued some $430\,$Myr
after the effective decay of that model. This allows the tidal tails
to be traced because the stellar particles continue to orbit the
Galaxy after the dissolution of the cluster model.

The 3D velocity dispersion in a
model is
\begin{equation}
 \sigma_{\rm 3D} = \left(
    {1\over n_{\rm tid}}     \left( \, \sum_{\rm i = 1}^{
        {n_{\rm tid}}} \, \vec{v}_{\rm i}^2 \right)
  \right)^{1\over2},
  \label{eq:sigma3D}
  \end{equation}
  where $\vec{v}_{\rm i}$ is the velocity vector of particle~i
  relative to the model cluster's centre.

\begin{figure}
\begin{center}
\includegraphics[width=0.48\textwidth]{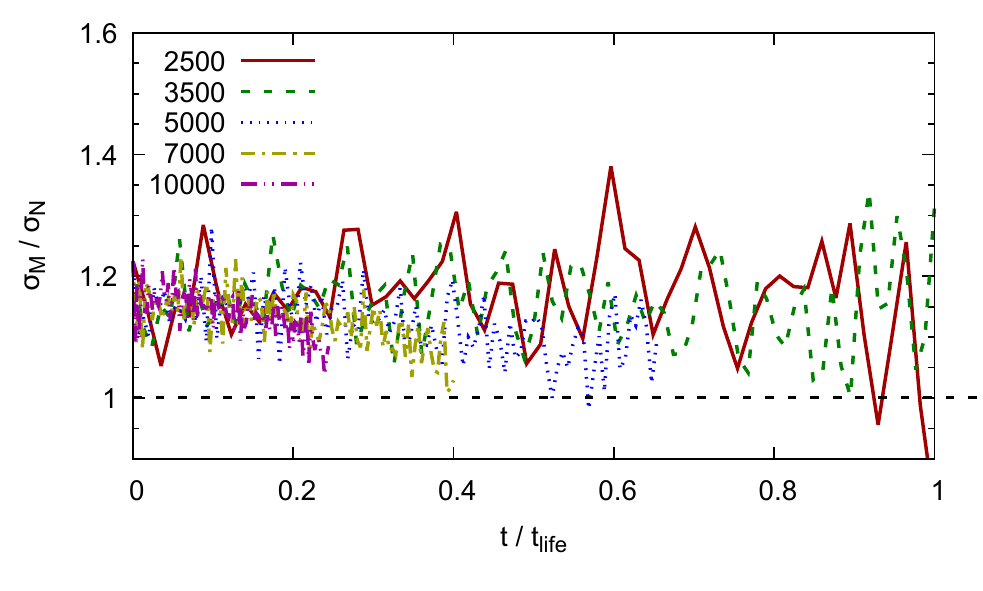}
\caption{The ratio of the Milgromian to Newtonian 3D velocity
  dispersion (Eq.~\ref{eq:sigma3D}), as a function of relative time
  for the five cluster models (the initial masses in $M_\odot$ of
  which are indicated in the legend). The relative time is
  $t/t_{\rm life}=1$ when the model has lost 90~per cent of its
  stellar particles. The data are truncated to the lifetime to focus
  on the most meaningful data range. }
\label{fig:sigmarat}
\end{center}
\end{figure}

In Fig.~\ref{fig:sigmarat}, the ratio of $\sigma_{\rm 3D}$ for the
Milgromian and Newtonian models is shown as a function of time. The
Milgromian models consistently have a velocity dispersion which is
about 20~per cent larger than the Newtonian models with excursions to
ratios as large as~1.4. The tendency is that the less-massive models
show a larger ratio, but given the limitation of the numerical method
applied, it is not possible to assess this ratio for a model with a
mass as low as the observed open clusters in
Table~\ref{tab:clusterdata}.  The analytically calculated Milgromian
velocity dispersion documented in this table under the influence of
the EF provides a similar estimate (compare $\sigma_{\rm M, ef, los}$
with $\sigma_{\rm N, los}$).

Concerning the overall structure of the model tidal tails, snapshots
of the simulation with $M_{\rm oc, 0} = 5\,000\,M_\odot$ in Newtonian
and Milgromian dynamics are shown in Fig.~\ref{fig:5000_600myr} as an
overview (left panels) and zoomed-in (right panels). Note the apparent
asymmetry in the number of particles: the leading (rightwards) tidal
tail has more particles (see the bottom left panel) relative to the
number in the trailing tail in the Milgromian model compared to the
Newtonian one that appears much more symmetrical as should be the case
in Newtonian gravitation (Fig.~\ref{fig:ql2t}). This asymmetry is
consistent with the basic consequence of Milgromian gravitation,
namely that the far-side (to the left) of the cluster has a larger
barrier against escape than the near side, as evident in
Fig.~\ref{fig:rforce}.

The visual appearance of the tail asymmetry near to the Milgromian
model clusters in Fig.~\ref{fig:5000_600myr} resembles the previously
published images of the tidal tails close to the Hyades, Coma
Berenices, Praesepe and COIN-Gaia~13 clusters
(Sec.~\ref{sec:theTails}).

\begin{figure*}
\begin{center}
\includegraphics[width=0.48\textwidth]{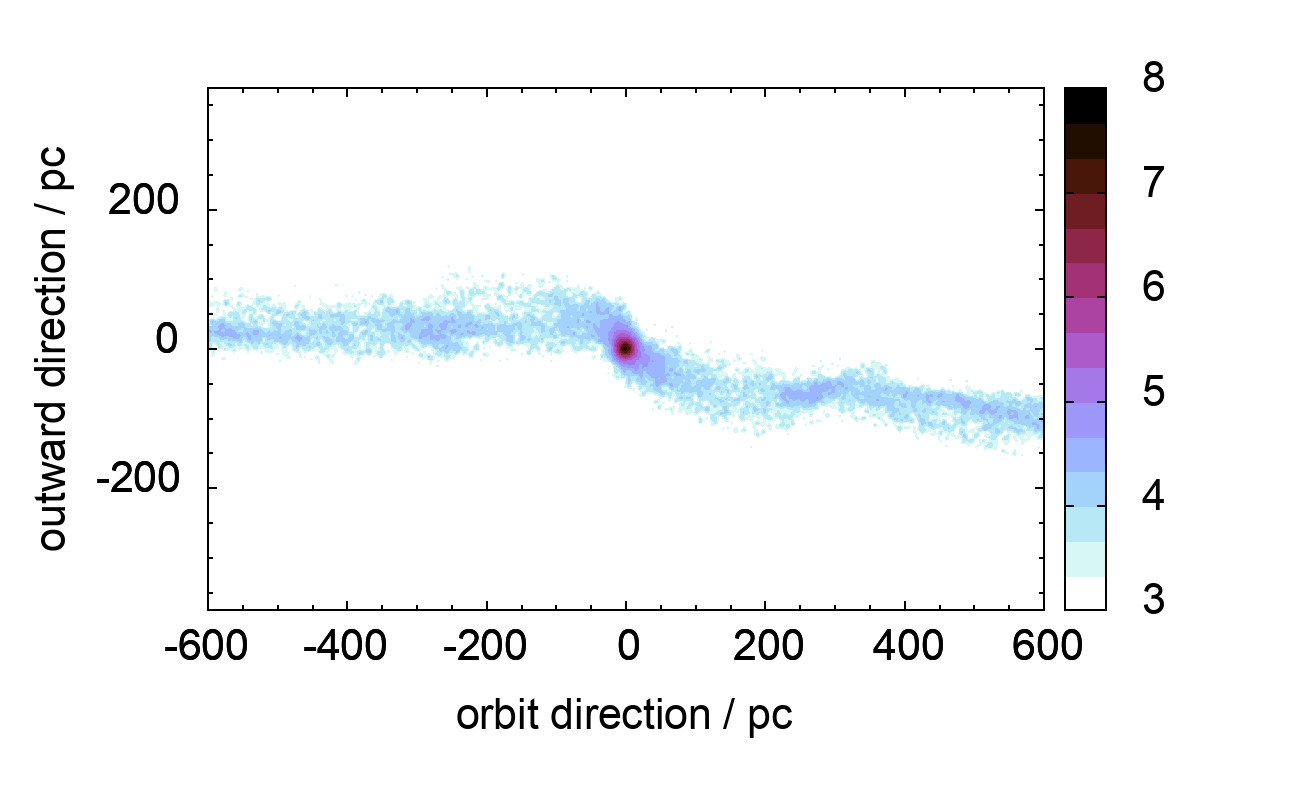}
\includegraphics[width=0.48\textwidth]{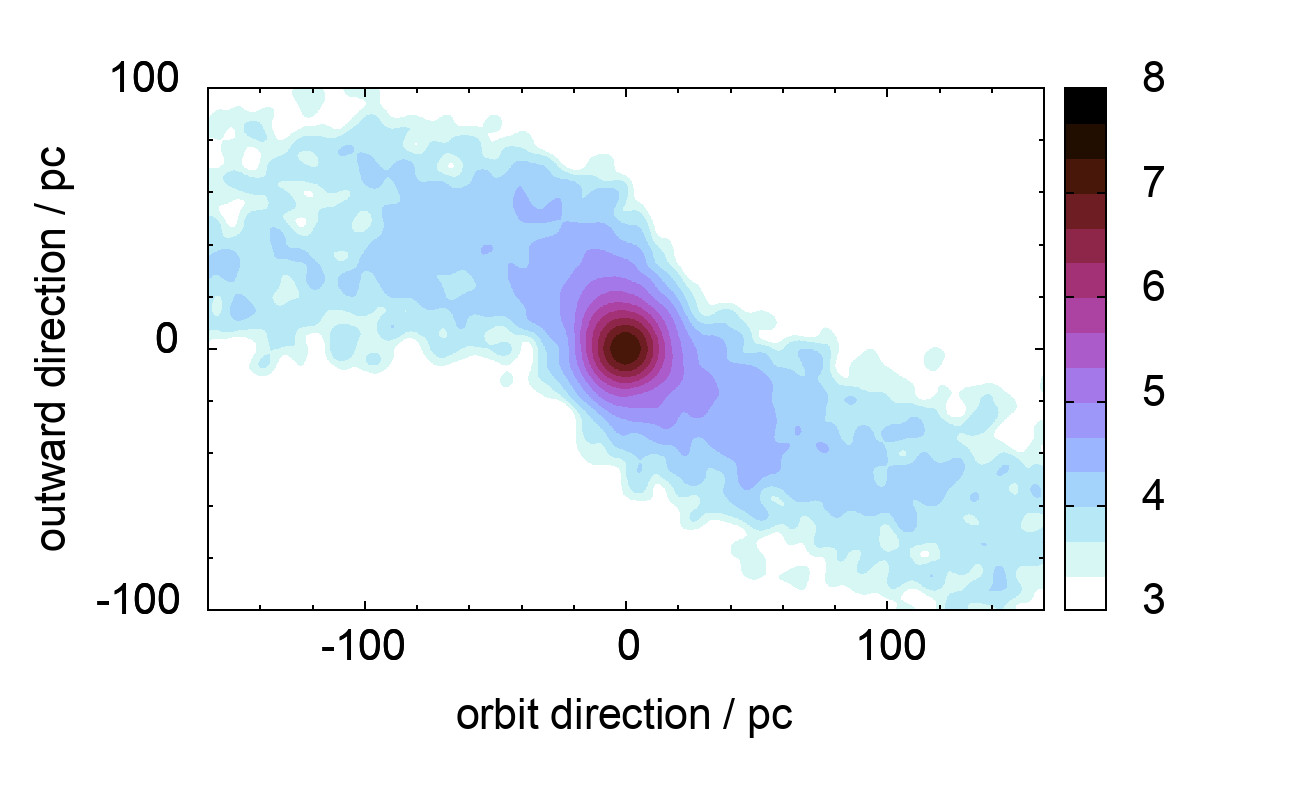}
\includegraphics[width=0.48\textwidth]{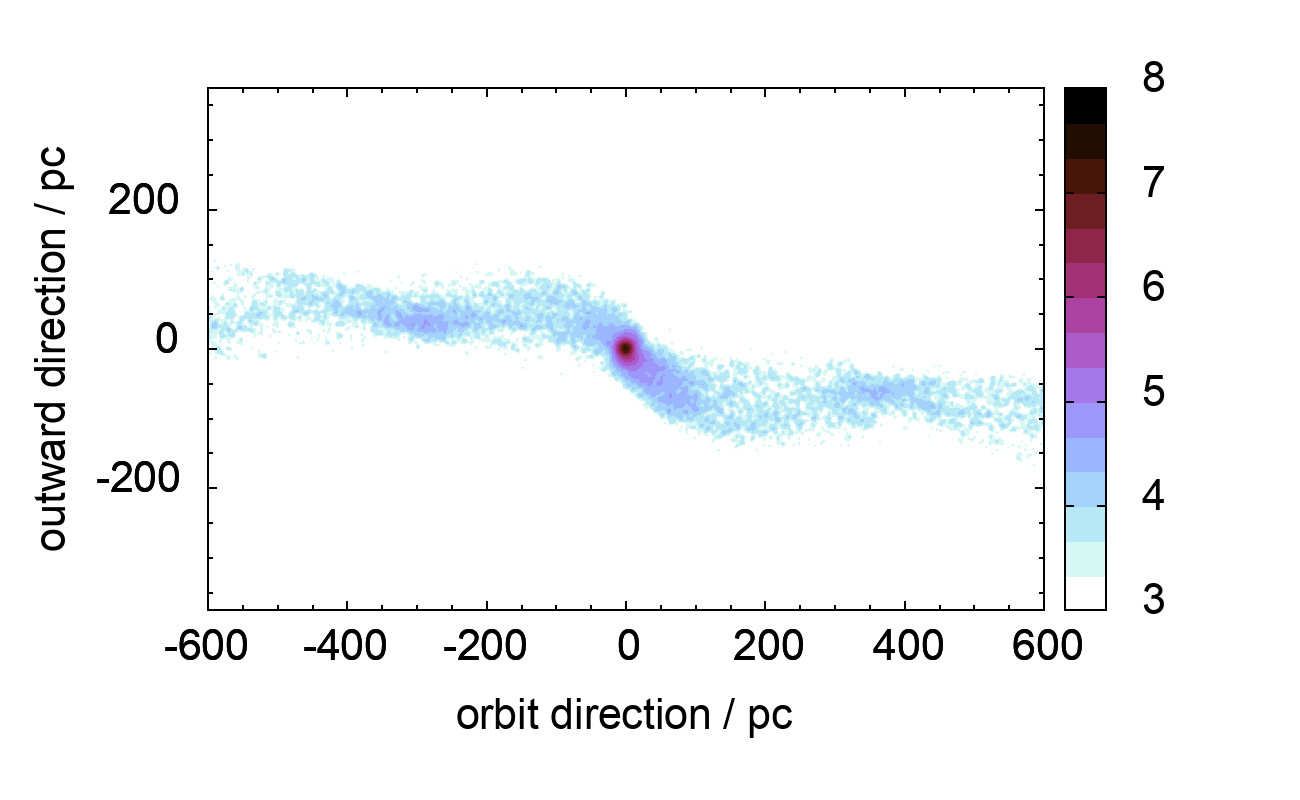}
\includegraphics[width=0.48\textwidth]{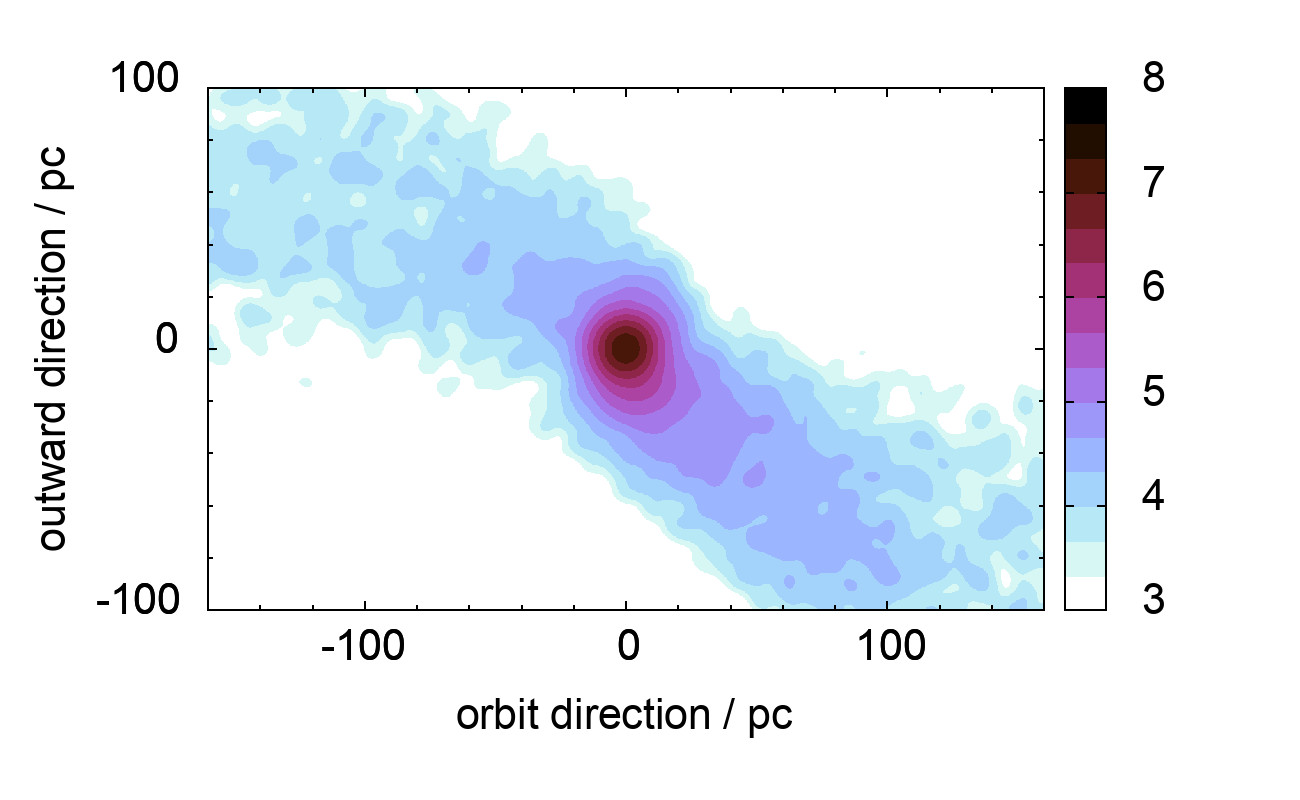}
\caption{Snapshot of the cluster model with initial mass
  $M_{\rm oc, 0}=5\,000\,M_\odot$ at 600~Myr in Newtonian (upper two
  panels) and Milgromian (lower two panels) dynamics.  The right
  panels are detailed views of the model's surroundings. Note the
  lopsided shape of the Milgromian model, which is consistent with
  previous work \citep{Wu+08, Wu+10, Wu+17, Thomas+18}.  The galactic
  center is towards the bottom of the plot and the orbital direction
  is to the right.  The surface density scale (key to the right of
  each panel) is in
  ${\rm log}_{10} \left( M_\odot\,{\rm kpc}^{-2}\right)$ and is for
  illustration only. The color scale has been divided into discrete
  steps to enhance the visibility.  }
\label{fig:5000_600myr}
\end{center}
\end{figure*}

In Newtonian dynamics the first K\"upper overdensities are equidistant
from the open cluster in the leading and trailing tail and move to a
smaller distance from the open cluster as the cluster evaporates
\citep{Kuepper+08, Kuepper+10, Kuepper+12}. The two distances between
the centre of the cluster and the two first overdensities can be used
to measure the effective-Newtonian gravitational mass of the cluster
\citep{Kuepper+15}. We refrain from doing so here since the computed
models are exploratory and we do not want to over-interpret them. The
models are useful for probing the asymmetry between the two distances,
as this may be an additional diagnostic for assessing whether
Newtonian or Milgromian gravitation is closer to reality.  Due to the
cluster-centric potential being asymmetrical in Milgromian gravitation
(Fig.~\ref{fig:rforce}), the distances between the overdensities are
expected to differ. Indeed, in this simulation, the first K\"upper
epicyclic overdensity is located at $\pm 270\,$pc in the Newtonian
models (Fig.~\ref{fig:5000_600myr}), while the Milgromian models have
the leading overdensity located at $+360\,$pc and the trailing one at
$-290 \,$pc. The larger cluster-centric distance in the leading tail
comes from the particles spilling across the pr\'ah into the leading
tail having larger velocities due to the smaller potential barrier at
the first Lagrange point. That both K\"upper overdensities are at
larger distances from the cluster centre in Milgromian dynamics than
in Newtonian dynamics is due to the larger effective-Newtonian
gravitational mass of the former.  This Milgromian asymmetry and
Newtonian symmetry appears to be evident in the observed and modelled
tidal tails of the globular cluster Pal~5 (respectively, fig.~7 and~8
in \citealt{Erkal+17}: the observed leading K\"upper overdensity being
at coordinate $\phi_1 = -0.7$ and the trailing one being at
$\phi_2=+0.6$ with the cluster at $\phi=0$), but its non-circular
orbit complicates the interpretation and no conclusive conclusion can
emanate. Noteworthy in this context is that modelling a satellite
galaxy in Newtonian gravitation, \cite{ReinosoFellhauer18} find the
position of the first K\"upper overdensity to be correlated with the
orbital distance and the mass of the satellite. They also find the
first K\"upper epicyclic overdensity to be closer to the satellite in
the leading tail than in the trailing tail (their fig.~5). The authors
argue that this asymmetry, which is inverted to the asymmetry of the
present Milgromian star-cluster models, ``could be explained by the
fact that we have a much more extended object than used in the
previous study.''  A follow-up study will address this aspect (the
relative distances of the K\"upper overdensities in the leading and
trailing tails) to probe how these can be employed as a test of
gravitational theory.

\subsection{Time-averaged leading-to-trailing tail number ratio,
  $\overline{q}_{\rm 50-200\,pc}$, in terms of model mass and velocity
dispersion}
\label{sec:q1}

As a next step in the analysis, the ratio of the number of particles
in the leading versus the trailing tail per output time, $q(t)$, is
computed for the Newtonian and Milgromian models.  To calculate $q(t)$
it is necessary to distinguish between particles from each tail.  The
coordinate system is first translated to the position of the density
centre of the model cluster. The coordinates refer to the galactic
coordinate system, i.e., the X-axis points towards the galactic centre
and~Y is along the orbital motion.  A dividing line is defined through
the density centre at an angle of 45 degrees (counterclockwise),
therewith lying at about a right angle relative to the inner tidal
tails (i.e., a diagonal from the lower left to the upper right passing
through the position of the cluster centre in
Fig.~\ref{fig:5000_600myr}). The number of particles to the left and
right of this line enumerates, respectively, the numbers in the
trailing and the leading arms.  As for the real clusters
(Sec.~\ref{sec:theTails}), $q_{\rm 50-200\,pc}(t)$ is calculated
between model cluster-centric radial limits of $d_{\rm cl}=50$ and
200~pc (but see Fig.~\ref{fig:qs_multi} for an exploration of
different ranges).  Since the tails consist of particles that drift
away from the models, $q_{\rm 50-200\,pc}(t)$ is here time-averaged
between 400 and 800~Myr, covering the ages of the observed clusters in
Table~\ref{tab:clusterdata}, yielding $\overline{q}_{\rm 50-200\,pc}$.

Fig.~\ref{fig:ql2t} shows $\overline{q}_{\rm 50-200\,pc}$ in
dependence of the model cluster mass. The Newtonian models have
$0.85 < \overline{q}_{\rm 50-200\,pc} < 1.03$, being close to unity
for the smaller model masses but showing an increasing departure
towards $\overline{q}_{\rm 50-200\,pc} <1$ for the more massive models
except for the most massive model.  The Milgromian models, on the
other hand, have $1.01< \overline{q}_{\rm 50-200\,pc} < 1.25$.  Note
the apparent systematic variation of the Milgromian
$\overline{q}_{\rm 50-200\,pc}$ from
$\overline{q}_{\rm 50-200\,pc}\approx 1.25$ for
$M_{\rm oc, 0} = 3\,500\,M_\odot$ to a minumum near
$M_{\rm oc, 0} = 8\,500\,M_\odot$ with an increase again towards
larger masses (the value of $\overline{q}_{\rm 50-200\,pc}$ for
$M_{\rm oc, 0} = 2\,500\,M_\odot$ may be affected by the limited
resolution). For the time being it remains unclear if this systematic
behaviour of $\overline{q}_{\rm 50-200\,pc}$ is a real feature.

\begin{figure}
\begin{center}
\includegraphics[width=0.48\textwidth]{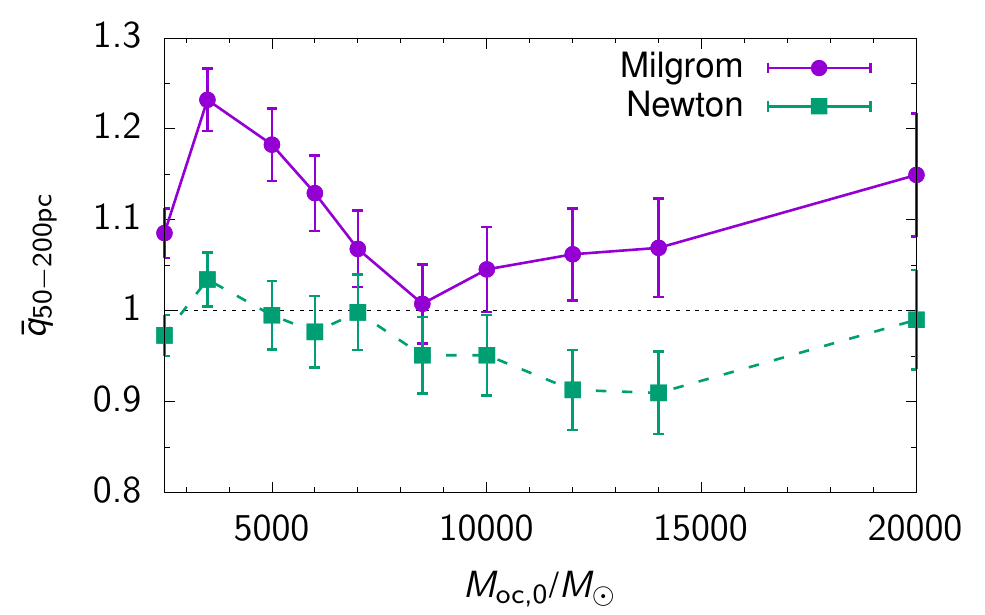}
\caption{The number ratio between the leading and trailing tidal arm
  in the cluster-centric distance range 50--200~pc and averaged over
  400--800~Myr as a function of the initial model cluster mass
  $M_{\rm oc, 0}$.  The green dashed line corresponds to the Newtonian
  models while the purple solid line shows the results for the
  Milgromian models. The errorbars indicate the 5-$\sigma$ Poisson
  errors. The dependence of the ratio on the velocity dispersion is
  shown in Fig.~\ref{fig:qs}.}
\label{fig:ql2t}
\end{center}
\end{figure}

Given the time-averaged systematic change of
$\overline{q}_{\rm 50-200\,pc}$ with $M_{\rm oc,0}$, this behaviour is
next probed in terms of the 3D velocity dispersion,
$\sigma_\mathrm{3D}$, of the cluster models. This might lead to
insights concerning the real open clusters since the models are, by
numerical necessity, much more massive than the real clusters. By
having larger radii, the velocity dispersion falls into the range of
the observed open clusters ($\approx 0.8\,$km/s).

For all cluster models in both Newtonian and Milgromian dynamics, a
Plummer sphere is fitted through the volume density distribution out
to a radius of 20~pc. The free parameters are the Plummer radius,
$\rpl$, and the cluster mass, $\mcl$ (e.g., Fig.~\ref{fig:dens}). The
velocity dispersion, $\sigma_\mathrm{3D}$, is calculated within the
initial tidal radius at $500\,$Myr.  The time of $500\,$Myr ensures
the clusters to be well virialised while the lowest-mass models are
still not too dissolved.  The velocity dispersions of the models is
shown in Fig.~\ref{fig:sigma}. As can be seen from the diagram, the
velocity dispersion in the numerical Milgromian case is about 20\%
higher than in the numerical Newtonian case and about 25\% higher than
in the corresponding analytical Plummer Newtonian models. This is also
evident in Fig.~\ref{fig:sigmarat}.

Fig.~\ref{fig:qs} shows $\overline{q}_{\rm 50-200\,pc}$ for the Newtonian
and Milgromian cases as a function of the respective velocity
dispersion. Both, the Newtonian and Milgromian models show a
comparable shape of the $\overline{q}_{\rm 50-200\,pc}$ vs
$\sigma_{\rm 3D}$ numerical data, with a minimum near
$\sigma_{\rm 3D} \approx 2\,$km/s, but the Milgromian models are
systematically asymmetrical with the leading tail containing
significantly more stellar particles than the trailing tail in the
distance range 50~to~200~pc from the cluster centre.
Fig.~\ref{fig:qs_multi} displays also the additional intervals
50--100~pc and 50--300pc, demonstrating that the
$\overline{q}_{\rm 50-200\,pc}>1.1$ asymmetry remains comparable in the
Milgromian models, while the Newtonian models have values of
$\overline{q}_{\rm 50-200\,pc}$ closer to~1 in all cases.

\begin{figure}
\begin{center}
\includegraphics[width=0.48\textwidth]{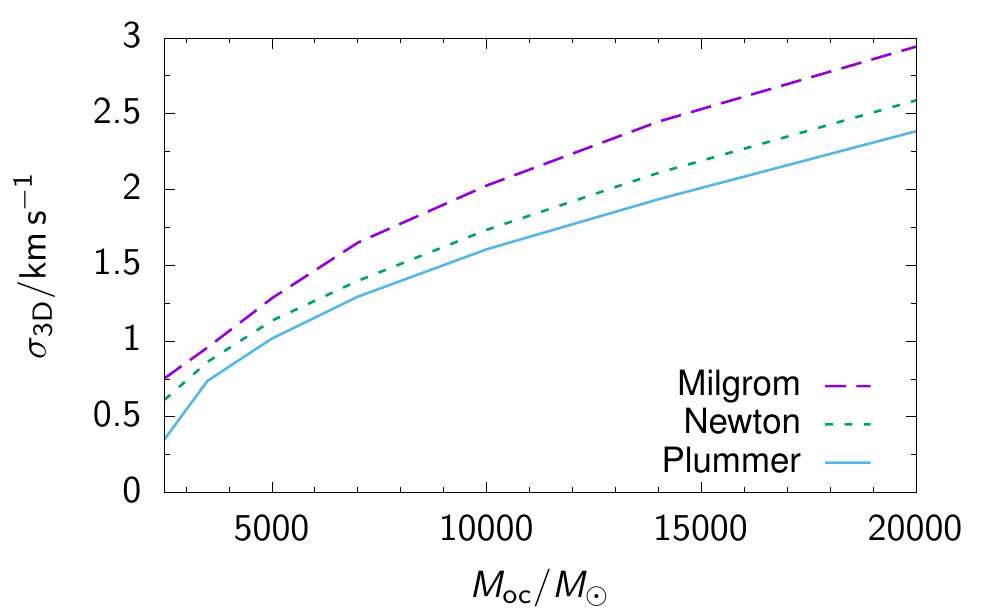}
\caption{The numerically calculated 3D velocity dispersion,
  $\sigma_\mathrm{3D}$, of all particles within the initial
  $r_{\rm tid}(0)$ of the cluster centre, is shown at 500~Myr for the
  Newtonian (short-dashed blue) and Milgromian (long-dashed purple)
  models.  The analytical Newtonian characteristic velocity
  dispersion, $\sigma_\mathrm{ch}$ (Eq.~\ref{eq:plummer}), is plotted
  as the solid blue line.}
\label{fig:sigma} 
\end{center}
\end{figure}

\begin{figure}
\begin{center}
\includegraphics[width=0.48\textwidth]{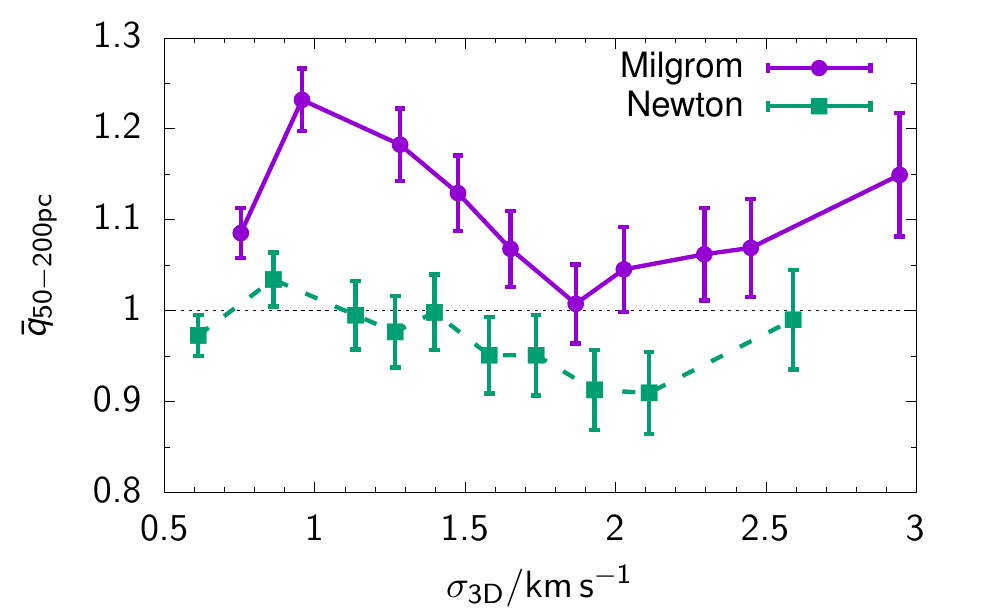}
\caption{Same as Fig.~\ref{fig:ql2t} (the models are the same from
  left to right), but showing $\overline{q}_{\rm 50-200\,pc}$ as a
  function of the velocity dispersion (Fig.~\ref{fig:sigma}) of the
  models at 500~Myr.  Note the same shape of the dependency as in
  Fig.~\ref{fig:ql2t} and that at the same mass, the Milgromian models
  have a larger velocity dispersion.  The observed Hyades and Praesepe
  have an observed 3D velocity dispersion near $0.8\,$km/s
  (Sec.~\ref{sec:oc_MOND}), which, according to these results, is near
  to where the systems achieve the maximum tail asymmetry in the
  Milgromian models.}
\label{fig:qs}
\end{center}
\end{figure}

\begin{figure}
\begin{center}
\includegraphics[width=0.48\textwidth]{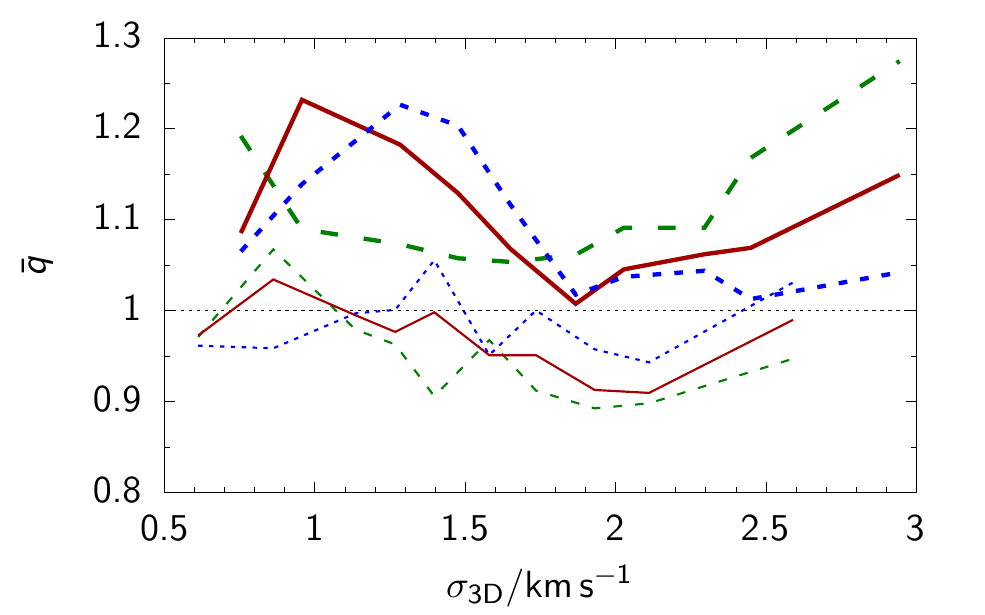}
\caption{Same as Fig.~\ref{fig:qs}, but for different radial averaging
  intervals. Milgromian (Newtonian) models are shown as thick (thin)
  lines. The solid line corresponds to the 50--200~pc interval, the
  dashed line to the 50--100~pc interval and the dotted line to the
  50--300~pc interval. The errorbars are omitted for better visibility
  and because the errors are similar for all averaging intervals.  }
\label{fig:qs_multi}
\end{center}
\end{figure}

\subsection{$q_{50-200\,{\rm pc}}$ as a function of time: models vs observations}
\label{sec:qtim}

How do these results compare with observations?

A direct comparison between the model- and
real-cluster-$q_{\rm 50-200\,pc}$ values needs to be made with due
caution because it is not clear if the values actually depend on the
mass or the velocity dispersion of the clusters, and because the
models suggest the Milgromian prediction of $q_{\rm 50-200\,pc}$ to
vary with the Galactocentric orbital phase (as discussed below), with
this information not being available for the real clusters.  But the
comparison with the observations is valuable because it shows that the
observed tidal tail asymmetry is not expected in Newtonian dynamics
while it can occur in the Milgromian framework.  The asymmetry ratio,
$q_{\rm 50-200\,pc}$, of the real clusters
(Table~\ref{tab:clusterdata}) cannot be plotted versus the theoretical
initial mass, $M_{\rm oc,0}$ (to be equivalent to Fig.~\ref{fig:ql2t}
for the models), because the present-day masses of the real clusters
are highly uncertain. These have not been well constrained,
propagating through to a significant uncertainty in their initial
masses, $M_{\rm oc, 0}$.  The ages of the real clusters are better
constrained though, and Fig.~\ref{fig:indq} plots $q_{\rm 50-200\,pc}$
vs age for these.

The Hyades (square) has $q_{\rm 50-200\,pc} = 2.53\pm 0.37$ and for
NGC~752 (upside down open triangle)
$q_{\rm 50-130\,pc} = 1.30\pm 0.24$, Coma Berenices (circle) has
$q_{\rm 50-200\,pc} = 1.2 \pm 0.15$ and the Praesepe (triangle) has
$q_{\rm 50-200\,pc} = 0.62\pm 0.08$.  The latter case appears to
contradict the expectation from the above theoretical analysis that
$q_{\rm 50-200\,pc} > 1$ always if Milgromian gravitation were to be
correct. Note that this apparent contradiction would not salvage
Newtonian gravitation because of the simultaneous (now on-going)
similar asymmetry of the Hyades, Coma Berenices and NGC~752 tails.  A
clue to this problem is obtained by noting that the data may suggest
an age sequence of $q_{\rm 50-200\,pc}$.

This is tested for in Fig.~\ref{fig:indq} by plotting the temporal
evolution of $q_{\rm 50-200\,pc}(t)$ for five models with
$\sigma_{\rm 3D}<2\,$km/s constituting computationally-reachable
approximate conformity with the observed open clusters.  For
$t \simless 100\,$Myr the models are evolving into an equilibrium
mass-loss rate across the $d_{\rm cl}=50\,$pc to~$200\,$pc distance
range.  After $\approx 400\,$Myr, the Newtonian models are limited to
$0.6<q_{\rm 50-200\,pc}(t)<1.4$, while the Milgromian models have
$q_{\rm 50-200\,pc}(t)$ oscillating near-periodically between~0.6
and~2.5.  The present computations indicate the clusters to show an
encouraging agreement with the Milgromian models within the
1$\,\sigma$ error ellipse since these rise to the level of the
observed asymmetry (Fig.~\ref{fig:indq}).  Noteworthy is that the
Milgromian models show oscillations in $q_{\rm 50-200\,pc}(t)$, with
it increasing to larger values when the models are near perigalacicon
to afterwards fall slightly below $q_{\rm 50-200\,pc} =1$.

As a caveat and reminder though: while consistency with the Milgromian
models is evident, the Newtonian PoR models are inconsistent at more
than 5sigma confidence only with the Hyades datum, with the other
three clusters being also consistent with the Newtonian PoR
models. Before reaching final conclusions on which theory of
gravitation is valid, more observational work is needed to improve the
quantification of the tidal tail asymmetry.

\begin{figure*}
\begin{center}
\includegraphics[width=1.0\textwidth]{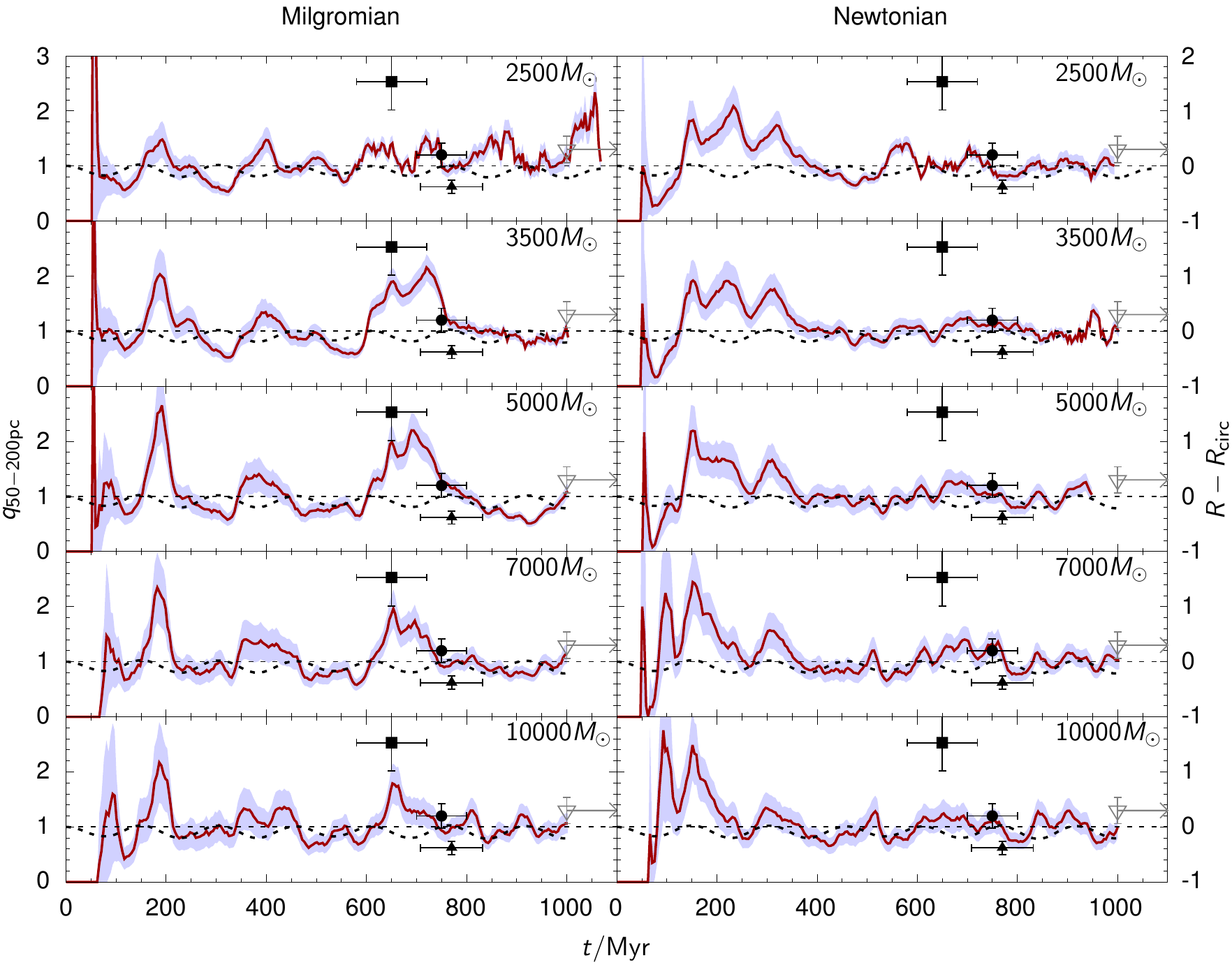}
\caption{The number ratio, $q_{\rm 50-200\,pc}(t)$, for five
  Milgromian (left panels) and five Newtonian (right panels) models as
  a function of time. The horizontal dashed line marks
  $q_{\rm 50-200\,pc}(t)=1$ (values $>1$ correspond to more stars in
  the front tail while values $<1$ correspond to the trailing tail
  having more stars, i.e., the dashed horizontal line is a marker for
  perfect symmetry in both tails for the left y-axis, and marks the
  Galactocentric radius of a circular orbit at~8.3~kpc for the right
  y-axis) and the initial masses, $M_{\rm oc, 0}$, are written into
  the panels. The shaded region corresponds to the Poisson $1\,\sigma$
  uncertainty range at each time.  The data for the four open clusters
  listed in Table~\ref{tab:clusterdata}, for which the extended tidal
  tails have been extracted using Gaia eDR3 with the Jerabkova-CCP
  method (Sec.~\ref{sec:theTails}), are shown as the symbols (square:
  Hyades, circle: Coma Berenices, triangle: Praesepe, upside down open
  triangle: NGC~752). For each cluster, the error-bar on
  $q_{\rm 50-200\,pc}$ is the $1\,\sigma$ Poisson uncertainty and the
  error-bar on the age is the age range given in
  Table~\ref{tab:clusterdata}, except for NGC~752 which has an age of
  $\approx 1.75\,$Gyr not reached by the present simulations. For
  NGC~752, $q_{\rm 50-130\,pc}$ is shown.  Note that the Newtonian
  models fluctuate erratically around the value of
  $q_{\rm 50-200\,pc}(t) \approx 1$ while the Milgromian models show
  quasi-periodic excursions to $q_{\rm 50-200\,pc}(t) > 1$ and
  $q_{\rm 50-200\,pc}(t) <1$ with maxima near perigalacticon. The
  dotted line is the galactocentric distance in kpc
  ($R=|^{{{\rm j}}=6}\vec{R}|$, Eq.~\ref{eq:RGC}) relative to a
  circular orbit at $R_{\rm circ}=8.3\,$kpc. The model cluster orbits
  thus have apo-galactica at $\approx 8.3\,$kpc and peri-galactica at
  $\approx 8.1\,$kpc.  The model clusters are on approximate rosette
  orbits such that the apo-galactica are $<360^\circ$ distant from
  each other and occur about every $150\,$Myr while the orbital period
  is about 209~Myr (Fig.~\ref{fig:rc}).  The computations indicate
  that the Milgromian models show a pronounced asymmetry in the tidal
  tails when the model clusters are near peri-galacticon.}
\label{fig:indq}
\end{center}
\end{figure*}

\subsection{Why does $q_{50-200\,{\rm pc}} (t)$ oscillate?}
\label{sec:qosc}

With the aim to shed some light on the question why the Milgromian
models have an oscillating $q_{\rm 50-200\,pc}(t)$ with maxima
occurring near the cluster's peri-galactica, the velocity dispersion
and spin angular momentum of the cluster models are studied next. The
notion is that groups of stellar particles might be moving in a
correlated manner within the cluster to exit across the pr\'ah
together as the Milgromian cluster potential adjusts, causing the
momentary flare-like increase of $q_{\rm 50-200\,pc}(t)$.

The bulk 3D velocity dispersion of the stellar particles within the
tidal radius, $\sigma_{\rm 3D}(t)$ (Eq.~\ref{eq:sigma3D}), is plotted
in Fig.~\ref{fig:indsigma}.  Neither the Milgromian nor the Newtonian
models show a near-periodic change in $\sigma_{\rm 3D}(t)$ that
resembles the quasi-periodic evolution of $q_{\rm 50-200\,pc}(t)$ in
Fig.~\ref{fig:indq}, such that the notion that the Milgromian cluster
might experience an internal kinematically instability cannot,
herewith, be affirmed.

\begin{figure*}
\begin{center}
\includegraphics[width=1.0\textwidth]{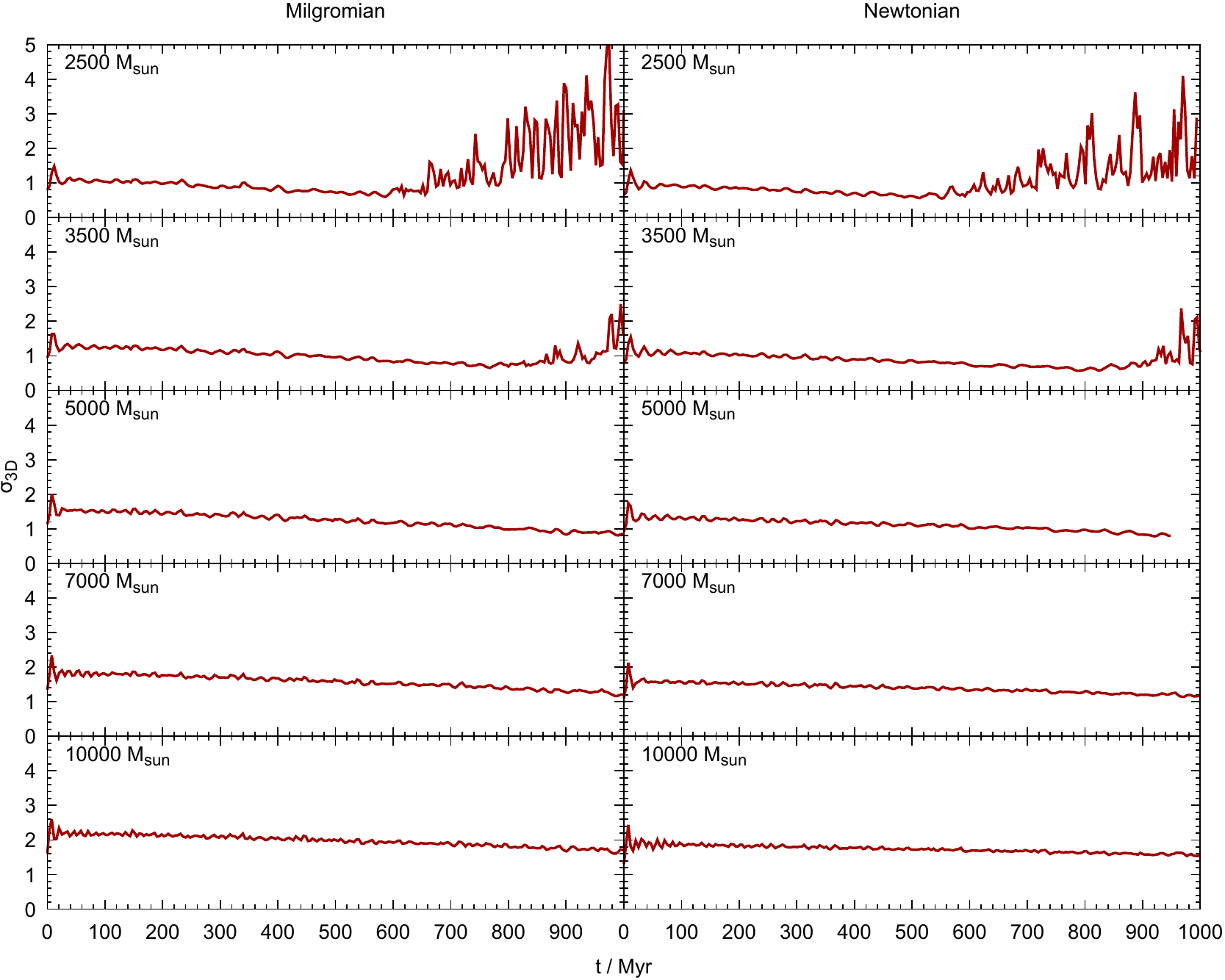}
\caption{As Fig.~\ref{fig:indq} but plotting (w/o the Poisson
  uncertainties) the 3D velocity dispersion of particles within the
  initial tidal radius (Eq.~\ref{eq:sigma3D}). The velocity dispersion
  decreases as the models loose particles and the initially lighter 
  models show erratic fluctuations in $\sigma_{\rm 3D}(t)$ after they
  have largely dissolved. }
\label{fig:indsigma}
\end{center}
\end{figure*}

The specific angular momentum per stellar particle for particles within the
initial tidal radius, $r_{\rm tid, 0}$ (Eq.~\ref{eq:rtid}), of their
model cluster centre is calculated as
\begin{equation}
  \vec{L}_{/n} =
          {1\over n_{\rm tid}}     \left( \sum_{\rm i = 1}^{n_{\rm
                tid}}
            \vec{r}_{\rm i} \times \vec{v}_{\rm i} \right),
  \label{eq:L}  
\end{equation}
where $\vec{r}_{\rm i} , \vec{v}_{\rm i}$ are, respectively, the
position and velocity vectors of particle~i relative to the model
cluster's centre.  The Z-axis points towards the galactic north pole
and the model clusters orbit anticlockwise about the galactic
centre. The specific inner angular momentum per particle of a model,
$\vec{L}_{/n2}$, is calculated for all particles within half of
$r_{\rm tid}$.

\begin{figure*}
\begin{center}
\includegraphics[width=1.0\textwidth]{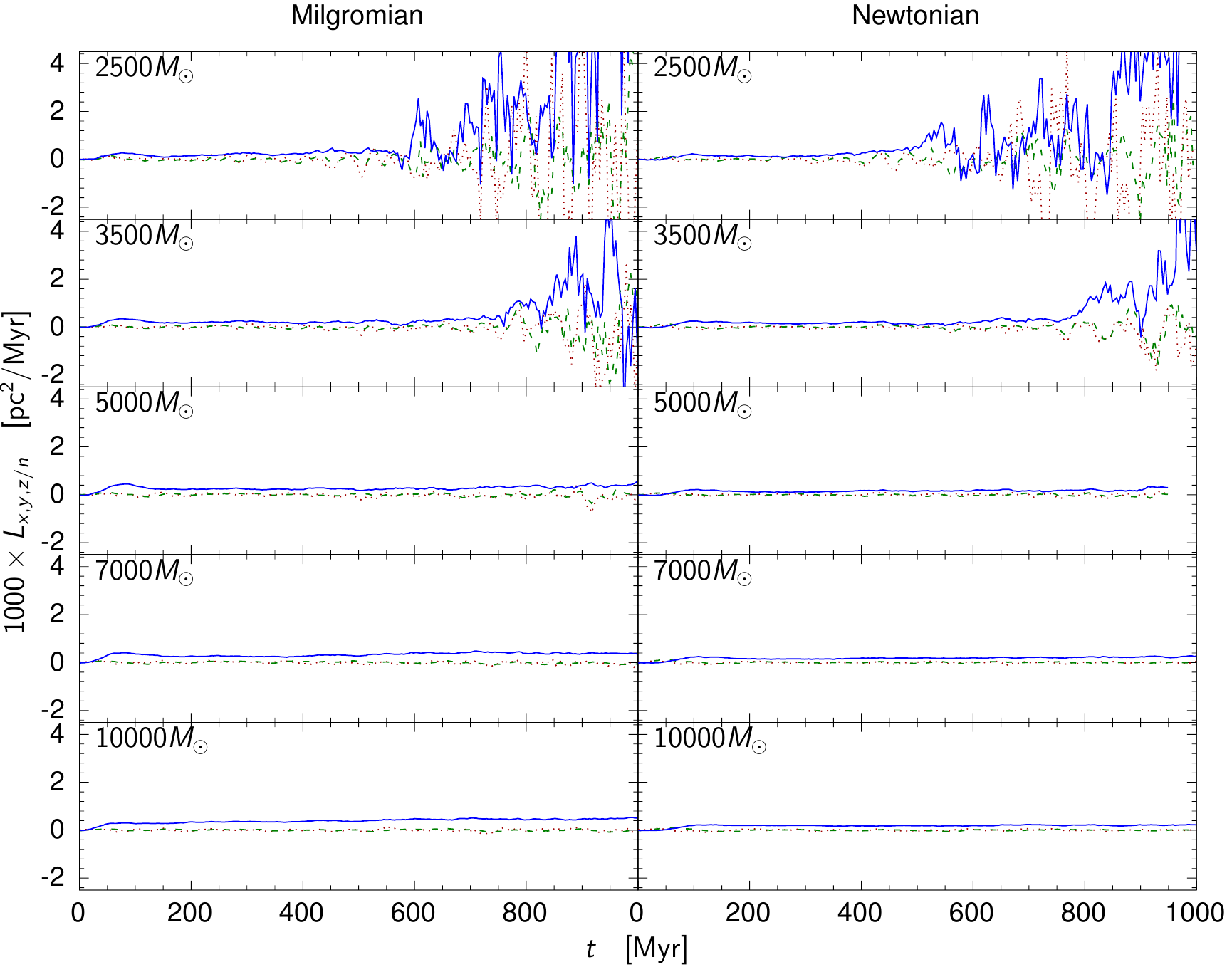}
\caption{As Fig.~\ref{fig:indq} but plotting (w/o the Poisson
  uncertainties) the X- (dotted red line), the Y- (dashed green line)
  and the Z-component (solid blue line) of the time-evolution of the
  cluster's specific spin angular momentum per particle, $\vec{L}/n$
  (Eq.~\ref{eq:L}, using all particles within the initial
  $r_{\rm tid}$ of each model). A positive $Z$-component is
  antiparallel to the angular momentum of the cluster's orbit around
  the Galaxy. Note that the Milgromian models typically have larger
  spins which is consistent with their larger mass loss and the
    argument of cluster spin-up in Sec.~\ref{sec:qosc}. The
    time-derivative of $L_{{\rm z}/n}$ is plotted in
    Fig.~\ref{fig:dLz}.  }
\label{fig:indL}
\end{center}
\end{figure*}
The temporal evolution of the components of $\vec{L}_{/n}$ is plotted
in Fig.~\ref{fig:indL}. It is evident that the models are initially
non rotating, but that rotation develops with the loss of stellar
particles. The Newtonian and Milgromian models follow a similar
evolution. All models spin up rapdily within the first
$\approx 100\,$Myr due to the inital settling phase into the galactic
potential which is associated with enhanced loss of particles.

This spin-up of initially not-rotating open clusters that fill their
tidal radii is well understood and has been found to be the case also
for the evolution of the Milky Way satellite galaxies
\citep{Kroupa97}: The escaping stellar particles preferably stem
  from the pro-grade population within the cluster as this is
  energetically favoured \citep{Henon70, Read+06}. An unbound star on
  a pro-grade cluster-centric orbit experiences a smaller tidal radius
  than other stars. It may overtake the cluster slightly but by
  retarding from it towards a larger galactocentric distance it will
  fall behind and populate the trailing tail. If it transits the
  pr\'ah near the inner Lagrange point at smaller galactocentric
  distance it will initially fall behind the cluster but will move
  ahead of it due to the increasing angular orbital frequency with
  smaller galactocentric distance, populating the leading tail.  This
  pro-grade population can be re-populated after its loss from the
  cluster through artificial grid-relaxation in the models here and
  two-body relaxation in real clusters. From Fig.~\ref{fig:indL} it
  can be seen that the Milgromian models have a faster spin up which
  continues over time, while the Newtonian models show a near
  negligible rise in the spin after the the first $\approx
  100\,$Myr. This indicates that artificial grid relaxation is
  negligible (the Newtonian models do not re-populate their pro-grade
  stellar population significantly at the expense of the retrograde
  population) confirming the collision-less nature of these models.
  That the Milgromian models continue to spin up over time (the blue
  lines have a small positive gradient) indicates that another process
  is at work that re-populates the pro-grade population within the
  models at the expense of the retro-grade orbits, and that this
  process accelerates. It needs to accelerate because the pro-grade
  population needs to be lost more rapidly than the retrograde
  population is depleted into the pro-grade one, in order for the
  cluster spin to increase.  Since the equivalent Newtonian models
  rule out artificial grid relaxation to be significant, this process
  is probably related to the precession of stellar orbits within the
  Milgromian clusters through the external field.

Indeed, in Milgromian gravitation, the EF leads to a star's orbit
within the EF-dominated cluster to be torqued and thus to precess with the rate
$\dot{\Omega}$.  An estimate of the precession rate can be obtained by
adapting eq.~34 in \cite{Banik+20}, although it is not clear that it
is fully applicable to the present fully EF-dominated case,
\begin{equation}
	\dot{\Omega} = { f_{\rm g} \, G\, M_{\rm oc} \,a_0  \,  r_{\rm
            st} \, {\rm sin}(\theta) \, {\rm cos}(\theta)
        \over  2\, v_{\rm st} \, r_{\rm efe}^3 }  .
	\label{eq:precession}
\end{equation}
Assuming $M_{\rm oc} = 275\, M_\odot$, its
$r_{\rm efe} \approx 0.29\,$pc (eq.~31 in \citealt{Banik+20}:
$r_{\rm efe} = \sqrt{G\,a_0\,M_{\rm oc}}/a_{\rm ext, kin}$ is the
radius in the cluster beyond which the EF dominates;
$a_{\rm ext, kin}\approx 7.5\,$pc/Myr$^2$ from
Table~\ref{tab:clusterdata}), the inclination between the direction of
the EF and the orbital angular momentum of the star in the cluster,
$\theta=45^{\rm o}$, maximises the sin-cos product, and
$f_{\rm g} = 1/2$ is a geometric factor that accounts for azimuthal
averaging. Eq.~\ref{eq:precession} suggests a precession rate of
$\approx 240 \, {\rm rad}$/Myr for a star at $r_{\rm st} = 5\,$pc
moving with a circular speed of $v_{\rm st}= 0.5\,$pc/Myr. This is
extremely rapid, implying a precession-induced orbital instability.
Note that
\begin{equation}
  \dot{\Omega} \propto M_{\rm oc}^{-0.5}\, ,
	\label{eq:fasterprecession}
\end{equation}
such that the precession speeds up as the cluster looses mass. This
may be the reason for the increasing spin of the Milgromian models
noted above. Also,
\begin{equation}
\dot{\Omega} \propto a_{\rm ext, kin}^3 \, ,
\end{equation}
suggests that the process of precession of cluster-centric stellar
orbits increases significantly at peri-galacticon, such that the
evaporation rate may be more sensitive to the orbital eccentricity of
the open cluster in Milgromian than in Newtonian gravitation.

The above estimate for $\dot{\Omega}$ is rough and other stars will
experience significantly different precession rates, but it indicates
that the EF is likely to have a highly significant systematic effect
on the orbital structures and orbital angular momenta orientations of
stars in an open cluster and may thus constitute an additional
important contribution to the energy-redistribution process in
Milgromian gravitation. This {\it EF-relaxation process} demands
further investigation as it should also be important in globular
clusters.

Every star cluster on a near-circular galactocentric orbit that is
older than $t\approx 100\,$Myr (conservatively $200\,$Myr), that had
revirialised after gas expulsion with a negligible spin (but see
\citealt{Mapelli17} for models of forming rotating clusters and
\citealt{Henault-Brunet+12} for observational evidence for rotation in
a very young massive cluster) will thus spin-up to rotate with a spin
oppositely directed to its orbital angular momentum around the
galactic centre.\footnote{Globular clusters are typically on chaotic
  rosette-type orbits about the Galactic centre such that their spins
  are not likely to be well correlated with their Galactocentric
  orbital angular momenta.  The interested reader is referred to the
  comprehensive Newtonian--$n$-body study of rotating globular
  clusters by \cite{Tiongco+16, Tiongco+17, Tiongco+18, Tiongco+21,
    Tiongco+22}. \cite{Bianchini+18} find significant evidence for a
  non-negligible spin of 11~out of~51 globular clusters but do not
  place this in relation to the orbital angular momentum of the
  clusters.  \cite{Sollima20} analyses Gaia-selected member stars
  around 18~globular clusters, finding evidence for tidal tails in
  seven of them with five having asymmetric tails (the directions of
  motion of the clusters are not given though). This asymmetry may be
  related to the asymmetry discussed here, but is likely affected
  significantly by the eccentricities of the cluster orbits. Flat
  outer velocity dispersion profiles around some globular clusters
  have been reported \citep{Scarpa+11}.}  This is shown in
Fig.~\ref{fig:indL} in that the z-component of $\vec{L}_{/n}$,
$L_{{\rm z}\;/n}$, increases and is positive while the orbital angular
momentum of the cluster about the Galaxy is directed towards the
negative $Z-$direction: if the initially non-rotating open cluster
orbits the Galaxy in an anti-clockwise direction, then the open
cluster will begin to rotate clockwise.

According to Fig.~\ref{fig:indL}, the evolution of $\vec{L}_{/n}$
  does not show fluctuations that would support the notion that
  correlated particle motions lead to the near-periodic increases and
  decreases of $q_{\rm 50-200\,pc}(t)$. The rate of change of
  $L_{{\rm z}/n}(t)$ and of $L_{{\rm z}/n2}(t)$ (using only particles
  within half of the initial $r_{\rm tid}$) are plotted in
  Fig.~\ref{fig:dLz} and~\ref{fig:dLz2}, respectively. The models with
  $M_{\rm oc, 0}\simless 4000\,M_\odot$ are barely resolved given the
  small density contrast of the cluster relative to the field, and the
  more reliable more massive models indicate that the Milgromian
  clusters keep spinning up at a constant rate,
  ${\rm d}L_{{\rm z}/n}/{\rm d}t \approx 0.22 \times 10^{-6} \, {\rm
    pc}^2 / {\rm Myr}^2$, independently of their initial mass (similarly
  for the inner part), while the Newtonian models experience a
  significantly smaller spin-up over time which decreases with
  increasing model mass (see \citealt{Tiongco+16} for an in-depth
  $n$-body study).  Given the explorative and grid-based approximative
  nature of this work, we do not further analyse the rate of spin up
  in relation to cluster mass loss rate.

\begin{figure}
\begin{center}
\includegraphics[width=0.48\textwidth]{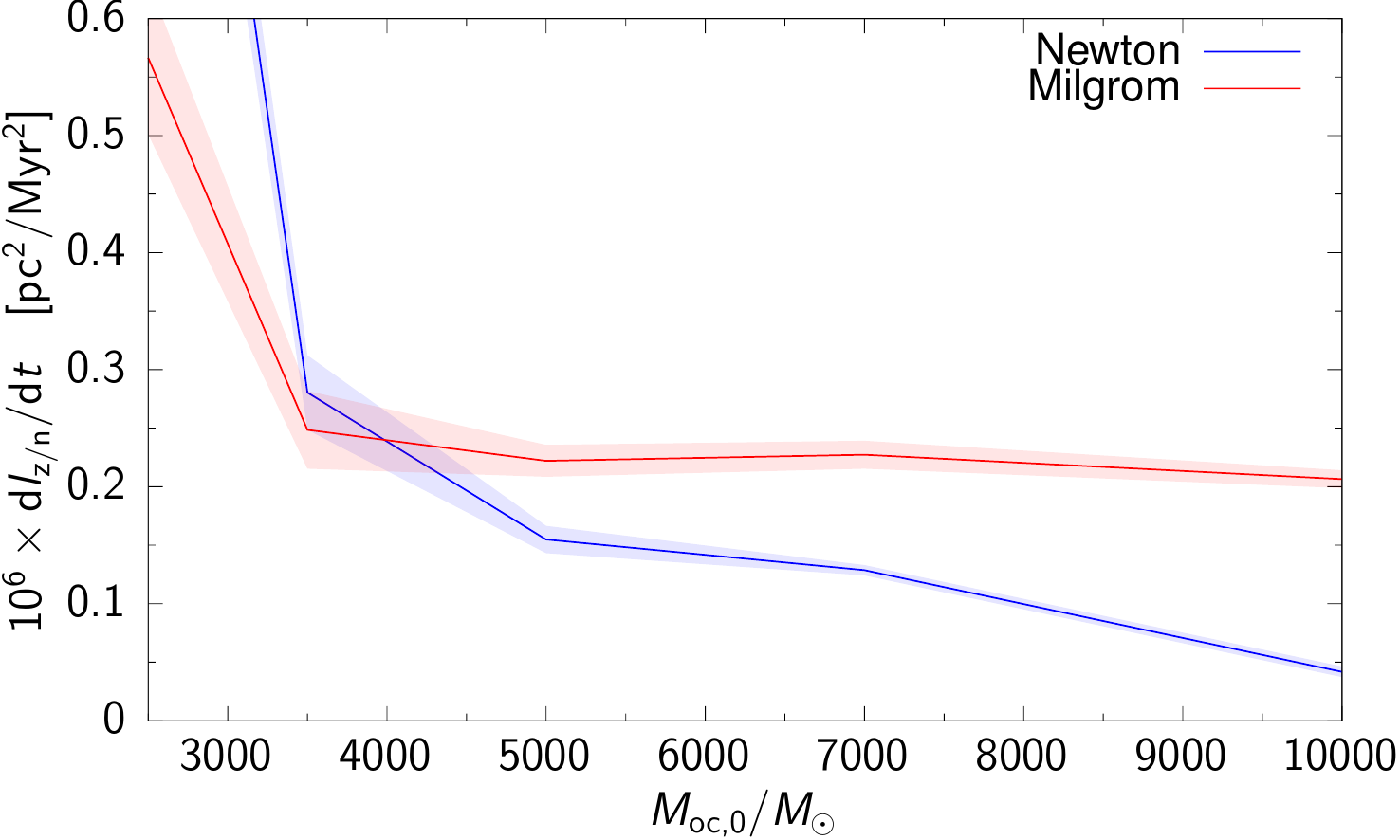}
\caption{The rate of change of the Z-component of the specific spin
  angular momentum per particle of the model clusters,
  ${\rm d}L_{{\rm z}/n} /{\rm d}t$, in dependence of the initial model
  mass, $M_{\rm oc, 0}$, for all particles within the initial tidal
  radius.  The shaded regions corresponding to the $1\,\sigma$
  uncertainty in the linear fit to $L_{{\rm z}/n}(t)$
  (Fig.~\ref{fig:indL}). For this purpose, the linear fit is
  restricted to the following time intervals: 200--500~Myr
  ($M_{\rm oc, 0}=2500\,M_\odot$), 200--750~Myr
  ($M_{\rm oc, 0}=3500\,M_\odot$), 200--1000~Myr
  ($M_{\rm oc, 0}>3500\,M_\odot$).  Fig.~\ref{fig:dLz2} shows the time
  derivative of $L_{\rm z, /n2}$ for particles within half the tidal
  radius.}
\label{fig:dLz}
\end{center}
\end{figure}

\begin{figure}
\begin{center}
\includegraphics[width=0.48\textwidth]{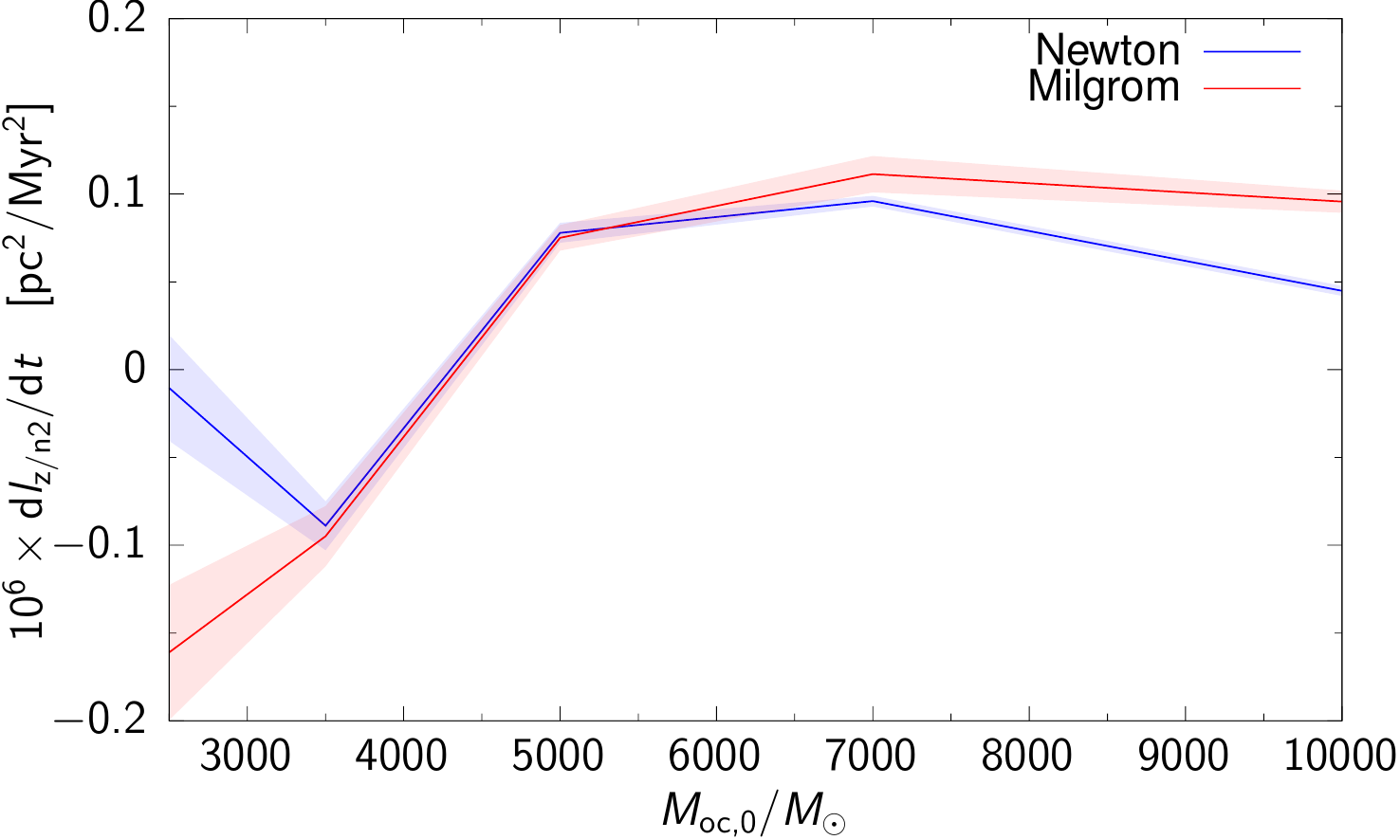}
\caption{As Fig.,~\ref{fig:dLz} but for all particles within half the
    initial tidal radius. }
\label{fig:dLz2}
\end{center}
\end{figure}

The explorative calculations performed so far show Milgromian
  models to preferentially loose stars across their pr\'ahs through
  the leading tidal tail. This leads to them spinning up with time
  (Fig.~\ref{fig:indL}) and also leads to growth of their orbital
  eccentricity (Eq.~\ref{eq:ecc}) since the clusters decelerate
  through this one-sided out-sourcing of their stars. The initial
  orbital eccentricities of the present models are $e\approx 0.012$,
  and using the method of Appendix~A to calculate the momentary value
  of $e(t)=e_{\rm snap}$ from the position and velocity vectors of the
  model centres (Eq.~\ref{eq:esnap}), the time evolution of
  $e_{\rm snap}(t)$ can be studied.

  The evolution of $e_{\rm snap}(t)$ is shown for the Milgromian model
  with $M_{\rm oc,0}=5000\,M_\odot$ in Fig.~\ref{fig:e_5000}. It is
  evident that, when the model looses more stars across its pr\'ah
  into the leading tail, then $e_{\rm snap}$ increases as a
  consequence of the loss of momentum of the cluster, while, when the
  trailing tail receives more stars, $e_{\rm snap}$ decreases
  again. Due to the time-averaged asymmetry the clusters loose more
  stars into the leading tail (Fig.~\ref{fig:ql2t}, \ref{fig:qs}
  and~\ref{fig:qs_multi}) and thus continuously decelerate therewith
  steadily increasing their overall $e(t)$ with time.

\begin{figure}
\begin{center}
\includegraphics[width=0.48\textwidth]{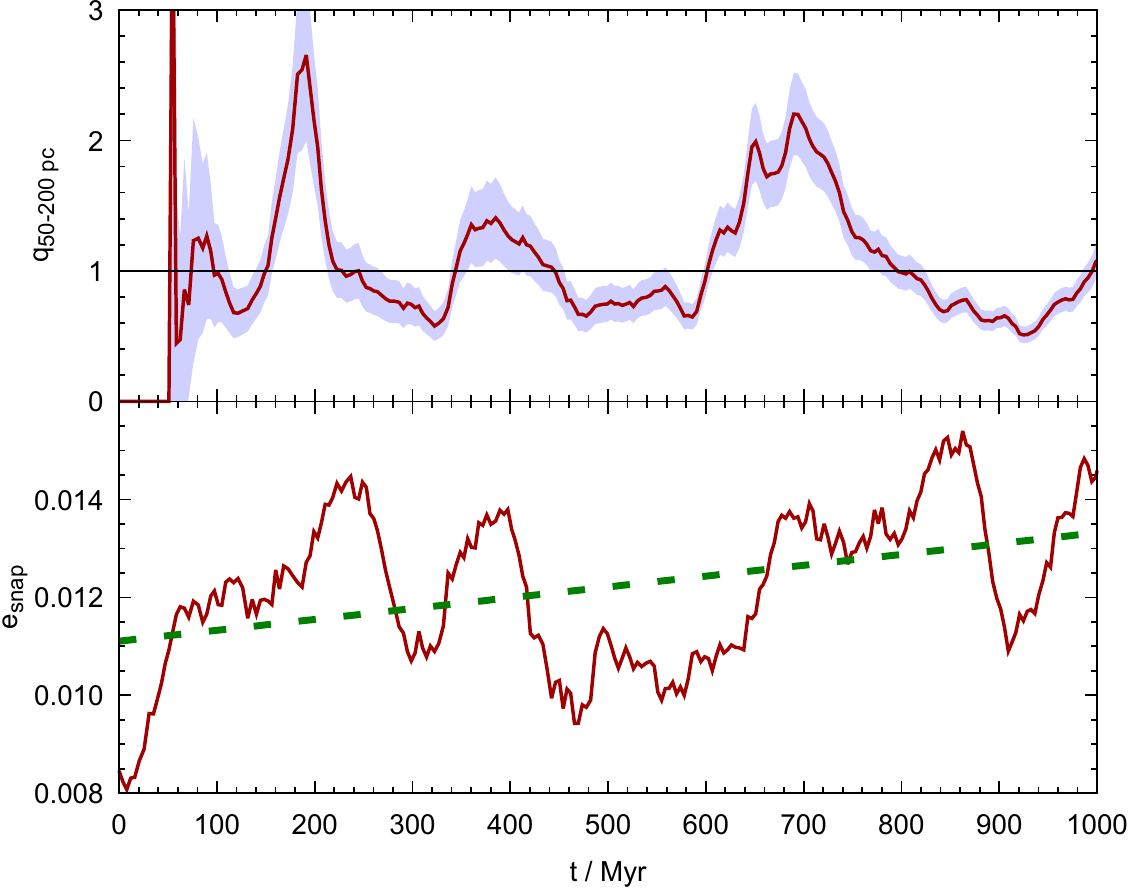}
\caption{The evolution of the orbital eccentricity, $e_{\rm snap}(t)$,
  of the Milgromian model with $M_{\rm oc,0}=5000\,M_\odot$ (lower
  panel) in comparison with the evolution of $q_{\rm 50-200\,pc}(t)$
  (upper panel, same as in Fig.~\ref{fig:indq}). The orbital
  eccentricity varies as $q_{\rm 50-200\,pc}$ does, but is retarded by
  about 40~Myr for the first two maxima. The dashed line shows the
  overall linear trend of $e_{\rm snap}(t)$.}
\label{fig:e_5000}
\end{center}
\end{figure}

While the Newtonian models approximately retain their initial orbital
eccentricity, the Milgromian models thus show an approximately
linearly increasing $e_{\rm snap}(t)$. By performing a linear
regression on the $e_{\rm snap}(t)$ data (as shown for example in
Fig.~\ref{fig:e_5000}), the gradient, ${\rm d}e_{\rm snap}/{\rm d}t$
is evaluated and shown in Fig.~\ref{fig:decc}.  Interestingly, the
Milgromian models have a similar time-gradient
${\rm d}e_{\rm snap}/{\rm d}t \approx 1.8 \times 10^{-6} \, {\rm
  Myr}^{-1}$ at more than $5\,\sigma$ confidence, while the Newtonian
models have a gradient consistent within $2\,\sigma$ with zero. The
present Milgromian models thus increase their orbital eccentricity by
20~percent over a time of 1~Gyr.

This confirms the supposition that the one-sided loss of stars
decelerates the Milgromian models such that their orbital eccentricity
increases with time. A more detailed analysis of this problem is
beyond this explorative study, given the limited access to
computational resources, and is relegated to a follow-up study.  This
explorative work does, however, suggest that in Milgromian dynamics,
open star clusters would self-destroy as they spin up. The
asymmetrical loss of stars dominantly into the leading tail leads to a
deceleration of the cluster, a decreasing peri-Galacticon which
increases the asymmetrical loss to a potentially catastrophic level as
the process appears to be subject to positive dynamical
feedback. Clearly, further research on this process of open-cluster
suicide is needed to reach more secure conclusions.

\begin{figure}
\begin{center}
\includegraphics[width=0.48\textwidth]{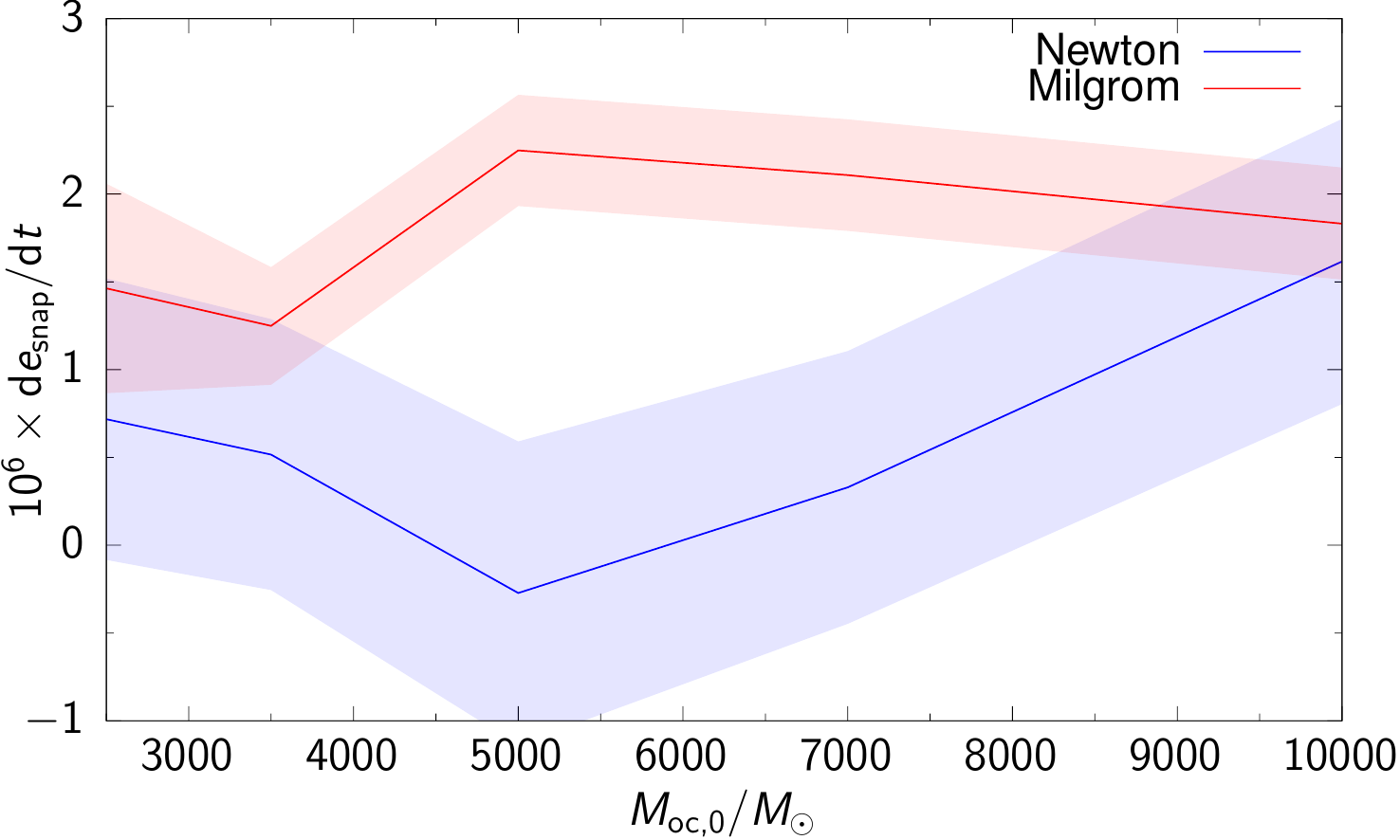}
\caption{The rate of change of the orbital eccentricity in dependence
  of the initial model mass, $M_{\rm oc,0}$. The shaded regions
  correspond to the $1\,\sigma$ uncertainty in the linear fit to
  $e_{\rm sn}(t)$.  }
\label{fig:decc}
\end{center}
\end{figure}

\subsection{Emergence from molecular cloud and lifetimes}
\label{sec:efmix}

In Milgromian gravitation, galactic disks have a stronger
self-gravitation than in Newtonian plus dark-matter-halo models,
leading to an enhanced star-formation rate per unit gas mass and also
to star-formation extending to larger galactocentric radii
\citep{Zonoozi+21}. The emergence of star clusters from their
molecular clouds of birth involves a phase transition from the
Newtonian deeply embedded cluster phase to the Milgromian open cluster
configuration.

From the half-mass-radius--embedded-cluster-mass-in-stars relation,
$r_{\rm h}/{\rm pc} = 0.10\,\left(M_{\rm ecl}/M_\odot \right)^{0.13}$
\citep{MarksKroupa12}, it follows, together with Eq.~\ref{eq:rmond},
that $r_{\rm M} > r_{\rm h}$ for $M_{\rm ecl} > 4\,M_\odot$ for a
star-formation efficiency of 33~per cent.  It is interesting to note
that $M_{\rm ecl}\approx 5\,M_\odot$ corresponds to the least-massive
embedded clusters that are observed in nearby molecular clouds
\citep{KroupaBouvier03, Joncour+18}.  This also means that the
precursors of all open and globular star clusters form in the
Newtonian regime.  But, the least-massive embedded clusters have a
MOND radius comparable to their half-mass radius, suggesting they
would form in the EF-dominated Milgromian gravitational regime, i.e.,
with an effectively larger gravitational constant, $G_{\rm eff}$.  The
emergence from the natal molecular cloud and the re-virialisation of a
part of the embedded cluster will occur mostly in the Milgromian EF
dominated spatially asymmetrical regime (Fig.~\ref{fig:rforce}) such
that the presently known limits on bound fractions of this process
(e.g., \citealt{BK03a, BK03b, Brinkmann+17, Farias+18, Dom+21}) may
need adjustment.  This problem has been studied by \cite{WuKroupa18,
  WuKroupa19} indicating the rich stellar-dynamical evolution and
possible outcomes. A larger fraction of stars can remain bound to the
freshly formed open cluster due to the larger $G_{\rm eff}$ with a
reduced fraction of stars to be found in tidal tail~I
(Sec.~\ref{sec:introd}) compared to Newtonian models. These processes
will need future attention to better understand the kinematics and
shapes of natal cocoons.

Observations of extragalactic open star clusters appear to show
  these to dissolve unexpectedly quickly
  \citep{FallChandar12,Chandar+17}. The dissolution appears to be
  faster than expected from Newtonian $n$-body simulations in smooth
  galactic potentials \citep{ Dinnbier+22a}. The lifetimes of real
  open clusters are subject to assumptions on their orbits, the IMF,
  binary fraction, their initial sizes and masses (e.g.,
  \citealt{Mapelli17, Ballone+20, Semadeni20, Ballone+21}).  The
  calculated models are thus degenerate to various combinations of the
  parameters (unless all open clusters form following the same
  mass-radius relation, the same IMF and initial binary population and
  the Galactic potential is smooth, cf. \citealt{MarksKroupa12,
    Dinnbier+22a}). Therefore, the reported {\it short life-time
    problem} needs more research and should not, for the time being,
  be taken as conclusive evidence in the one or other direction. Given
  this situation, it is relevant to ask: Is it possible that
Milgromian open clusters dissolve faster than Newtonian ones?

In Sec.~\ref{sec:mass_veldisp} it was found that the particle-mesh
Milgromian models dissolve more rapidly than the Newtonian models of
the same mass due to Milgromian clusters becoming more unbound per
unit mass loss.  In addition to this general process, the
energy-equipartition-driven stellar loss is likely to be faster for
Milgromian clusters than for Newtonian clusters.  \cite{Ciotti+04}
showed that the two-body relaxation time-scale is significantly
shorter in Milgromian clusters, although reliable values are not
available and their computation is only valid for clusters in
isolation. Ignoring this (truly fundamental) limitation for the
moment, we can compute the relaxation time ignoring the external
field.  The following two estimates can be made: The lifetime of a
Newtonian open cluster (e.g., \citealt{BT}),
$T_{\rm diss, N} = \gamma \, t_{\rm relax}$, where $\gamma \approx 19$
when the cluster is in the field of the Solar neighbourhood and
$t_{\rm relax}$ is the median two-body relaxation time
scale. According to \cite{Ciotti+04}, the ratio of the two-body
relaxation time in Milgromian to Newtonian gravitation is
\begin{equation}
{t_{\rm relax, M} \over t_{\rm relax,N}} = 1.4\, \left(1 +
  \cal{R} \right)^{-{5/2}},
\label{eq:citotti}
\end{equation}
where
${\cal R} = \left(M_{\rm oc, grav} - M_{\rm oc}\right) / M_{\rm
  oc}$. For example, for the Hyades, $M_{\rm oc}=275\,M_\odot$ is the
stellar mass (Table~\ref{tab:clusterdata}) while the
effective-Newtonian gravitational mass is about a factor of four
larger \citep{Roeser+11}. Thus, for the Hyades,
$t_{\rm relax, M} / t_{\rm relax, N} \approx 0.044$ such that
Milgromian open clusters would dissolve about 23~times more rapidly
than Newtonian ones (taking the Hyades as representative).  A
twenty-fold shortening of the life-times is ruled out by the
observation that Hyades-type clusters with a birth mass near
$1300 \,M_\odot$ \citep{Jerabkova+21a} have life-times longer than
600~Myr while the Newtonian expectation is 2~Gyr
(Sec.~\ref{sec:fillup}). The empirical evidence suggests a shortening
at most by not much more than a factor of two.

Another estimate can be obtained by remembering that
$t_{\rm relax, N} \approx \left(2\,N / {\rm ln}\left(N/2\right)
\right)\,t_{\rm cross}$, where
$t_{\rm cross} = 2\,r_{\rm h} / \sigma_{\rm 3D}$ is the half-mass
crossing time.  If the boost in velocity dispersion is only 25~per
cent using the EF-dominated MOND estimates
(Sec.~\ref{sec:mass_veldisp}), then the lifetimes would be shorter by
only about 20~per cent.  Assuming that, for some yet unknown reasons,
the collisional treatment would reproduce the observed velocity
dispersion, then the relaxation time and thus the life-time would be
reduced by a factor of two for the Hyades.

Another estimate for the shortening of the lifetimes of Milgromian vs
Newtonian clusters can be obtained as follows: the lifetime of a
Newtonian cluster, $T_{\rm diss, N}$, is given by Eq.~\ref{eq:lifet}.
As noted in Sec.~\ref{sec:mass_veldisp}, open clusters are in the
EF-dominated regime such that they follow Newtonian dynamics. If
$T_{\rm diss, M}$ is the lifetime of the Milgromian cluster, then
$T_{\rm diss, M}/T_{\rm diss, N} = \left( G / G_{\rm eff}
\right)^{1/2} = \sqrt{\mu} \approx 0.8$ (Eq.~\ref{eq:mu}). This
estimate thus implies EF-dominated clusters to have a lifetime which
is about 80~per cent that of the corresponding Newtonian cases.

The above estimates thus indicate that the lifetimes of open clusters
in Milgromian gravitation are between 20 to~50~per cent of those in
Newtonian gravitation.  It remains to be studied how star-cluster
populations evolve in a Milgromian galaxy, noting that observational
evidence suggests that the dissolution rate of star clusters may be
independent of their mass \citep{FallChandar12, Chandar+17}. From the
results of Sec.~\ref{sec:mass_veldisp} it is tentatively suggested
that
$t_{\rm life, M}/ t_{\rm life, N} \approx 7.4\,\left( M_{\rm oc} /
  M_\odot \right)^{-0.25}$. While the lifetimes of Newtonian star
clusters in the Solar neighbourhood lengthen with increasing
$M_{\rm oc}$ due to the increasing two-body relaxation time
approximately according to $T_{\rm diss} \propto M_{\rm oc}^{0.79}$
(Eq.~\ref{eq:lifet}), this suggests that the lifetimes of EF-dominated
Milgromian clusters may at the same time be reduced due to the larger
loss in binding energy per unit mass loss. The combination of these
two effects may lead to life-times of EF-dominated star
clusters being less-dependent on the cluster mass,
$t_{\rm life, M} \simprop \left(M_{\rm oc}\right)^{0.5}$.

The rapidity with which open clusters dissolve is dependent on how
quickly the orbital energies of the stars in the cluster are
redistributed through weak gravitational encounters. The observation
in Sec.~\ref{sec:qosc} that the Milgromian models spin-up over time
while the Newtonian ones do not may indicate another contribution to
the more rapid dissolution of Milgromian clusters: the rapid
precession of stellar orbits in Milgromian clusters leads to a more
rapid depopulation of the cluster-centric retro-grade orbits which
precess into pro-grade orbits that are preferentially lost. This
process accelerates with time (Eq.~\ref{eq:fasterprecession}) as the
cluster's mass decreases, leading to cluster suicide.

For completeness, further results on the richness of the Milgromian
gravitational dynamics of open star clusters, globular clusters and
dwarf galaxies can be found in \cite{BradaMilgrom2000} who study the
dynamical influence of the EF on orbiting satellite galaxies. Thus the
asymmetry of an ultra-diffuse dwarf galaxy's potential (see
Fig.~\ref{fig:rforce}) will affect the morphology of its tidal
features with implications for the interpretation of its dark matter
content in Newtonian gravitation (cases in point being the
dark-matter-lacking dwarf galaxies NGC1052-DF2/DF4, \citealt{FMF18,
  Kroupa+18, Keim+21, Montes+21}).  \cite{WuKroupa13} analyse the
phase transition a massive star cluster experiences on a radial orbit
when moving from the Newtonian into the outer Milgromian regime, and
\cite{Thomas+18} demonstrate that the asymmetry of a cluster's
potential leads to asymmetrically long tidal tails~II explaining the
unequal lengths of the observed tidal tails of the globular cluster
Palomar~5.

\section{Summary and Conclusion}
\label{sec:concs}

Twenty per cent to a half of all stars in a galaxy pass through a
classical tidal tail~II which is fed through the evaporation process
of their star cluster of origin. The remainder are lost from their
embedded clusters through gas expulsion forming, together with the
stars from other embedded clusters that formed in the same molecular
cloud, dispersing natal cocoons about individual revirialised open
clusters. After $\approx 200\,$Myr the classical tidal tail~II becomes
the dominant coeval population within a few hundred~pc of the open
cluster. In Newtonian dynamics and for a smooth galactic potential,
the stars cross the pr\'ah of their cluster symmetrically at the inner
and outer Lagrange points leading to symmetrical classical tidal tails
within Poisson noise (Pflamm-Altenburg et al., in prep.). But for all
open clusters for which tail data are available the leading tail
contains more stars than the trailing tail within a cluster-centric
distance of $d_{\rm cl}\approx 50\,$pc (Sec.~\ref{sec:theTails}).

With the introduction of the compact convergent point method by
\cite{Jerabkova+21a}, it has now become possible to map-out the
extended ($d_{\rm cl}>50\,$pc) phase-space distribution of coeval
stars in the classical tidal tail~II around nearby open clusters using
Gaia eDR3. In the case of the Hyades, the tidal tail is asymmetrical
with the leading tail containing significantly more stars than the
trailing tail (Fig.~\ref{fig:realtails}). This was interpreted by
\cite{Jerabkova+21a} to possibly be due to a recent encounter with a
massive perturber, which would have damaged the trailing tail, thereby
also leading to dynamical heating of the Hyades which would
consequently dissolve within a few cluster crossing times. This is
consistent with the observed super-virial state of the real cluster (a
factor of~four missing in mass assuming Newtonian gravitation,
\citealt{Roeser+11, OE20}). A problem with this scenario is that the
Sun and Hyades are in the Local Bubble such that a sufficiently
massive molecular cloud does not exist nearby to the Hyades and the
existence of dark matter sub-halos remains to be speculative
(Sec.~\ref{sec:properties}). Furthermore, the about three times older
open cluster NGC~752 shows a comparable, albeit much less significant,
asymmetry (\citealt{Boffin+22}, Fig.~\ref{fig:indq}) requiring a
similar encounter, making this explanation unlikely.

Can the asymmetry in the tidal tails be explained if gravitation is
Milgromian rather than Newtonian?  \cite{Mil83} noted that the open
star clusters would be in the Milgromian regime with their internal
gravitational forces being dominated by the external field from the
Galaxy (Sec.~\ref{sec:oc_MOND}).  The work of \cite{Wu+08, Wu+10,
  Wu+17} showed the equipotential surfaces of satellites (dwarf
galaxies or star clusters) to be lopsided in Milgromian gravitation,
therewith breaking the symmetry between the inner and outer Lagrange
points. A star moving away from the centre of its hosting cluster will
experience a larger radial backwards force when moving away from the
galactic centre (Fig.~\ref{fig:rforce}).  This may be the reason for
the number of stars in the leading tidal tail being larger than in the
trailing tail for $d_{\rm cl}\simless 50\,$pc for all five clusters
for which tail data exist (Sec.~\ref{sec:theTails}).  But, the older
open cluster Coma Berenices has a nearly symmetrical tidal tail in the
distance range $50 < d_{\rm cl}/{\rm pc}< 200$ while the Praesepe has
slightly fewer stars in the leading tail in this same distance range
from the cluster (Fig.~\ref{fig:indq}). This appears to contradict a
Milgromian interpretation of the data without salving the disagreement
with Newtonian gravitation for $d_{\rm cl}\simless 50\,$pc for all
five clusters and for $d_{\rm cl}>50\,$pc for the Hyades and NGC~752.

Given the non-existence of a relaxational Milgromian $n$-body code,
the symmetry-problem of tidal tails is here approached by applying the
existing collision-less adaptive-mesh refinement PoR code
(Sec.~\ref{sec:results}) to follow how stellar particles leak across
the pr{\'a}h of their model cluster. The equivalent Newtonian
computations verify that the tails are largely symmetrical, while the
Milgromian calculations demonstrate them to be significantly
asymmetrical. This confirms the expectation that a Milgromian star
cluster orbiting the Galaxy on a near-circular orbit in the MW disk
looses more stars through the leading tidal tail (Sec.~\ref{sec:q1}).

The calculations show the ratio between the number of stellar
particles in the leading to the trailing tail to oscillate
near-periodically, reaching the high observed values of the Hyades and
decreasing to the Newtonian symmetry reaching a small but reversed
asymmetry (Fig.~\ref{fig:indq}).  The maximum asymmetry is reached
when the models are near peri-galacticon while the asymmetry nearly
disappears at apo-galacticon.  The observed tidal tails of the Hyades,
Coma Berenices and the Praesepe appear to follow the theoretical time
evolution of the asymmetry ratio (Fig.~\ref{fig:indq}).  While the
asymmetry and ages of the models and real clusters agree, it is not
clear if the real Hyades and NGC~752 are in periastron, and if
Coma~Ber and Praesepe are near apo-galacticon. An interesting problem
to consider would be to search for a Galactic potential which places
the Hyades and NGC~752 near peri-galacticon and Coma~Ber and the
Praesepe near apo-galacticon, given the observational constraint on
their velocity vectors. In future work the contribution from the
Galactic bar needs to be taken into account, as well as orbital
oscillations about the Galactic mid-plane.

The calculations performed for this explorative study indicate a
possible physical mechanism for the oscillating behaviour of the
50--200~pc tail asymmetry. While the Newtonian models retain an
approximately constant orbital eccentricity, it grows with time for
the Milgromian models because these loose more stars across their
pr\'ah into the leading tail than the trailing tail. The physical
process of the loss of stars (Sec.~\ref{sec:qosc}) may be driven by
the rapid precession of stellar orbits within the cluster due to the
EF, leading to an increasingly rapid EF-relaxational redistribution of
orbits as the cluster looses mass, therewith growing the orbital
eccentricity.  This process, if true, would accelerate cluster death.

A prediction of this work is that open star clusters that are
initially non-rotating and older than about~200~Myr show a spin which
is opposite to the orbital angular momentum of the cluster
(Sec.~\ref{sec:qosc}). Another prediction is that in Milgromian
  gravitation the first K\"upper epicyclic overdensity lies further
  from the cluster in the leading tail than the trailing tail, both
  being more distant from the cluster centre than in Newtonian
  dynamics (Sec.~\ref{sec:mass_veldisp}).

The estimates in Sec.~\ref{sec:efmix} suggest Milgromian open clusters
to have life times that are~20 to 50~per cent of those of Newtonian
open clusters of the same initial mass because the two-body relaxation
process is faster. This may have a bearing upon the observation
\citep{FallChandar12} that open clusters appear to dissolve more
rapidly than expected from Newtonian $n$-body computations.


In the future it will be important to verify and better measure the
tidal tails for open star clusters covering a larger range of
ages. This will allow an elaboration on the present findings and an
assessment of the evaporation rate of stars into the leading and
trailing tidal tails as well as the measurement of the locations of
the K\"upper epicyclic overdensities as more refined tests of
Milgromian gravitation.  Given the present results based on a
collision-less method and the result that two-body relation is much
more significant in Milgromian systems than in Newtonian ones
\citep{Ciotti+04}, and to achieve theoretical advances, it will be
necessary to develop a collisional $n$-body code in Milgromian
gravitation to allow open star clusters to be evolved
self-consistently. This is a very major mathematical and computational
challenge, with currently no clear solution in sight. An ansatz could
be to discretise the generalised Poisson equation
(Sec.~\ref{sec:genPois}) and possibly to use an iterative procedure to
calculate the instantaneous effective-Newtonian gravitational mass of
each star in the cluster, and/or to just use the original definition
of the MONDian force equation \citep{Mil83} with the usual particle
summation over the {\it particle-distance}$^{-2}$ mass terms
(Pflamm-Altenburg, in prep.).  Close encounters between stars and
multiple systems and their perturbations will need to be treated
according to regularisation methods \citep{Heggie74, MikkolaAarseth90,
  MikkolaAarseth93, Funato+96, MikkolaAarseth96} and stellar-evolution
will need to be implemented through fast look-up tables
\citep{Hurley+00, Hurley+02, Banerjee+20}.  The disks of galaxies are
entirely self-gravitating in MOND, enhancing the star-formation rate,
as opposed to the Newtonian case where the dark matter halo dominates
the potential \citep{Zonoozi+21}. A Milgromian $n$-body simulation
code will be mandatory to address the emergence through gas expulsion
of open and globular clusters from their deeply-embedded Newtonian
state (Sec.~\ref{sec:efmix}), the rate with which stars mass-segregate
and evaporate, how long open star clusters live and how their orbits
evolve.

Gravitation remains the least-understood physical
phenomenon. Different interpretations of this phenomenon have been
proposed, as a geometrical distortion of space-time through matter
\citep{Einstein16}, as an emergent property related to the information
content of space \citep{Verlinde11}, or it being related to the
wave-nature of particles \citep{Stadtler+21}. The present contribution
suggests that departures at low accelerations from Newtonian/Einstein
gravitation might be already evident on the pc-scale. To advance our
understanding of this phenomenon, it will be important to achieve more
direct tests on the sub-pc scale. Three methods have been proposed to
achieve headway:
\begin{enumerate}
\item Using very wide binary systems to test the law of gravitation
  (\citealt{Hernandez2012b, Scarpa+17, Banik+21a,
    PittordisSutherland22}, but see \citealt{Clarke20,Loeb22a}).

\item Tracking the orbital motion of Proxima Cen with high-precision
astrometry to uncover the expected Milgromian departures from the
Newtonian trajectory \citep{BanikKroupa19b}.

\item Using an ensemble of small space craft to map out the
  acceleration field surrounding the Sun which will allow to detect
  the Milgromian departures from the Newtonian force law on the scale
  of the outer Solar System \citep{BanikKroupa19a}.
\end{enumerate}

\section*{Acknowledgements}

We thank an anonymous referee for very helpful suggestions, and
Franti\v{s}ek Dinnbier for useful comments. Tereza Jerabkova was an
ESA Research Fellow when this work was begun.  J\"org Dabringhausen,
Jaroslav Haas, Pavel Kroupa and Ladislav \v{S}ubr acknowledge support
through the Grant Agency of the Czech Republic under grant number
20-21855S and the DAAD-East-European-Exchange programme at the
University of Bonn. Benoit Famaey acknowledges funding from the Agence
Nationale de la Recherche (ANR projects ANR-18- CE31-0006 and
ANR-19-CE31-0017), and from the European Research Council (ERC) under
the European Union's Horizon 2020 Framework programme (grant agreement
number 834148).  Guillaume Thomas acknowledges support from the
Agencia Estatal de Investigaci\'on (AEI) of the Ministerio de Ciencia
e Innovaci\'on (MCINN) under grant FJC2018-037323-I.  The restaurant
Havelsk\'a Koruna (``knedl\'{i}karna'') in Prague was an essential
location throughout this work -- Knedl\'{i}k\r{u}m zdar!

The word 'pr{\'a}h' means threshold in Czech.  Prague (`Praha' in
Czech) derives its mythological name from pr{\'a}h alluding to the
location being the threshold to another world: on princess Libu\v{s}e
having a vision of a great new city at the Vltava, Praha was founded
with the pr\'ah to a new house.  We use this as a general term
referring to the true tidal threshold instead of calling it the
`Jacobi radius', which is only an approximation for perfectly
circular orbits.


\section*{Data Availability}
The results are based on calculated models and the observational data
are available as described within this manuscript. 


\bibliographystyle{mnras}
\interlinepenalty=10000
\bibliography{refs_tailpap}

\newpage
\vspace{20mm}
\noindent{\bf \Large Appendix~A: Estimation of the orbital
  eccentricity from one snapshot}

\noindent
The periods of the oscillation of the Galactocentric distance of the
open star clusters of interest are of the order of 100~Myr.  Within the
current age of the open star clusters ($\lesssim\,800\,$Myr) only a
few apo- and perigalacticon passages occur, and thus only a few
orbital eccentricities can be calculated directly from
Eq.~\ref{eq:ecc}. In order to better track the evolution of the radial
eccentricity in the simulations, it is desirable to estimate the
eccentricity at each time step, given only the actual position
$\vec{R}=(X,Y,Z)$ and velocity vector $\vec{V}=(V_{X},V_{Y},V_{Z})$ of
the star cluster centre in the Galactic rest frame.
 
The eccentricity at a certain snapshot of the orbit is defined by
\begin{equation}
  e_{\rm snap}
  =\frac{R_\mathrm{aposn}-R_\mathrm{perisn}}{R_\mathrm{aposn}+R_\mathrm{perisn}}\,,
\label{eq:esnap}
\end{equation}
where $R_\mathrm{perisn}$ is the perigalactic and $R_\mathrm{aposn}$ the
apo-galactic distance estimated in the following from the snapshot.
By having flat rotation curves, galaxies are observationally inferred
to be sourcing a logarithmic potential.
The Hamiltonian in polar coordinates of a particle with mass $m$ moving
in the $X$-$Y$-plane of a logarithmic potential with
circular velocity, $v_\mathrm{circ}$, is
\begin{equation}
  H = \frac{P_{R}^2}{2\,m}+\frac{P_\varphi^2}{2\,m\,R^2}+m\,v_{\rm circ}^2
  {\rm ln} (R)\,,
\end{equation}
where $R$ is the radial distance to the Galactic centre, $P_{R}$ the
radial momentum, and $P_\varphi$ the azimuthal momentum.
As $H$ is time independent $H$ is conserved. After 
mass-normalisation we get
\begin{equation}
\frac{H}{m} = \frac{(P_{R}/m)^2}{2}+\frac{(P_\varphi/m)^2}{2R^2}+v_{\rm circ}^2
{\rm ln}(R)
=\mathrm{const}\;.
\end{equation}
Because $P_\varphi$ is cyclic the $Z$-component of the angular momentum, $L_{\rm z}$, is
conserved,
\begin{equation}\label{eq:app_l0}
  P_\varphi/m = L_{\rm Z}/m = X \, V_{\rm Y} - Y \, V_{\rm X} = \mathrm{const}\,.
\end{equation}
Expressing the radial momentum by
\begin{equation}
  P_{\rm R}/m = \dot{R} = \frac{\vec{R}\bullet\vec{V}}{R}\;,
\end{equation}
we obtain for the two turning points $R_\mathrm{t}$ with $\dot{R}_\mathrm{t}=0$
\begin{equation}
\frac{\dot{R}^2}{2}+\frac{(L_{\rm Z}/m)^2}{2\,R^2} + v_{\rm circ}^2{\rm ln}(R)=
\mathrm{const.}=\frac{(L_{\rm Z}/m)^2}{2\,R_\mathrm{t}^2}+v_{\rm circ}^2{\rm ln}(R_\mathrm{t}).
\end{equation}
The solution of this trancendental equation requires a numerical
method. As there are two turning points (except for the case of a
cirular orbit) a numerical root finding algorithm requires appropriate
starting points for the iteration. These points can be obtained by
fitting this equation locally up to second order, and can be also used
as approximations of the turning points.

The Galactocentric distance, $R_{\rm t}$, of each turning point is
expressed as the sum of the current Galactocentric distance, $R$, and
the radial distance, $\delta R$, from the current position to the
Galactocentric distance of the respective turning point,
$R_{\rm t} = R + \delta R$ along the line Galactic centre to cluster,
where $\delta R$ is positive in case of the apo-galactic turning point
and negative in case of the peri-galactic turning point.  After
normalisation of the radial distance, $x = \delta R/R$, i.e.,
\begin{equation}\label{eq:app_R_x}
  R_{\rm t} = R(1+x)\,,
\end{equation}
we get
\begin{equation}\label{eq:app_full}
\frac{\dot{R}^2}{2}+\frac{(L_{\rm Z}/m)^2}{2\, R^2}=
\frac{(L_{\rm Z}/m)^2}{2\, R^2}(1+x)^{-2}+v_{\rm circ}^2
{\rm ln}(1+x) \, .
\end{equation}
This yields a solution for $x$, a positive one for the peri-galactic
and a negative for the apo-galactic turning point.

Given that $|\delta R| \ll R$, the full solution (Eq.~\ref{eq:app_full})
can be simplified: For a star cluster of current interest the
Galacticentric distance is $R\approx 8.3\,$kpc and $|\delta R|$ is of
the order of $\lesssim$ few hundred~pc. Thus, as $|x|\ll 1$, both
$x$-terms can be Taylor expanded,
\begin{equation}
(1+x)^{-2} = 1-2x+3x^2-\ldots\;\;\;,\mathrm{for} |x|<1\,,
\end{equation}
and
\begin{equation}
{\rm ln}(1+x) = x-\frac{1}{2}x^2+\ldots\;\;\;,\mathrm{for} |x|<1\,,
\end{equation}
and are truncated after the second order. Introducing the abbreviations
\begin{equation}\label{eq:app_eta_mu}
  \eta = \frac{\dot{R}^2}{2}
  \;\;\;\mathrm{and}\;\;\;
\mu=\frac{(L_{\rm Z}/m)^2}{2\,R^2}\,,
\end{equation}
a quadratic equation for $x$ emerges,
\begin{equation}
\eta+\mu = \mu(1-2x+3x^2) +v_{\rm circ}^2(x-\frac{1}{2}x^2)\;,
\end{equation}
with two solutions
\begin{equation}\label{eq:app_x_1_2}
x_{1/2}=\frac{-b\pm\sqrt{b^2-4\,a\eta}}{2\,a}\;,
\end{equation}
and parameters
\begin{equation}\label{eq:app_a_b}
a=\frac{v_{\rm circ}^2}{2}-3\mu
\;\;\;\mathrm{and}\;\;\;
b=2\mu-v_{\rm circ}^2\;.
\end{equation}
Using Eq.~\ref{eq:app_R_x} the first solution for $x$ having the plus
sign in front of the square root determines the Galactocentric
distance of the apo-galactic turning point, the second solution for
$x$ the Galactocentric distance of the peri-galactic turning point.

We test the accuracy of the approximating method for a logarithmic
potential with a rotational velocity of $v_{\rm circ}=250\,\rm km/s$
(this is for consistency with the models of Sec.~~\ref{sec:models}
and~\ref{sec:results}, while the data in Table~\ref{tab:clusterdata}
assume $v_{\rm circ}=220\,$km/s) and local radial velocity of
$10\,\rm km/s$ at a galactocentric distance of 8.5~kpc, which are
typical data of solar neighbourhood star clusters.  Assume the actual
position and velocity vectors in the Galactic rest frame are
$\vec{R}=(8500,\,0,\,0)\,$pc and $\vec{V}=(10.23,\,255.65,\,0)\,$\rm
pc/Myr.  The complete solution of Eq.~\ref{eq:app_full} leads to peri-
and apocentre distances of $R_\mathrm{aposn}=8740.4\,$pc and
$R_\mathrm{perisn}=8259.6\,$pc and an orbital eccentricity
$e_{\rm snap}=0.0282823$.

With the approximative solution (Eqs.~\ref{eq:app_l0},
\ref{eq:app_eta_mu}, \ref{eq:app_a_b}, \ref{eq:app_x_1_2}, and
\ref{eq:app_R_x}, in this order) we obtain
$R_\mathrm{aposn}=8746.2\,$pc and $R_\mathrm{perisn}=8265.1\,$pc
giving an orbital eccentricity of $e_{\rm snap}=0.0282812$.  In this
case the relative error of the eccentricity is $4\times 10^{-5}$.

These approximative solutions of the two turning points
\begin{equation}
  R_{\mathrm{t}_1} = R(1+x_1)\;\;\;,\;\;\;R_{\mathrm{t}_2} = R(1+x_2)
\end{equation}
can be used as starting points $R_1 = R_{\mathrm{t}_1}$
and $R_1 = R_{\mathrm{t}_2}$ for the Newton-Raphson root finding algorithm
with iterative step
\begin{equation}
  R_{n+1} = R_n - \frac{f(R_n)}{f^\prime(R_n)}\,,
\end{equation}
where the function, of which the roots are to be found, is
\begin{equation}
  f(R_n) = \frac{\dot{R}^2}{2}+\frac{(L_{\rm Z}/m)^2}{2\,R^2} + V_{\rm circ}^2{\rm ln}(R)-\frac{(L_{\rm Z}/m)^2}{2\,R_n^2}-V_{\rm circ}^2{\rm ln}(R_n)
\end{equation}
and its primitive
\begin{equation}
  f^\prime(R_n) =
  \frac{(L_{\rm Z}/m)^2}{R_n^3}-\frac{V_{\rm circ}}{R_n}\,.
\end{equation}
It turns out, that two or three steps are sufficient.

\bsp	
\label{lastpage}
\end{document}